\documentclass[%
 aip,
 amsmath,amssymb,
 reprint,%
]{revtex4-1}

\usepackage{graphicx,color}% Include figure files
\usepackage{dcolumn}% Align table columns on decimal point
\usepackage{bm}% bold math
\usepackage[utf8]{inputenc}
\usepackage[T1]{fontenc}
\usepackage{mathptmx}

\usepackage{braket}

\begin{document}

\preprint{AIP/123-QED}

\title{Perspective: Toward large-scale fault-tolerant universal photonic quantum computing}
% Force line breaks with \\

\author{S. Takeda}
 \altaffiliation[Also at ]{Institute of Engineering Innovation, Graduate School of Engineering, The University of Tokyo, 7-3-1 Hongo, Bunkyo-ku, Tokyo 113-8656, Japan.}
 \altaffiliation[Also at ]{JST, PRESTO, 4-1-8 Honcho, Kawaguchi, Saitama, 332-0012, Japan.}
 \email{takeda@ap.t.u-tokyo.ac.jp}
\author{A. Furusawa}%
 \email{akiraf@ap.t.u-tokyo.ac.jp}
 \affiliation{Department of Applied Physics, School of Engineering, The University of Tokyo,\\
7-3-1 Hongo, Bunkyo-ku, Tokyo 113-8656, Japan
}%

\date{\today}% It is always \today, today,

\begin{abstract}
Photonic quantum computing is one of the leading approaches to universal quantum computation.
However, large-scale implementation of photonic quantum computing has been hindered by its intrinsic difficulties,
such as probabilistic entangling gates for photonic qubits
and lack of scalable ways to build photonic circuits.
Here we discuss how to overcome these limitations
by taking advantage of two key ideas which have recently emerged.
One is a hybrid qubit-continuous variable approach
for realizing a deterministic universal gate set for photonic qubits.
The other is time-domain multiplexing technique to perform arbitrarily large-scale quantum computing
without changing the configuration of photonic circuits.
These ideas together will enable scalable implementation of universal photonic quantum computers
in which hardware-efficient error correcting codes can be incorporated.
Furthermore, all-optical implementation of such systems can increase the operational bandwidth beyond THz in principle,
utimately enabling large-scale fault-tolerant universal quantum computers with ultra-high operation frequency.
\end{abstract}

\maketitle

%\begin{quotation}
%\end{quotation}

\section{\label{sec:level1}Introduction}

\begin{figure*}
\includegraphics[width=0.8\linewidth,clip]{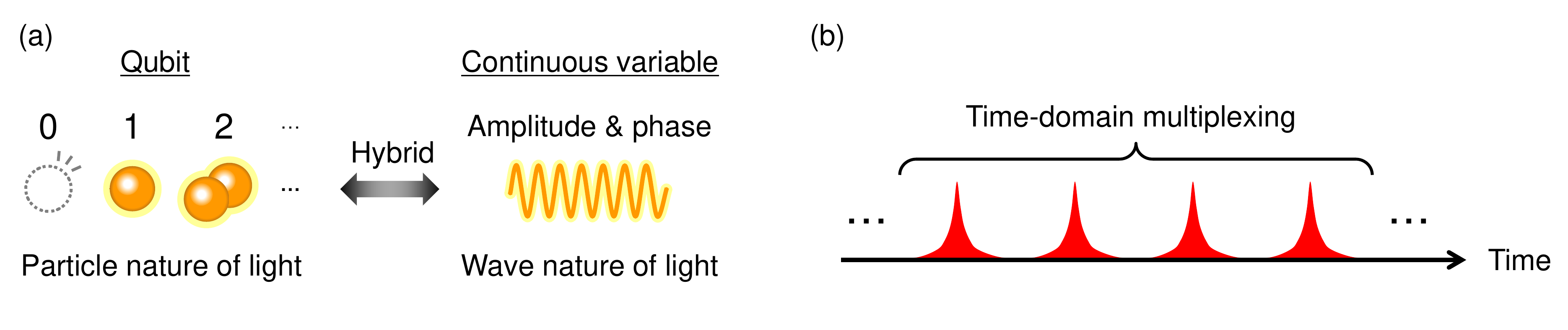}% Here is how to import EPS art
\caption{Two key ideas for large-scale photonic quantum computing.
(a) Hybrid approach. (b) Time-domain multiplexing.}
\label{fig:TwoKeyIdeas}
\end{figure*}

With the promise of performing previously impossible computing tasks, quantum computing has received a lot of public attention.
Today quantum processors are implemented with a variety of physical systems~\cite{10Nielsen,10Ladd},
and quantum processors with tens of qubits have been already reported~\cite{18Yuanhao,18Friis}.
The leading physical systems for quantum computing include
superconducting circuits, trapped ions, silicon quantum dots and so on.
However, scalable implementation of fault-tolerant quantum computers is still a major challenge for any physical system
due to the inherent fragility of quantum states.
In order to protect fragile quantum states from disturbance,
most of these physical systems need to be fully isolated from external environment
by keeping the systems at cryogenic temperature in dilution refrigerators 
or in vacuum environment inside metal chambers.

In contrast, photonic systems have several unique and advantageous features.
First, quantum states of photons are maintained without vacuum or cooling systems
due to their extremely weak interaction with the external environment.
In other words, photonic quantum computers can work in atmospheric environment at room temperature.
Second, photons are an optimal information carrier for quantum communication,
since they propagate at the speed of light and offer large bandwidth for a high data transmission capacity.
Therefore, photonic quantum computers are completely compatible with quantum communication.
The large bandwidth of photons also provides high-speed (high clock frequency) operation in photonic quantum computers.
These advantageous features, together with mature technologies
to prepare and manipulate photonic quantum states
with linear optical elements and nonlinear crystals,
have made photonic systems one of the leading approaches to building quantum computers~\cite{12Pan,07OBrien,05Braunstein,09OBrien}.

However, these unique features of photons, at the same time,
introduce intrinsic difficulties in quantum computing.
Since photons do not interact with each other,
it is difficult to implement two-qubit entangling gates which require interaction between photons.
In addition, since photons propagate at the speed of light and do not stay at the same position,
many optical components have to be arranged along the optical path of photons
to sequentially process photonic qubits.
As a result, large-scale photonic circuits are required for large scale quantum computing, which is not efficient.
It is also pointed out that photonic circuits are often designed to perform specific quantum computing tasks,
and the design of the circuits has to be modified to perform different tasks.
In the case of general classical computers,
users only need to change the program (software), not the hardware, to perform different computing tasks.
However, standard photonic quantum computers don't have such programmability,
and users are required to change the circuit (hardware) itself.
These problems are unique to the photonic systems.
For other systems like superconducting circuits and trapped ions,
the physical systems are processed by injecting microwave or laser pulses into the systems from external devices
(not by building any physical circuits like photonic circuits).
In this case, it is easy to sequentially process qubits only by sequentially injecting the pulses
and reprogram the quantum computers only by changing the control sequence of the pulses.

Despite these intrinsic difficulties,
promising routes to large-scale photonic quantum computing
have recently emerged thanks to the progress in theory and technology.
In this perspective, we explain these promising routes
by focusing on two innovative ideas in photonic quantum computing.
The first idea is a ``hybrid'' approach combining two complementary approaches.
As shown in Fig.~\ref{fig:TwoKeyIdeas}(a),
photonic quantum computing has traditionally been developed
by two approaches, qubits and continuous variables (CVs),
each exploiting only one aspect of the wave-particle duality of light.
However, recent progress in combining these two approaches
has shown that it is more powerful to take advantage of the both aspects~\cite{11vanLoock,11Furusawa,15Andersen}.
This hybrid qubit-CV approach potentially enables deterministic and robust quantum computing,
which is hard to achieve by either qubit of CV approach alone.
The second idea is time-domain multiplexing in Fig.~\ref{fig:TwoKeyIdeas}(b),
where many units of information are encoded in a string of optical pulses sharing the same optical path.
This idea itself has already been used to efficiently increase the number of optical modes
for quantum computation and communication.
However, it has recently been discovered that
the time-domain multiplexing is even more powerful
when combined with specific quantum computing schemes;
time-domain multiplexed one-way quantum computation~\cite{13Yokoyama,16Yoshikawa}
and a loop-based architecture for photonic quantum computing~\cite{17Takeda,18Takeda}.
These two schemes enable us to programmably perform arbitrarily large-scale quantum computing
without changing the configuration of optical circuits.
Recent experiments based on these schemes~\cite{13Yokoyama,16Yoshikawa,18Takeda}
clearly show superior performance to conventional schemes in scaling up photonic quantum computing.

These schemes also offer several unique advantages to photonic quantum computing. For example, nonlinearity is often required for photonic quantum gates, but nonlinear optical systems often introduce unwanted distortion of optical pulses and crosstalk between pulses. In contrast, the schemes presented in this perspective are based only on linear optical components, and nonlinearity is fed from external sources as ancillary optical pulses only when required~\cite{11Marek,16Miyata,18Marek}. This feature is advantageous to scale up quantum computers without introducing any additional sources of errors. These schemes are also compatible with hardware-efficient error correction codes where one optical pulse represents one logical qubit~\cite{01Gottesman,99Cochrane,13Leghtas,16Michael},
in contrast to standard codes where many pulses represent one logical qubit~\cite{95Shor,96Steane}. Finally, these scheme can in principle be realized all-optically~\cite{99Ralph}, i.e., without using electrical circuits. Therefore, electronics never limit the bandwidth of the system, ultimately enabling ultra-large bandwidth (ultra-fast clock frequency) of orders of THz in principle.

Below, we describe the two key ideas in Fig.~\ref{fig:TwoKeyIdeas}
for large-scale quantum computing in more detail.
Section~\ref{sec:hybrid} deals with the idea of the hybrid approach.
Here, we first give a brief review over existing qubit and CV approaches,
and then introduce the advantages and recent development of the hybrid approach.
Section~\ref{sec:timedomain} deals with the idea of time-domain multiplexing.
Here, we explain the two schemes for large-scale quantum computing with time-domain multiplexing,
while mentioning related experimental progress and technical challenges.
Finally, Sec.~\ref{sec:conclusion} summarizes this perspective.

\section{Hybrid quantum computing}\label{sec:hybrid}

There have been two major approaches for photonic quantum computing,
qubits and CVs.
Here we first review these two approaches
and then describe why and how the hybrid approach is promising.
The comparison between qubit and CV quantum information processing
is summarized in Table~\ref{tab:qubit_CV}.

\begin{table*}
\caption{Comparison between qubit and CV photonic quantum information processing (QIP).}
\label{tab:qubit_CV}
\begin{tabular}{p{2cm}p{5cm}p{9cm}}
\hline
 & Qubit QIP & Continuous-variable QIP \\
\hline \hline \noalign{\smallskip}
Carrier & Degrees of freedom of a photon  & Quadratures of a light field  \\ \hline
Basis & Photon number basis: $\{\ket{n}\}$ & Quadrature basis: $\{\ket{x}\}$ or $\{\ket{p}\}$\\ \hline
Encoding & $\ket{\psi}=\alpha\ket{1}\ket{0}+\beta\ket{0}\ket{1}$ & $\ket{\psi}=\int_{-\infty}^{\infty}\psi(x)\ket{x}dx$ \\ \hline
%Source & Photons by PDC (weak pump) & Squeezed light by PDC (strong pump)\\ \hline
%Detector & Photon detector (measures $\hat{n}$) & Homodyne detector (measures $\hat{x}$ or $\hat{p}$) \\ \hline
Easy gates & One-qubit rotation gate & Gaussian gate (displacement, phase shift, beam splitter, squeezing) \\ \hline
Difficult gates & Two-qubit gate (\textit{e.g.} CNOT gate) & Non-Gaussian gate (\textit{e.g.} cubic phase gate) \\ \hline
%\svhline
\end{tabular}
\end{table*}

\subsection{Qubit approach}

In classical digital information processing,
the basic unit of information is a bit, which takes only one of two values, `0' or `1'.
The basic unit of operation on bits is called logic gates,
which transform input bits to output bits according to given rules.
Examples of the logic gates are one-bit NOT gate and two-bit AND gate,
and it is known that arbitrary logic operation can be constructed by NOT and AND gates.

When it comes to quantum computing,
the quantum analogue of the classical bit is called a quantum bit or qubit, which is a superposition of two states, $\ket{\bar0}$ and $\ket{\bar1}$, given by $\ket{\psi}=\alpha\ket{\bar0}+\beta\ket{\bar1}$
($|\alpha|^2+|\beta|^2=1$, $\bar0$ and $\bar1$ denote logical `0' or `1').
Here the information is encoded in the complex coefficients $\alpha$ and $\beta$.
For qubits,
two types of quantum logic gates are necessary to construct arbitrary quantum computation~\cite{10Nielsen}.
One is one-qubit rotation gates to convert the coefficients $\alpha$ and $\beta$,
corresponding to the rotation of the qubit in the Bloch sphere.
The other is two-qubit entangling gates,
such as a controlled-NOT gate
which flips the state of a target qubit ($\ket{\bar0}\leftrightarrow\ket{\bar1}$)
only if the control qubit is in the state $\ket{\bar1}$.

In photonic quantum information processing, information of a qubit is typically encoded in any of several degrees of freedom of a single photon, such as polarization, propagation direction (path), and arrival time~\cite{12Pan,07OBrien,09OBrien}.
For example, polarization of a single photon can represent a qubit by
$\alpha\ket{\bar0}+\beta\ket{\bar1}=\alpha\ket{1}_\text{V}\ket{0}_\text{H}+\beta\ket{0}_\text{V}\ket{1}_\text{H}$,
where ``V'' and ``H'' denote vertical and horizontal polarization, respectively,
and 0 and 1 represent the number of photons.
In this polarization encoding,
one-qubit gates physically mean the rotation of polarization of a photon,
which can be implemented easily with a series of wave plates.
The main difficulty in photonic quantum computation lies in the implementation of two-qubit gates.
For example, the photonic controlled-NOT gate
physically means that the polarization of a target photon is flipped
only if a control photon is horizontally polarized.
Here, flipping polarization is equivalent to introducing a $\pi$ phase shift between two diagonal polarizations.
Therefore, the operation of the controlled-NOT gate
corresponds to a $\pi$ phase shift of a photon conditioned by the existence of another photon.
This phenomenon can be realized by an optical Kerr effect;
it is a third-order nonlinear effect which varies the refractive index of a medium depending on the input light power, thereby introducing a phase shift.
However, no known nonlinear optical material has a nonlinearity strong enough to implement this
conditional $\pi$ phase shift by single photons.

At an early stage of developing photonic quantum computers,
a lot of effort has been devoted to theoretical and experimental investigation on
how to efficiently implement the photonic controlled-NOT gate.
In 2001, Knill, Laflamme, and Milburn (KLM) have discovered a method for scalable
photonic quantum computation with only
single photon sources, detectors, and linear optics (without any nonlinear medium)~\cite{01Knill}.
They proposed a probabilistic controlled-NOT gate
based on ancillary photons, beam splitters, and photon detection.
Furthermore, the success probability is shown to be increased
based on the technique of quantum teleportation~\cite{93Bennett,99Gottesman}, a process whereby an unknown state of a qubit
is transferred to another qubit.
However, quantum teleportation of photonic qubits is fundamentally probabilistic by itself~\cite{97Bauwmeester}
because so-called Bell measurement required for the teleportation protocol cannot be deterministic with linear optics~\cite{01Calsamiglia}.
In order to avoid this probabilistic nature and make the controlled-NOT gate deterministic,
infinitely large number of ancillary photons are required.
Therefore, deterministic controlled-NOT gate based on this approach
is still too demanding, even though the KLM scheme is in principle scalable.

The proposal by KLM was followed by several experimental demonstrations of
probabilistic two-qubit gates~\cite{03OBrien,03Pittman,04Gasparoni,11Okamoto}.
Even though these two-qubit gates are probabilistic,
a set of quantum logic gates necessary for universal photonic quantum computation has become completed.
This enabled several proof-of-principle demonstrations of small-scale quantum algorithms with photonic quantum computers,
such as Shor's factoring algorithm~\cite{07Lu, 07Lanyon}, quantum chemistry calculations~\cite{10Lanyon,14Peruzzo},
and quantum error correction algorithms~\cite{05Pittman,05OBrien,08Lu}.
In addition, an alternative quantum computation scheme called one-way quantum computation~\cite{01Raussendorf}
has been proposed in 2001 and shown to have several advantages~\cite{04Nielsen}.
In this scheme, a large-scale entangled state called a cluster state is 
prepared first by applying entangling gates to qubits.
This state serves as a universal resource for quantum computation, and 
a suitable sequence of single-qubit measurements on the state
can perform any quantum computation
(the idea of one-way quantum computation is described in more detail in Sec.~\ref{sec:cv}).
This proposal was soon followed by experimental demonstrations~\cite{05Walther,05Kiesel,07Prevedel}.

However, in any cases,
the low success probability of the two-qubit gates
makes larger-scale quantum computation almost impractical.
In fact, probabilistic two-qubit gates do not enable scalable quantum computation
since the probability that a quantum computing task succeeds decreases exponentially
with the number of the two-qubit gates.
Deterministic two-qubit gates are also being pursued by other approaches,
especially by interacting single photons with a single atom in high-finesse optical cavities~\cite{95Turchette,04Duan,16Hacker}.
However, this approach also introduces additional difficulties for
satisfying strong atom-photon coupling condition in a cavity,
converting freely-propagating photons to an intra-cavity photons with high efficiency,
and avoiding spectral distortion of photons due to nonlinearity.
Therefore, the approaches based on only linear optics still seem to be the leading approach.

\subsection{CV approach}\label{sec:cv}

\begin{figure*}
\includegraphics[width=0.6\linewidth,clip]{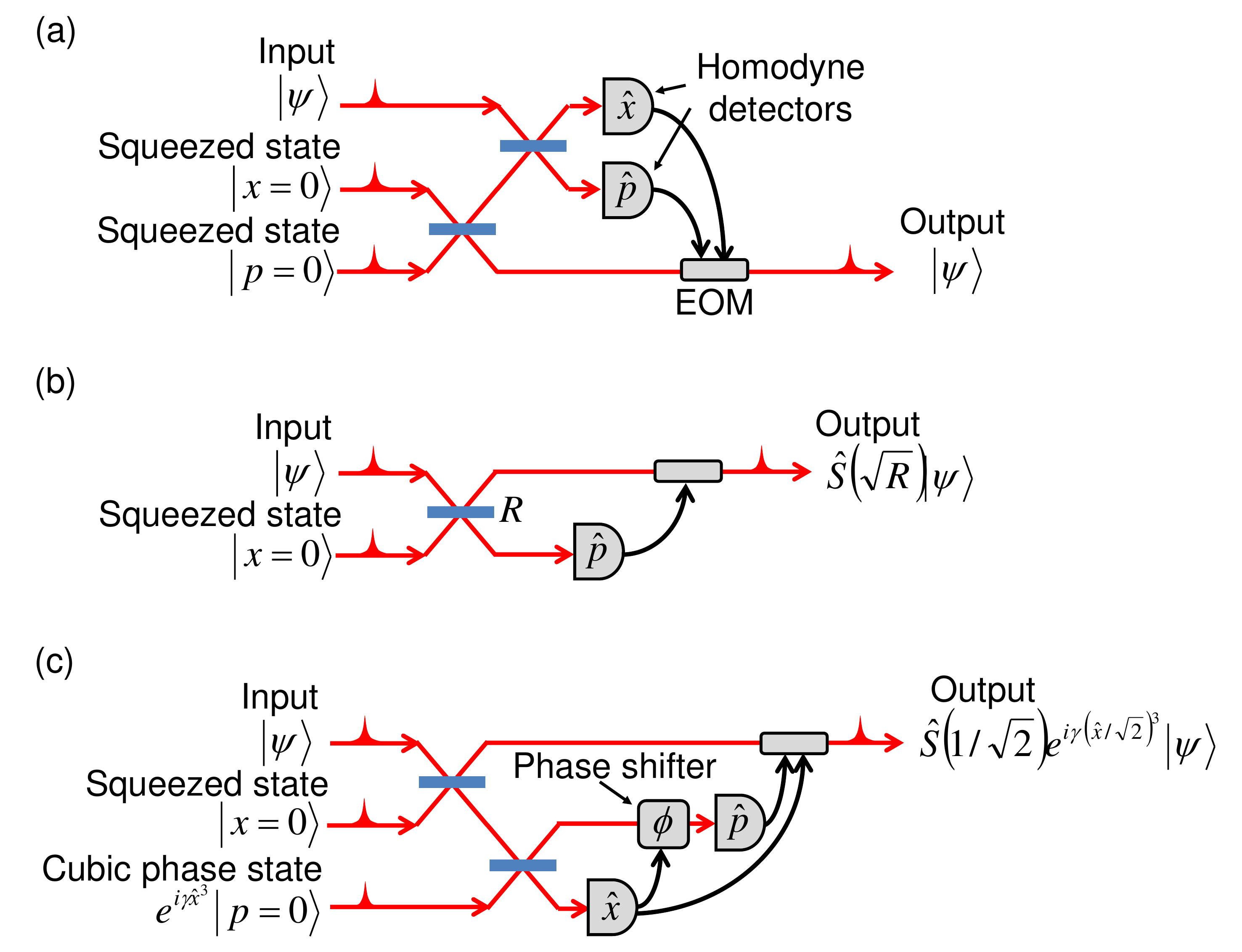}% Here is how to import EPS art
\caption{CV quantum teleportation and its extension to quantum gates.
(a) CV quantum teleportation~\cite{94Vaidman,98Furusawa}.
(b) Teleportation-based squeezing gate~\cite{05Filip}.
(c) Teleportation-based cubic phase gate~\cite{16Miyata}.
All beam splitters in (a) and (c) have 50\% reflectivity.
$\hat{S}(y)$ ($y\in\mathbb{R}$)
is a squeezing operator transforming quadrature operators $\hat{x}$ and $\hat{p}$ to
$\hat{S}^\dagger(y)\hat{x}\hat{S}(y)=y\hat{x}$ and
$\hat{S}^\dagger(y)\hat{p}\hat{S}(y)=\hat{p}/y$, respectively.
}
\label{fig:CVgates}
\end{figure*}

In the case of qubits, the unit of quantum information is a superposition of two discrete values `0' and `1'.
Such information is encoded in single photons,
and the state of photonic qubits can be described in the discrete photon-number basis.
There is an alternative approach~\cite{05Braunstein} where the unit of quantum information is a superposition of any continuous real value $x$ (CVs).
This type of information can be represented by utilizing continuous degrees of freedom of light,
such as amplitude and phase quadratures $\hat{x}$ and $\hat{p}$ of a field mode.
In this case, quantum information can be described by
$\ket{\psi}=\int^\infty_{-\infty}\psi(x)\ket{x}dx$,
where $\ket{x}$ is an eigenstate of $\hat{x}$ ($\hat{x}\ket{x}=x\ket{x}$)
and the information is encoded in the function $\psi(x)$.
Note that this state can also be expanded in the photon number basis as
$\ket{\psi}=\sum_{n=0}^\infty c_n\ket{n}$ with $c_n=\braket{n|\psi}$.
Therefore, quantum computing with photonic qubits uses only the zero- and one-photon subspace
of the originally infinite dimensional Hilbert space of a light mode,
and CV quantum computing includes qubit quantum computing as a special case.

Quantum logic gates for CVs can be written as a unitary operator $\hat{U}$
which transforms the initial superposition of CVs
$\ket{\psi}=\int^\infty_{-\infty}\psi(x)\ket{x}$
into another superposition of CVs
$\hat{U}\ket{\psi}=\int^\infty_{-\infty}\psi(x)\hat{U}\ket{x}$.
In order to construct an arbitrary unitary transformation
$\hat{U}=\exp(-i\hat{H}t/\hbar)$, Hamiltonians $\hat{H}$ of arbitrary polynomials of
$\hat{x}$ and $\hat{p}$ are required.
Unitary transformations which involves Hamiltonians
of linear or quadratic in $\hat{x}$ and $\hat{p}$ are called Gaussian gates.
It is known that an arbitrary Gaussian gate and at least one non-Gaussian gate which involves a higher order Hamiltonian
are required to construct arbitrary unitary transformation (universal CV quantum computation)~\cite{99Lloyd}.

In photonic systems, easily implementable gates include a displacement operation by
amplitude and phase modulation of optical beams with an electro-optic modulator (EOM)
($\hat{H}\propto a\hat{x}-b\hat{p}$),
a phase shift of optical beams
($\hat{H}\propto \hat{x}^2+\hat{p}^2$),
and an interference of two optical beams at a beam splitter
($\hat{H}\propto \hat{x}_1\hat{p}_2-\hat{p}_1\hat{x}_2$; 1 and 2 represent the mode index).
An arbitrary Gaussian gate also requires
a squeezing gate based on a second-order nonlinear effect
($\hat{H}\propto \hat{x}\hat{p}+\hat{p}\hat{x}$).
In addition, as an example of non-Gaussian gates,
a cubic phase gate based on a third-order nonlinear effect is required
($\hat{H}\propto \hat{x}^3$).
The last two gates require nonlinear effects,
and especially the cubic phase gate requires third order nonlinearity which
is hard to achieve for very weak light at the quantum regime;
This difficulty is the same as in the case of qubits
where the controlled-NOT gate requires impractically gigantic third order nonlinearity (Kerr effect).
Therefore, CV quantum computing seems to
share the same difficulty as in qubit quantum computing at first glance.

\begin{figure*}
\includegraphics[width=0.9\linewidth,clip]{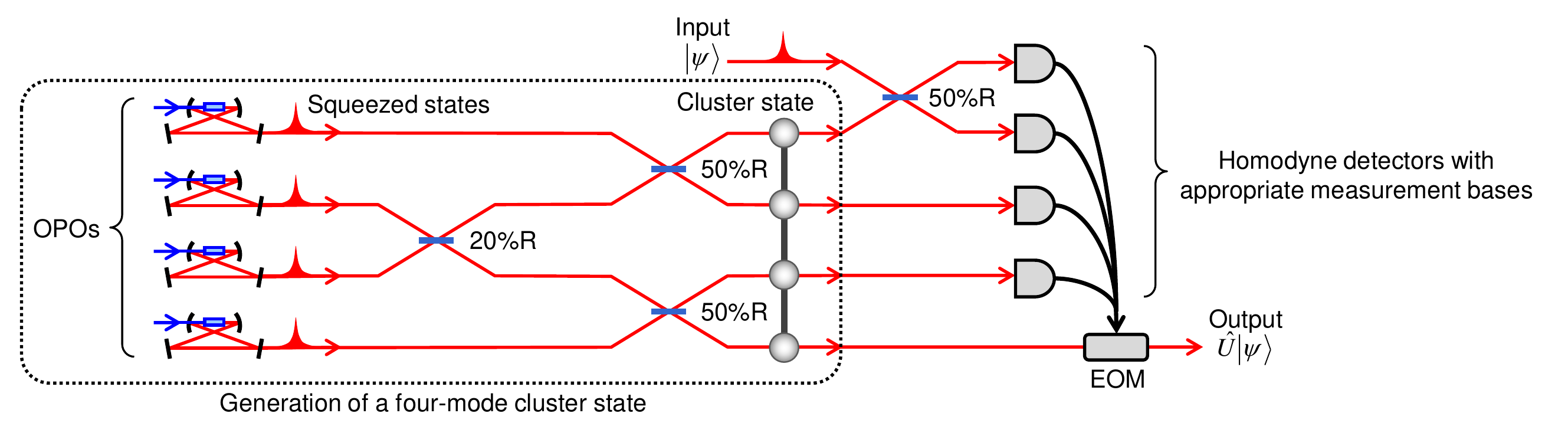}% Here is how to import EPS art
\caption{One-way quantum computation with a four-mode cluster state~\cite{11Ukai1}.
OPO, optical parametric oscillator; R, reflectivity.
Spheres and links between them represent optical modes and entanglement, respectively.
}
\label{fig:ClusterComputation}
\end{figure*}

However, the important advantage of CVs is that 
quantum logic gates based on quantum teleportation~\cite{94Vaidman,98Furusawa},
which is inevitably probabilisitic in the qubit approach,
can be implemented deterministically.
Figure~\ref{fig:CVgates}(a) shows the basic circuit for CV quantum teleportation,
which transfers an unknown quantum state $\ket{\psi}$
from the input port to the output port.
In this circuit, two ancillary squeezed light beams are first generated by
squeezing one quadrature (for example, $\hat{x}$) of a vacuum state in a second-order nonlinear medium
so that its quantum fluctuation ($\Delta x$) is reduced below the vacuum fluctuation
(infinite squeezing $\Delta x\to0$ gives the state $\ket{x=0}$).
Except for this part,
the circuit itself is linear; the input beam is first mixed with two squeezed light beams by beam splitters,
then two beams are sent to homodyne detectors measuring $\hat{x}$ and $\hat{p}$,
and finally the last beam is displaced with an EOM
by an amount determined by the measurement results.
This CV teleportation always succeeds
since all the procedures,
including the preparation of ancillary squeezed light beams,
simultaneous measurement of $\hat{x}$ and $\hat{p}$ (Bell measurement), and operation depending on the measurement results,
are deterministic.
This is in contrast to the photonic qubit teleportation,
which is always probabilistic since the qubit-version of Bell measurement is probabilistic in principle~\cite{01Calsamiglia}.
However, the major disadvantage of the CV teleportation is limited transfer fidelity,
since perfect fidelity requires infinite squeezing and thus infinite energy
(this disadvantage can be overcome by taking the hybrid approach
and introducing appropriate error-correcting codes, as described in Sec.~\ref{subsec:hybrid}).

This CV teleportation circuit is a one-input one-output identity operation where the output state is equivalent to the input state.
However, once the types of ancillary states and/or the configuration of measurement and feedforward operations are slightly altered,
this circuit can be transformed into a one-input one-output quantum gate which applies a certain unitary operation
to the input state and sends it to the output port~\cite{03Bartlett}.
This is the idea of quantum logic gates based on quantum teleportation.
Typical examples of such gates are
the squeezing gate~\cite{05Filip} in Fig.~\ref{fig:CVgates}(b)
and the cubic phase gate~\cite{11Marek,16Miyata} in Fig.~\ref{fig:CVgates}(c).
If these gates have to be directly performed on the input state,
the state has to be sent to nonlinear materials with a sufficiently strong second or third order nonlinear effect.
However, especially the third order nonlinear effect is too small for very weak light at the quantum regime.
In the teleportation-based gates,
the task of directly applying nonlinear effects to arbitrary states is replaced by an easier task of preparing specific ancillary states prior to the actual gate.
In this case, the ancillary states may be prepared in probabilistic (heralding) ways;
the production of the ancillary state can be repeated until it succeeds,
and when the state is produced, the state is stored in optical quantum memories and
subsequently injected into the teleportation-based gates at a proper time.
As a result, the nonlinear effect is deterministically teleported from the ancillary state to the input state,
thus one can indirectly apply the gate to an arbitrary input state in a deterministic way. 
The same method can be extended to other non-Gaussian gates, such as higher-order phase gates~\cite{18Marek} ($\hat{H}\propto \hat{x}^n$ with $n\ge4$).
In this way, the squeezing gate and even non-Gaussian gates can be performed deterministically,
and thus all gates necessary for universal CV quantum computation can be deterministically achieved.

After theoretical proposals of  these teleportation-based CV gates~\cite{05Filip,11Marek,16Miyata,18Marek},
several teleportation-based Gaussian gates,
such as a squeezing gate~\cite{07Yoshikawa}
and a quantum-non-demolition sum gate~\cite{08Yoshikawa}($\hat{H}\propto\hat{x}_1\hat{p}_2$),
have been experimentally demonstrated.
Teleportation-based non-Gaussian gates have not been demonstrated yet
since they require exotic ancillary states
and more complicated configuration of measurement and feedforward operations~\cite{11Marek,16Miyata,18Marek}.
However, steady progress has been made towards the realization of the cubic phase gate,
such as preparation of approximated cubic phase states~\cite{13Yukawa},
development of a quantum memory for such states~\cite{13Yoshikawa, 18Hashimoto},
and evaluation of a feedforward system for the cubic phase gate~\cite{14Miyata}.
Therefore, all the components essential for the deterministic cubic phase gate
have become available in principle, awaiting their future integration.

As an alternative approach,
the one-way quantum computation scheme based on CV quantum teleportation~\cite{06Menicucci,07vanLoock}
is also recognized as a promising route to
perform universal quantum computation with CVs.
The CV teleportation circuit in Fig.~\ref{fig:CVgates}(a) can apply quantum gates to the input state
only by changing the measurement basis, without changing the ancillary states.
Therefore, one can repeatedly apply quantum gates by cascading many CV teleportation circuits
and choosing an appropriate measurement basis for each step.
This cascaded teleportation circuit is the essence of one-way quantum computation,
which can be understood in the following (Fig.~\ref{fig:ClusterComputation}).
First, a specific multimode entangled state (cluster state) is prepared by mixing squeezed light beams.
Then the input state is coupled to the cluster state,
and the quantum computation is performed
by repeated measurement and feedforward operations.
The advantage of one-way quantum computation is that
different quantum computing tasks can be performed by simply choosing a different measurement basis,
without changing the setup for preparing cluster states.
In this case, non-Gaussian gates, such as a cubic phase gate,
can be implemented by performing photon counting measurement to the cluster state~\cite{06Menicucci,09Gu}
or injecting ancillary cubic phase states~\cite{18Alexander}.
Based on these proposals,
generation of small-scale CV cluster states~\cite{08Yukawa} and
basic quantum gates based on the cluster states~\cite{11Ukai1,11Ukai2}
have already been experimentally demonstrated.

\subsection{Hybrid qubit-CV approach}\label{subsec:hybrid}

\begin{figure*}
\includegraphics[width=0.8\linewidth,clip]{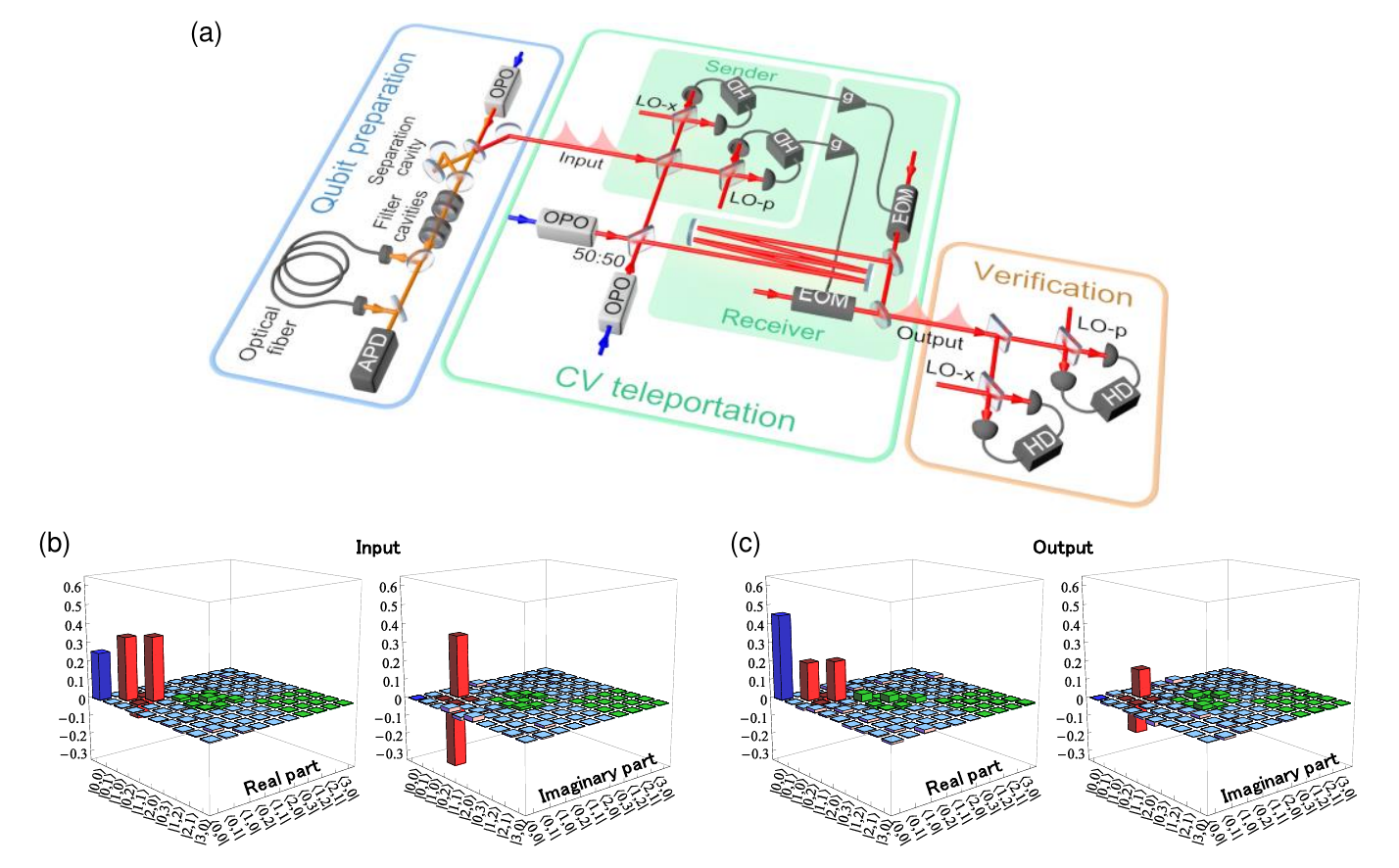}% Here is how to import EPS art
\caption{CV quantum teleportation of photonic qubits~\cite{13Takeda2}.
(a) Experimental setup.
(b) Reconstructed density matrix of an input qubit $\ket{\psi}=(\ket{0,1}-i\ket{1,0})/\sqrt{2})$.
(c) Reconstructed density matrix of an output qubit.
Density matrices are expanded in the photon number basis,
$\hat{\rho}=\sum_{k,l,m,n=0}^{\infty}\rho_{klmn}\ket{k,l}\bra{m,n}$.
APD, avalanche photo diode; HD, homodyne detector; LO, local oscillator.}
\label{fig:HybridTeleportation}
\end{figure*}

Until recently, the qubit and CV approaches to photonic quantum computing
have been pursued separately.
As mentioned above,
the advantage of the CV approach lies in deterministic teleportation-based gates
which is essential for scalable photonic quantum computation.
However, teleportation-based gates have limited fidelity due to finite squeezing,
thereby destroying fragile CV quantum information with only a few steps.
In contrast, information of qubits is more robust and
can be protected against errors by means of several error-correcting codes~\cite{95Shor,96Steane}.
Therefore, the best strategy should be a hybrid approach~\cite{11vanLoock,15Andersen,11Furusawa}
which combines robust qubit encoding and deterministic CV gates.
Below we focus on this type of approach,
but it should be noted that
there are several types of hybrid qubit-CV approach,
such as combination of CV encoding and qubit operations~\cite{13Andersen}.

Let us more specifically discuss
how to implement the universal gate set for qubits from CV gates.
In general, CV quantum gates can be applied to any quantum state $\ket{\psi}$,
let alone single-photon based qubits
$\alpha\ket{1}\ket{0}+\beta\ket{0}\ket{1}$.
One-qubit gates for such states can be directly performed with only beam-splitter operations and phase shifts.
As an example of two-qubit entangling gates, the controlled-phase gate
corresponds to the unitary transformation
$\exp(i\pi\hat{a}_1^\dagger\hat{a}_1\hat{a}_2^\dagger\hat{a}_2)\ket{k}\ket{l}=(-1)^{kl}\ket{k}\ket{l}$
($k,l=0,1$).
This unitary transformation
is known to be decomposed into a sequence
of several cubic phase gates and other Gaussian gates~\cite{13Sefi}.
Since each CV gate can be deterministically performed,
a deterministic controlled-phase gate can be implemented in principle.

Recently much progress has been made to realize the hybrid qubit-CV approach.
The important first step should be the combination of photonic qubits and CV teleportation.
However, this combination had been not straightforward for the following reason.
Photonic qubits are usually defined in pulsed wave packet modes
and thus have broad frequency spectrums;
such qubits are not compatible with the conventional CV quantum teleportation device,
which works only for narrow sideband frequency modes~\cite{98Furusawa}.
This technical hurdle has been overcome by development of a broadband CV teleportation device~\cite{11Lee,12Takeda}
and a narrowband photonic qubit compatible with the teleportation device~\cite{13Takeda1}.
Finally these technologies were combined,
thereby enabling deterministic quantum teleportation of photonic qubits for the first time~\cite{13Takeda2} (Fig.~\ref{fig:HybridTeleportation}).
Later several related hybrid teleportation experiments have been reported, such as
CV teleportation of two-mode photonic qubit entanglement~\cite{15Takeda} and
teleportation-based deterministic squeezing gates on single photons~\cite{14Miwa}.

CV quantum gates are applicable not only to single photons,
but also to any quantum states with higher photon number components.
Therefore, the hybrid approach is not restricted to single-photon based qubits;
we can take advantage of the infinite dimensional Hilbert space of CVs
to encode quantum information beyond qubits (such as qudits).
This possibility is already demonstrated in an experiment
where two-photon two-mode qutrits $\alpha\ket{2}\ket{0}+\beta\ket{1}\ket{1}+\gamma\ket{0}\ket{2}$
were teleported by the CV teleportation device~\cite{17Okada}.
The infinite dimensional Hilbert space also enables us to
redundantly encode a qubit in a single optical mode for quantum error correction.
Examples of such error correction codes are
the Gottesman-Kitaev-Preskill (GKP) code~\cite{01Gottesman}, cat code~\cite{99Cochrane,13Leghtas}, and binomial code~\cite{16Michael}.
The advantage of these codes are as follows.
For typical error correcting codes~\cite{95Shor,96Steane},
one logical qubit is encoded in many physical qubits to obtain such redundancy.
However, this approach is technically challenging for several reasons.
First, the number of possible errors
increases with the number of qubits, and the correction of errors become more difficult.
Furthermore, such encoding requires nonlocal
gates between many physical qubits for logical operations.
Finally, preparation of such a large number of qubits is still a hard task by itself.
Compared to such typical error correction codes, the GKP, cat, and binomial codes only use
a single optical mode for encoding one logical qubit, making the logical operation and error correction much simpler
and enabling hardware-efficient implementation.

\begin{figure*}
\includegraphics[width=0.6\linewidth,clip]{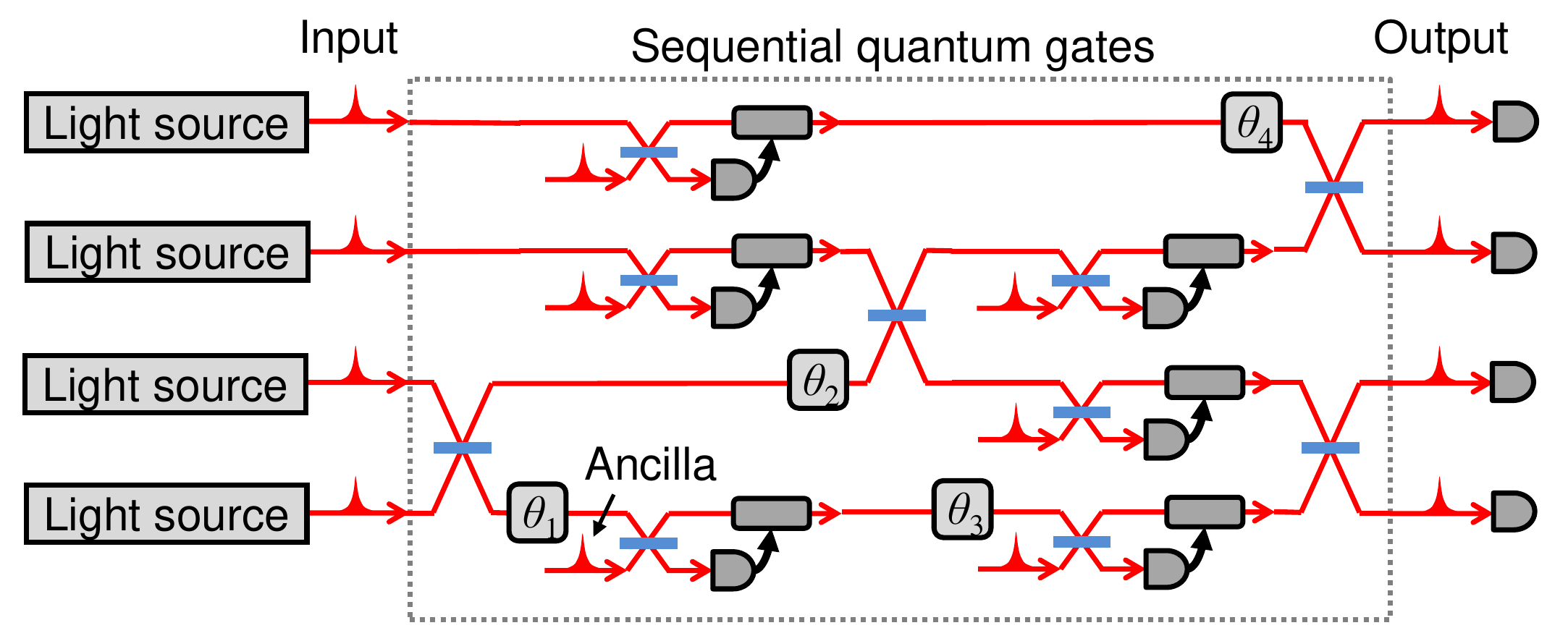}% Here is how to import EPS art
\caption{Typical photonic circuit for quantum computing.}
\label{Fig:ConventionalCircuit}
\end{figure*}

In photonic systems, 
the dominant error channel is photon loss.
Among the proposed error correction codes described above,
the GKP code is shown to significantly outperform other codes under the photon loss channel in most cases~\cite{18Albert}.
In this code, logical $\ket{\bar 0}$ and $\ket{\bar 1}$ states are defined as superpositions of $\hat{x}$-eigenstates,
$\ket{\bar j}=\sum_{s\in\mathbb{Z}}\ket{x=\sqrt{\pi}(2s+j)}$ ($j=0, 1$).
This qubit can be protected against
sufficiently small phase-space displacement errors
and photon-loss errors~\cite{19Noh}.
Furthermore, error correction and
logical qubit operations can be easily implemented with CV gates based only on homodyne detection~\cite{01Gottesman}.
Although fidelity of CV teleportation-based gates is limited by finite squeezing, 
it has been proven that
there is a fault-tolerant threshold for squeezing level (conservative upper bound is 20.5 dB)
for quantum computation with the GKP qubits and CV one-way quantum computation~\cite{14Menicucci,19Walshe}.
Therefore, fault-tolerant quantum computation is possible with proper encoding of a qubit
and finite level of squeezing.

Thus far there has been much
experimental effort to increase the squeezing level,
and up to 15 dB of optical squeezing has been reported~\cite{16Vahlbruch}.
At the same time, theoretical proposals
to reduce the fault-tolerant threshold have also been made recently~\cite{17Fukui,18Fukui}.
The next key technology in the hybrid approach should be the production of the GKP states
and implementation of quantum error correction with these states.
Several methods to generate approximated GKP states in the optical regime are known~\cite{10Vasconcelos,18Weigand,17Motes,19Eaton,18Arrazola},
awaiting experimental demonstration.

\section{Strategy for large-scale quantum computing}\label{sec:timedomain}

Here we explain promising architectures for large-scale photonic quantum computing
which can perform sequential CV gates on many qubits.
We first describe problems of typical architectures for photonic quantum computing.
We then introduce two specific architectures,
time-domain multiplexed one-way quantum computation
and loop-based architectures for sequential CV gates,
and discuss their technical challenges.

\subsection{Typical architecture for photonic quantum computing}

\begin{figure*}
\includegraphics[width=0.8\linewidth,clip]{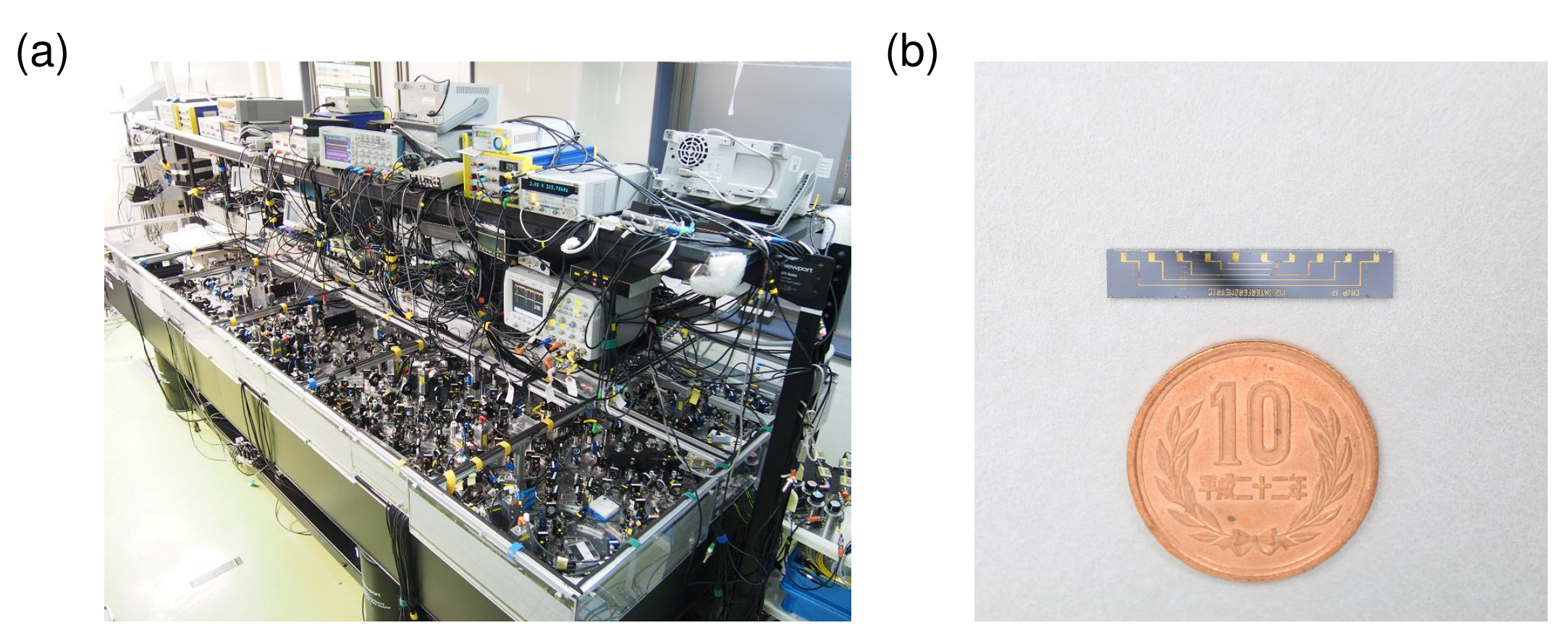}% Here is how to import EPS art
\caption{Photographs of actual photonic circuits.
(a) Free-space photoinc circuit for CV teleportation experiment~\cite{13Takeda2}.
The size of the optical table is 4.2 m $\times$ 1.5 m.
(b) Photonic chip for generating CV entangled beams~\cite{15Masada}.
The size of the chip is 26 mm $\times$ 4 mm.
}
\label{Fig:Experimental setup}
\end{figure*}

As discussed in Sec.~\ref{subsec:hybrid},
the hybrid approach is shown to provide error-correctable qubit encoding and deterministic quantum gates.
The next step would be to consider how to construct photonic quantum computers
in a scalable manner based on this hybrid approach.
The most well-established way of building photonic circuits
is to use one beam for one qubit, as shown in Fig.~\ref{Fig:ConventionalCircuit}.
Here, arrays of light sources (such as single photon sources) are operating in parallel,
and optical components to perform quantum gates are installed sequentially along with each optical path.
This configuration is convenient for small-scale photonic quantum computing,
but not suitable for large-scale quantum computing for two reasons.
One reason is that the size of the optical circuit increases with increasing number of qubits and gates.
Figure~\ref{Fig:Experimental setup}(a) shows the setup for a single-step CV quantum teleportation experiment,
which is built by putting more than 500 mirrors and beam splitters on an optical table.
The setup is already very complicated, and construction of larger optical circuits in this way is impractical.
The other reason is the lack of programmability of photonic circuits;
one optical circuit realizes one specific quantum computing task,
and the optical circuit has to be modified for the other tasks.
It is more desirable to be able to change quantum computing tasks
without changing the optical circuit itself.

For scalable and programmable quantum computing,
integrated photonic chips have been developed to miniaturize and scale up
photonic circuits both in qubit~\cite{08Politi,15Carolan} and CV~\cite{15Masada,18Lenzini} quantum computing
[Fig.~\ref{Fig:Experimental setup}(b)].
Ultimately, all necessary components for photonic quantum computing,
including nonlinear optical materials, beam splitters, EOMs,
and detectors, can be integrated on small photonic chips.
Furthermore, parameters of photonic circuits,
such as the amount of phase shift and beam splitter transmissivity,
can be externally controllable.
Therefore, the photonic circuits become programmable.
Such chips are also expected to enhance the fidelity of operations
by improving spatial mode matching (quality of interference) between optical beams
and phase stability of interferometers.
However, the photonic chip itself does not overcome the fundamental problem
that larger optical circuits are required for larger-scale quantum computing.
In fact, photonic chips might limit the maximum size of photonic circuits
since optical elements and their control elements have a certain minimal area footprint
and also the area of the chips is limited.
Therefore, although development of the integrated photonic chips is quite useful,
some other approach is required to overcome the fundamental problem
and fully scale up photonic quantum computing.

\subsection{Time-domain multiplexed one-way quantum computation}

\begin{figure*}
\includegraphics[width=0.8\linewidth,clip]{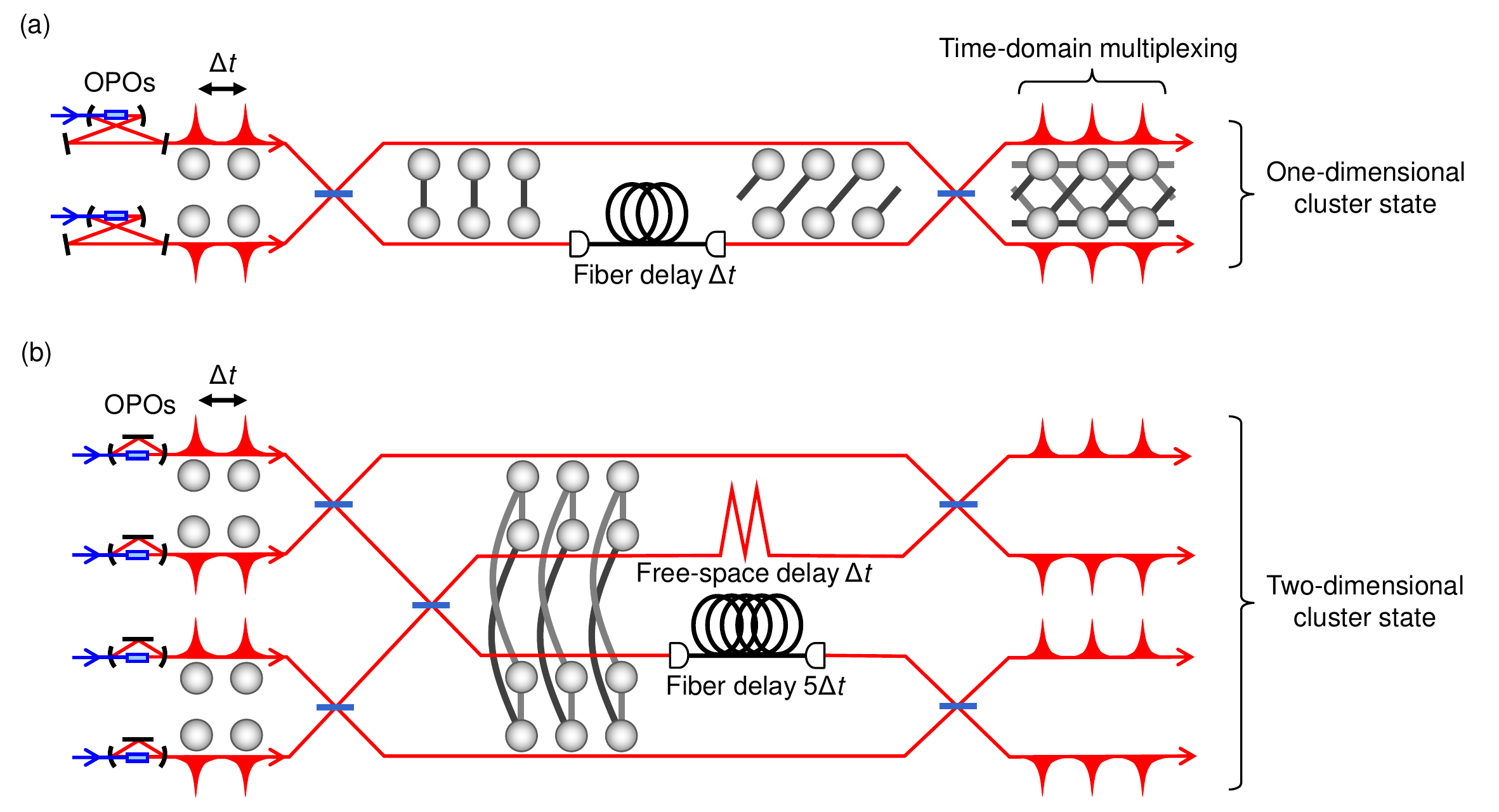}% Here is how to import EPS art
\caption{Generation of time-domain multiplexed cluster states.
(a) One-dimensional cluster state~\cite{13Yokoyama,16Yoshikawa}.
(b) Two-dimensional cluster state~\cite{19Asavanant}.
All beam splitters have 50\% reflectivity.
}
\label{fig:TimeDomainClusterGeneration}
\end{figure*}

In order to scale up photonic quantum computers,
an efficient and scalable method to increase the number of qubits and operations is needed.
Fortunately, by exploiting rich degrees of freedom of light,
we can encode a lot of qubits in a single optical beam
and perform quantum computation more efficiently.
In the CV approach,
several such approaches have been pursued,
such as time-domain multiplexing~\cite{10Menicucci,11Menicucci,13Yokoyama,16Yoshikawa,19Asavanant,17Takeda,18Takeda},
frequency-domain multiplexing~\cite{08Menicucci,11Pysher,14Chen,13Roslund},
and spatial-mode multiplexing~\cite{09Lassen,12Armstrong}.
In the case of time-domain multiplexing,
we can use a train of a lot of optical pulses propagating in a single (or a few) optical path(s)
to encode arbitrary number of qubits.
Furthermore, all of these qubits are individually accessible and easily controllable
by using a small number of optical components at different times.
Therefore, time-domain multiplexing may be a reasonable choice to realize scalable photonic quantum computers
which performs arbitrarily large-scale quantum computation with a constant number of optical components.

Another problem of the typical photonic quantum computing architecture in Fig.~\ref{Fig:ConventionalCircuit}
is the lack of programmability.
Fortunately, one solution to this problem is already known: one-way quantum computation.
As we explained in Sec.~\ref{sec:cv},
a specific type of a large-scale entangled state (cluster state) is sufficient for universal quantum computation in this scheme,
and different quantum computing tasks can be performed by simply choosing different measurement bases.
Therefore, once a sufficiently large cluster state can be produced,
it enables arbitrary quantum computation in a programmable way.

Recently, it has been discovered that
ultra-large-scale CV cluster states can be deterministically generated
by the time-domain multiplexing approach~\cite{10Menicucci,11Menicucci,13Yokoyama,16Yoshikawa}.
For the typical architecture, generation of $n$-mode cluster states
requires one to prepare $n$ squeezed light sources
and let the squeezed light beams interfere with each other at beam splitters, as shown in Fig.~\ref{fig:ClusterComputation}.
However, in the time-domain multiplexing approach in Fig.~\ref{fig:TimeDomainClusterGeneration},
continuously produced squeezed light beams are artificially divided into
time bins to define independent squeezed light modes,
and these modes are coupled with each other
by appropriate delay lines and beam splitters.
In the setup of Fig.~\ref{fig:TimeDomainClusterGeneration}(a),
large-scale one-dimensional cluster states, i.e.,
cluster states where modes are entangled in one-dimensional chain fashion,
were experimentally generated by using two squeezed light sources and one delay line~\cite{13Yokoyama,16Yoshikawa}.
This method was later extended to generation of large-scale two-dimensional cluster states
by using four squeezed light sources and two optical delay lines with different length~\cite{19Asavanant},
as shown in Fig.~\ref{fig:TimeDomainClusterGeneration}(b).
The generated two-dimensional cluster state is known to be a universal resource for
5-input 5-output quantum information processing~\cite{10Menicucci,11Menicucci,18Alexander}.
In these experimental schemes, the cluster states are sequentially generated and soon measured,
so the number of modes is never limited by the fundamental coherence time of the laser
and infinite in principle.
In the actual experiments, one-dimensional cluster states up to one million modes~\cite{16Yoshikawa}
and two-dimensional cluster states up to 5$\times$5000 modes~\cite{19Asavanant}
are verified
by the time-domain multiplexing schemes;
these are in fact the largest entangled states demonstrated to date among any physical system
(such as superconducting circuits, trapped ions, etc.).
Note that generation of large-scale optical cluster states has also been pursued
in other multiplexing schemes, such as frequency multiplexing~\cite{08Menicucci,11Pysher,14Chen}
and spatial mode multiplexing~\cite{12Armstrong}.

As already mentioned in Sec.~\ref{subsec:hybrid}, when CV cluster states with a squeezing level above a certain threshold
are prepared, fault-tolerant quantum computation is possible with the GKP qubits.
Therefore, time-domain multiplexed one-way quantum computation
should be a promising route to scalable, universal, and fault-tolerant photonic quantum computing.

\subsection{Loop-based architecture for photonic quantum computing}

\begin{figure*}
\includegraphics[width=0.9\linewidth,clip]{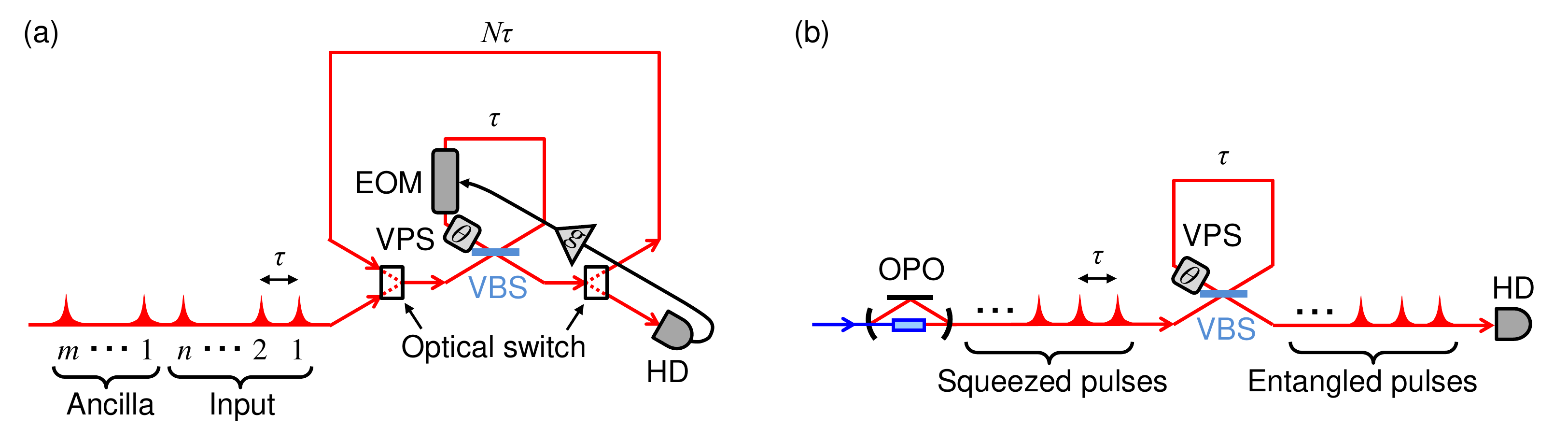}% Here is how to import EPS art
\caption{Loop-based architecture for photonic quantum computing.
(a) Loop-based architecture for universal quantum computation~\cite{17Takeda}.
(b) Loop-based entangled state generation~\cite{18Takeda}.
VPS, variable phase shifter; VBS, variable beam splitter; HD, homodyne detector.
The measurement bases of the homodyne detectors are variable.
}
\label{fig:LoopCircuit}
\end{figure*}

In one-way quantum computation,
the initial universal cluster state has to be reshaped and converted into a modified,
smaller cluster state suitable for a specific quantum computing task
by appropriately decoupling the modes unnecessary for the computation~\cite{18Alexander,10Miwa}.
In this sense, sequentially applying only necessary gates
to the input state is more straightforward and requires less calculation steps
than one-way quantum computation.
One useful idea to perform sequential quantum gates
without increasing the number of optical components is to introduce optical loops
and use the same optical components repeatedly.
Especially, if this loop configuration is combined with time-domain multiplexing,
the number of optical components for large-scale quantum computation can be dramatically reduced.
For photonic qubits,
quantum computation schemes based on time-domain multiplexing and a loop-based architecture
have been proposed~\cite{14Motes,15Rohde} and related experiments have been reported~\cite{12Schreiber,17He}.
Recently, these ideas are also extended to CVs,
and a loop-based architecture for universal quantum computation in Fig.~\ref{fig:LoopCircuit}(a) has been proposed~\cite{17Takeda}.
Below we focus on this architecture.

In this architecture, quantum information
encoded in a string of $n$ pulses of a single spatial mode are
sent to a nested loop circuit with the other $m$ ancilla pulses
which are used for teleportation-based quantum gates in Fig.~\ref{fig:CVgates}.
All pulses are first stored in the outer large loop by controlling optical switches.
This loop plays a role of a quantum memory,
and it can store quantum information of a lot of pulses
while these pulses circulate around the loop.
On the other hand, the inner small loop
is a processor which sequentially performs teleportation-based quantum gates
on pulses stored in the large loop.
The round-trip time for the inner loop ($\tau$)
is equivalent to the time interval between optical pulses,
enabling us to add tunable delay to a certain optical pulse
and let it interfere with any other pulses.
By dynamically changing system parameters such as 
beam splitter transmissivity, phase shift, feedforward gain, and measurement basis,
this processor can perform different types of gates for each pulses.
It can be shown that, once necessary ancillary states are prepared in the outer loop,
this system can perform both the teleportation-based squeezing gate and cubic phase gate in Fig.~\ref{fig:CVgates}.
Furthermore, the EOM, variable phase shifter, and variable beam splitter
enables direct implementation of the displacement operation, phase shift, and beam splitter interaction, respectively.
As a result, all gates necessary for universal CV quantum computation
can be deterministically performed in this architecture.

Ideally speaking, this architecture enables us to perform quantum gates on any number of modes and for any number of steps
with almost minimum resources by increasing the length of the outer loop and letting the optical pulses circulate there.
Furthermore, by changing the program to control the system parameters,
this architecture can perform different calculation without changing the photonic circuit,
thus it possesses programmability as well.
In the actual situation, however, optical losses caused by long delay lines and optical switches
can limit the performance of quantum computation.
Therefore, several proposals to reduce the effect of losses while maintaining the scalability have been made,
such as a chain-loop architecture composed of a chain of reconfigurable beam splitters and delay loops~\cite{18Qi}
and a hybrid architecture which simultaneously exploits spatial and temporal degrees of freedom~\cite{18Su}.

Recently, part of the loop-based architecture in Fig.~\ref{fig:LoopCircuit}(a) was demonstrated experimentally~\cite{18Takeda};
the setup contains one squeezed light source,
a single optical loop, a variable beam splitter, a variable phase shifter, and
a homodyne detector with tunable measurement basis, as shown in Fig.~\ref{fig:LoopCircuit}(b).
In this experiment, by dynamically controlling these system parameters,
this loop circuit was able to programmably generate various types of entangled states,
such as the Einstein-Podolsky-Rosen state, Greenberger-Horne-Zeilinger state, and cluster state.
This setup has been built with bulk optics in free space,
but in order to realize longer delay lines,
fiber-based optical circuits are also promising.
Recently there have been a few reports on fiber-based CV experiments
such as the fully guided-wave squeezing experiment~\cite{16Kaiser}
and entangled state generation with a fiber delay line and switching~\cite{18Larsen}.
These experimental efforts would open the possibility of building
large-scale photonic quantum processors.

\subsection{Technical challenges}

Finally, let us discuss technical issues to be overcome to scale up photonic quantum computers.
In the time-domain multiplexing approach mentioned above,
the number of processable qubits
is limited by the length of delay lines divided by the width of optical pulses.
This is because 
this value determines the number of input modes of two-dimensional cluster states in Fig.~\ref{fig:TimeDomainClusterGeneration}(b),
as well as the number of pulses stored in the optical loop in the loop-based architecture in Fig.~\ref{fig:LoopCircuit}(a).

The temporal width of optical pulses need to be shortened to increase the number of qubits.
The shorter pulse width in the time domain means the broader spectrum in the frequency domain.
The spectrum of pulses need to be covered with operational bandwidth of the optical/electrical components
which constitute photonic circuits.
Recent experiments on CV teleportation-based gates
have reported the bandwidth of up to 100 MHz~\cite{18Shiozawa},
and there the bandwidth is mainly limited by
the bandwidth of homodyne detectors~\cite{12Kumar} and
squeezed light sources~\cite{16Serikawa}.
In order to achieve high-fidelity operations in such systems,
the bandwidth of pulses need to be sufficiently narrower than 100 MHz,
and for this purpose the temporal width has been set to $\sim50$ ns in actual experiments~\cite{18Shiozawa,18Takeda, 19Asavanant}.

However, these values are not the fundamental limit,
and several approaches are known to increase the bandwidth of the system.
The bandwidth of the squeezed light sources
can be increased by replacing optical parametric oscillators (cavity-enhanced squeezers)
with single-path waveguide squeezers.
In this case the bandwidth is not limited by the bandwidth of the cavity,
but limited by the bandwidth of phase matching condition for the second-order nonlinear process, which is typically $\sim10$ THz.
Such squeezers have already been reported in several experiments~\cite{07Yoshino,18Lenzini}.
On the other hand,
the bandwidth of electronics is often MHz to GHz range,
and the bandwidth of homodyne detectors is often the most severe limitation.
Recently, this limitation has been overcome by
replacing a standard homodyne detector with a broadband parametric amplifier
which amplifies quadrature signals by optical means~\cite{18Shaked}.
This method has enabled the measurement of squeezing up to 55 THz.
In fact, it is ultimately possible to replace all the electronics
in the teleportation-based circuit with optical means,
thereby removing the bandwidth of electronics.
This idea is originally proposed as all-optical CV quantum teleportation~\cite{99Ralph}.
In this proposal, Bell measurement is performed by
optically amplifying quadrature signals by parametric amplification,
and the feedforward operation is performed directly by injecting the amplified optical signals
into a target optical beam.
This method can in principle increase the bandwidth of the system beyond THz
and decrease the pulse width by several orders of magnitude.

Realizing long delay lines is also necessary to increase the number of processable qubits.
The length of optical delay lines is manly limited by transmission losses and stability
(rather than the coherence length of light sources, which can be much longer~\cite{16Lifei}).
Previous experiments for time-domain multiplexed CV quantum information processing
have used free-space optical delay lines
or optical fiber delay lines of
a few tens of meters~\cite{13Yokoyama,16Yoshikawa,19Asavanant,18Takeda} at the wavelength of 860 nm.
For much longer delay lines with sufficient stability and low losses,
optical fibers at telecommunication wavelength are the reasonable choice
(even though kilometer-scale free-space optical delay lines are possible in principle~\cite{13LIGO}).
Considering the minimum transmission loss of $0.2$ dB/km in the fiber,
we can obtain $99.5\%$ transmission for a $100$-m fiber
and $95.5\%$ transmission for a $1$-km fiber, for example
(corresponding to $\sim10$ and $\sim100$ qubits for 50-ns pulse width, respectively).
In fact, CV quantum information processing experiments
using optical fibers of a few hundred meters or a few kilometers have recently been reported~\cite{18Huo,18Larsen}.
Therefore, $10^1-10^2$ qubits can be straightforwardly processed with the current technology,
and the number could be increased by several orders by
increasing the operational bandwidth and shortening the pulse width.
If the pulse width is shortened by several orders, the necessary length of delay lines should be much shorter,
and in this case stable free-space optical delay lines such as the Herriott delay line~\cite{64Herriott}
may be useful as well.

\section{Conclusion}\label{sec:conclusion}

Until recently, photonic quantum computers have intrinsic disadvantages
which make scalable implementation almost impractical,
even though it is in principle scalable as shown by KLM.
However, the two key ideas explained in this perspective
-- hybrid qubit-CV approach and time-domain multiplexing --
are opening a new era in the history of photonic quantum computing,
showing that scalable photonic quantum computing is actually possible.
The hybrid approach can take advantage of both
deterministic CV operations and robust qubit encoding.
Here, all gates for universal quantum computation can be deterministically performed by CV teleportation-based gates,
where the circuit itself is linear (easy to be scaled up without pulse distortion or crosstalk)
but nonlinearity required for some quantum gates is fed from external sources only when required.
The hybrid approach can also achieve fault-tolerant quantum computation
by introducing hardware efficient quantum error correcting codes such as the GKP qubits.
Furthermore, time-domain multiplexed quantum information processing
based on either one-way quantum computation or a loop-based architecture
dramatically increase the processable number of qubits
without increasing the number of optical components.
If such systems are constructed by all-optical means,
ultra-large-scale photonic quantum computing with ultra-high clock frequency of $\sim$THz
is possible in principle.
Of course there remain many hurdles to overcome before 
ultimate performance of photonic quantum computers is achieved,
but a promising route to large-scale photonic quantum computers has become clear.
We expect that these ideas will stimulate further theoretical and experimental research
in photonic quantum information processing.

\begin{acknowledgments}
This work was partly supported by JST PRESTO (JPMJPR1764) and
JSPS KAKENHI (18K14143).
S. T. acknowledges Kosuke Fukui for his useful comments on the manuscript.
\end{acknowledgments}

\nocite{*}
\bibliography{manuscript3}% Produces the bibliography via BibTeX.

\providecommand{\noopsort}[1]{}\providecommand{\singleletter}[1]{#1}%
\begin{thebibliography}{119}%
\makeatletter
\providecommand \@ifxundefined [1]{%
 \@ifx{#1\undefined}
}%
\providecommand \@ifnum [1]{%
 \ifnum #1\expandafter \@firstoftwo
 \else \expandafter \@secondoftwo
 \fi
}%
\providecommand \@ifx [1]{%
 \ifx #1\expandafter \@firstoftwo
 \else \expandafter \@secondoftwo
 \fi
}%
\providecommand \natexlab [1]{#1}%
\providecommand \enquote  [1]{``#1''}%
\providecommand \bibnamefont  [1]{#1}%
\providecommand \bibfnamefont [1]{#1}%
\providecommand \citenamefont [1]{#1}%
\providecommand \href@noop [0]{\@secondoftwo}%
\providecommand \href [0]{\begingroup \@sanitize@url \@href}%
\providecommand \@href[1]{\@@startlink{#1}\@@href}%
\providecommand \@@href[1]{\endgroup#1\@@endlink}%
\providecommand \@sanitize@url [0]{\catcode `\\12\catcode `\$12\catcode
  `\&12\catcode `\#12\catcode `\^12\catcode `\_12\catcode `\%12\relax}%
\providecommand \@@startlink[1]{}%
\providecommand \@@endlink[0]{}%
\providecommand \url  [0]{\begingroup\@sanitize@url \@url }%
\providecommand \@url [1]{\endgroup\@href {#1}{\urlprefix }}%
\providecommand \urlprefix  [0]{URL }%
\providecommand \Eprint [0]{\href }%
\providecommand \doibase [0]{http://dx.doi.org/}%
\providecommand \selectlanguage [0]{\@gobble}%
\providecommand \bibinfo  [0]{\@secondoftwo}%
\providecommand \bibfield  [0]{\@secondoftwo}%
\providecommand \translation [1]{[#1]}%
\providecommand \BibitemOpen [0]{}%
\providecommand \bibitemStop [0]{}%
\providecommand \bibitemNoStop [0]{.\EOS\space}%
\providecommand \EOS [0]{\spacefactor3000\relax}%
\providecommand \BibitemShut  [1]{\csname bibitem#1\endcsname}%
\let\auto@bib@innerbib\@empty
%</preamble>
\bibitem [{\citenamefont {Nielsen}\ and\ \citenamefont
  {Chuang}(2010)}]{10Nielsen}%
  \BibitemOpen
  \bibfield  {author} {\bibinfo {author} {\bibfnamefont {M.~A.}\ \bibnamefont
  {Nielsen}}\ and\ \bibinfo {author} {\bibfnamefont {I.~L.}\ \bibnamefont
  {Chuang}},\ }\href@noop {} {\emph {\bibinfo {title} {Quantum Computation and
  Quantum Information}}}\ (\bibinfo  {publisher} {Cambridge University Press},\
  \bibinfo {year} {2010})\BibitemShut {NoStop}%
\bibitem [{\citenamefont {Ladd}\ \emph {et~al.}(2010)\citenamefont {Ladd},
  \citenamefont {Jelezko}, \citenamefont {Laflamme}, \citenamefont {Nakamura},
  \citenamefont {Monroe},\ and\ \citenamefont {O'Brien}}]{10Ladd}%
  \BibitemOpen
  \bibfield  {author} {\bibinfo {author} {\bibfnamefont {T.~D.}\ \bibnamefont
  {Ladd}}, \bibinfo {author} {\bibfnamefont {F.}~\bibnamefont {Jelezko}},
  \bibinfo {author} {\bibfnamefont {R.}~\bibnamefont {Laflamme}}, \bibinfo
  {author} {\bibfnamefont {Y.}~\bibnamefont {Nakamura}}, \bibinfo {author}
  {\bibfnamefont {C.}~\bibnamefont {Monroe}}, \ and\ \bibinfo {author}
  {\bibfnamefont {J.~L.}\ \bibnamefont {O'Brien}},\ }\bibfield  {title}
  {\enquote {\bibinfo {title} {Quantum computers},}\ }\href@noop {} {\bibfield
  {journal} {\bibinfo  {journal} {Nature}\ }\textbf {\bibinfo {volume} {464}},\
  \bibinfo {pages} {45--53} (\bibinfo {year} {2010})}\BibitemShut {NoStop}%
\bibitem [{\citenamefont {Wang}\ \emph {et~al.}(2018)\citenamefont {Wang},
  \citenamefont {Li}, \citenamefont {Yin},\ and\ \citenamefont
  {Zeng}}]{18Yuanhao}%
  \BibitemOpen
  \bibfield  {author} {\bibinfo {author} {\bibfnamefont {Y.}~\bibnamefont
  {Wang}}, \bibinfo {author} {\bibfnamefont {Y.}~\bibnamefont {Li}}, \bibinfo
  {author} {\bibfnamefont {Z.-q.}\ \bibnamefont {Yin}}, \ and\ \bibinfo
  {author} {\bibfnamefont {B.}~\bibnamefont {Zeng}},\ }\bibfield  {title}
  {\enquote {\bibinfo {title} {16-qubit {IBM} universal quantum computer can be
  fully entangled},}\ }\href@noop {} {\bibfield  {journal} {\bibinfo  {journal}
  {npj Quantum Information}\ }\textbf {\bibinfo {volume} {4}},\ \bibinfo
  {pages} {46} (\bibinfo {year} {2018})}\BibitemShut {NoStop}%
\bibitem [{\citenamefont {Friis}\ \emph {et~al.}(2018)\citenamefont {Friis},
  \citenamefont {Marty}, \citenamefont {Maier}, \citenamefont {Hempel},
  \citenamefont {Holz\"apfel}, \citenamefont {Jurcevic}, \citenamefont
  {Plenio}, \citenamefont {Huber}, \citenamefont {Roos}, \citenamefont
  {Blatt},\ and\ \citenamefont {Lanyon}}]{18Friis}%
  \BibitemOpen
  \bibfield  {author} {\bibinfo {author} {\bibfnamefont {N.}~\bibnamefont
  {Friis}}, \bibinfo {author} {\bibfnamefont {O.}~\bibnamefont {Marty}},
  \bibinfo {author} {\bibfnamefont {C.}~\bibnamefont {Maier}}, \bibinfo
  {author} {\bibfnamefont {C.}~\bibnamefont {Hempel}}, \bibinfo {author}
  {\bibfnamefont {M.}~\bibnamefont {Holz\"apfel}}, \bibinfo {author}
  {\bibfnamefont {P.}~\bibnamefont {Jurcevic}}, \bibinfo {author}
  {\bibfnamefont {M.~B.}\ \bibnamefont {Plenio}}, \bibinfo {author}
  {\bibfnamefont {M.}~\bibnamefont {Huber}}, \bibinfo {author} {\bibfnamefont
  {C.}~\bibnamefont {Roos}}, \bibinfo {author} {\bibfnamefont {R.}~\bibnamefont
  {Blatt}}, \ and\ \bibinfo {author} {\bibfnamefont {B.}~\bibnamefont
  {Lanyon}},\ }\bibfield  {title} {\enquote {\bibinfo {title} {Observation of
  entangled states of a fully controlled 20-qubit system},}\ }\href {\doibase
  10.1103/PhysRevX.8.021012} {\bibfield  {journal} {\bibinfo  {journal} {Phys.
  Rev. X}\ }\textbf {\bibinfo {volume} {8}},\ \bibinfo {pages} {021012}
  (\bibinfo {year} {2018})}\BibitemShut {NoStop}%
\bibitem [{\citenamefont {Pan}\ \emph {et~al.}(2012)\citenamefont {Pan},
  \citenamefont {Chen}, \citenamefont {Lu}, \citenamefont {Weinfurter},
  \citenamefont {Zeilinger},\ and\ \citenamefont {\ifmmode~\dot{Z}\else
  \.{Z}\fi{}ukowski}}]{12Pan}%
  \BibitemOpen
  \bibfield  {author} {\bibinfo {author} {\bibfnamefont {J.-W.}\ \bibnamefont
  {Pan}}, \bibinfo {author} {\bibfnamefont {Z.-B.}\ \bibnamefont {Chen}},
  \bibinfo {author} {\bibfnamefont {C.-Y.}\ \bibnamefont {Lu}}, \bibinfo
  {author} {\bibfnamefont {H.}~\bibnamefont {Weinfurter}}, \bibinfo {author}
  {\bibfnamefont {A.}~\bibnamefont {Zeilinger}}, \ and\ \bibinfo {author}
  {\bibfnamefont {M.}~\bibnamefont {\ifmmode~\dot{Z}\else \.{Z}\fi{}ukowski}},\
  }\bibfield  {title} {\enquote {\bibinfo {title} {Multiphoton entanglement and
  interferometry},}\ }\href {\doibase 10.1103/RevModPhys.84.777} {\bibfield
  {journal} {\bibinfo  {journal} {Rev. Mod. Phys.}\ }\textbf {\bibinfo {volume}
  {84}},\ \bibinfo {pages} {777--838} (\bibinfo {year} {2012})}\BibitemShut
  {NoStop}%
\bibitem [{\citenamefont {O'Brien}(2007)}]{07OBrien}%
  \BibitemOpen
  \bibfield  {author} {\bibinfo {author} {\bibfnamefont {J.~L.}\ \bibnamefont
  {O'Brien}},\ }\bibfield  {title} {\enquote {\bibinfo {title} {Optical quantum
  computing},}\ }\href {\doibase 10.1126/science.1142892} {\bibfield  {journal}
  {\bibinfo  {journal} {Science}\ }\textbf {\bibinfo {volume} {318}},\ \bibinfo
  {pages} {1567--1570} (\bibinfo {year} {2007})}\BibitemShut {NoStop}%
\bibitem [{\citenamefont {Braunstein}\ and\ \citenamefont {van
  Loock}(2005)}]{05Braunstein}%
  \BibitemOpen
  \bibfield  {author} {\bibinfo {author} {\bibfnamefont {S.~L.}\ \bibnamefont
  {Braunstein}}\ and\ \bibinfo {author} {\bibfnamefont {P.}~\bibnamefont {van
  Loock}},\ }\bibfield  {title} {\enquote {\bibinfo {title} {Quantum
  information with continuous variables},}\ }\href {\doibase
  10.1103/RevModPhys.77.513} {\bibfield  {journal} {\bibinfo  {journal} {Rev.
  Mod. Phys.}\ }\textbf {\bibinfo {volume} {77}},\ \bibinfo {pages} {513--577}
  (\bibinfo {year} {2005})}\BibitemShut {NoStop}%
\bibitem [{\citenamefont {O'Brien}, \citenamefont {Furusawa},\ and\
  \citenamefont {Vu\v{c}kovi\'{c}}(2009)}]{09OBrien}%
  \BibitemOpen
  \bibfield  {author} {\bibinfo {author} {\bibfnamefont {J.~L.}\ \bibnamefont
  {O'Brien}}, \bibinfo {author} {\bibfnamefont {A.}~\bibnamefont {Furusawa}}, \
  and\ \bibinfo {author} {\bibfnamefont {J.}~\bibnamefont {Vu\v{c}kovi\'{c}}},\
  }\bibfield  {title} {\enquote {\bibinfo {title} {Photonic quantum
  technologies},}\ }\href@noop {} {\bibfield  {journal} {\bibinfo  {journal}
  {Nat. Photonics}\ }\textbf {\bibinfo {volume} {3}},\ \bibinfo {pages} {687}
  (\bibinfo {year} {2009})}\BibitemShut {NoStop}%
\bibitem [{\citenamefont {van Loock}(2011)}]{11vanLoock}%
  \BibitemOpen
  \bibfield  {author} {\bibinfo {author} {\bibfnamefont {P.}~\bibnamefont {van
  Loock}},\ }\bibfield  {title} {\enquote {\bibinfo {title} {Optical hybrid
  approaches to quantum information},}\ }\href {\doibase
  10.1002/lpor.201000005} {\bibfield  {journal} {\bibinfo  {journal} {Laser \&
  Photonics Rev.}\ }\textbf {\bibinfo {volume} {5}},\ \bibinfo {pages}
  {167--200} (\bibinfo {year} {2011})}\BibitemShut {NoStop}%
\bibitem [{\citenamefont {Furusawa}\ and\ \citenamefont {van
  Loock}(2011)}]{11Furusawa}%
  \BibitemOpen
  \bibfield  {author} {\bibinfo {author} {\bibfnamefont {A.}~\bibnamefont
  {Furusawa}}\ and\ \bibinfo {author} {\bibfnamefont {P.}~\bibnamefont {van
  Loock}},\ }\href@noop {} {\emph {\bibinfo {title} {Quantum Teleportation and
  Entanglement: A Hybrid Approach to Optical Quantum Information Processing}}}\
  (\bibinfo  {publisher} {Wiley-VCH},\ \bibinfo {year} {2011})\BibitemShut
  {NoStop}%
\bibitem [{\citenamefont {Andersen}\ \emph {et~al.}(2015)\citenamefont
  {Andersen}, \citenamefont {Neergaard-Nielsen}, \citenamefont {van Loock},\
  and\ \citenamefont {Furusawa}}]{15Andersen}%
  \BibitemOpen
  \bibfield  {author} {\bibinfo {author} {\bibfnamefont {U.~L.}\ \bibnamefont
  {Andersen}}, \bibinfo {author} {\bibfnamefont {J.~S.}\ \bibnamefont
  {Neergaard-Nielsen}}, \bibinfo {author} {\bibfnamefont {P.}~\bibnamefont {van
  Loock}}, \ and\ \bibinfo {author} {\bibfnamefont {A.}~\bibnamefont
  {Furusawa}},\ }\bibfield  {title} {\enquote {\bibinfo {title} {Hybrid
  discrete- and continuous-variable quantum information},}\ }\href@noop {}
  {\bibfield  {journal} {\bibinfo  {journal} {Nat. Phys.}\ }\textbf {\bibinfo
  {volume} {11}},\ \bibinfo {pages} {713--719} (\bibinfo {year}
  {2015})}\BibitemShut {NoStop}%
\bibitem [{\citenamefont {Yokoyama}\ \emph {et~al.}(2013)\citenamefont
  {Yokoyama}, \citenamefont {Ukai}, \citenamefont {Armstrong}, \citenamefont
  {Sornphiphatphong}, \citenamefont {Kaji}, \citenamefont {Suzuki},
  \citenamefont {Yoshikawa}, \citenamefont {Yonezawa}, \citenamefont
  {Menicucci},\ and\ \citenamefont {Furusawa}}]{13Yokoyama}%
  \BibitemOpen
  \bibfield  {author} {\bibinfo {author} {\bibfnamefont {S.}~\bibnamefont
  {Yokoyama}}, \bibinfo {author} {\bibfnamefont {R.}~\bibnamefont {Ukai}},
  \bibinfo {author} {\bibfnamefont {S.~C.}\ \bibnamefont {Armstrong}}, \bibinfo
  {author} {\bibfnamefont {C.}~\bibnamefont {Sornphiphatphong}}, \bibinfo
  {author} {\bibfnamefont {T.}~\bibnamefont {Kaji}}, \bibinfo {author}
  {\bibfnamefont {S.}~\bibnamefont {Suzuki}}, \bibinfo {author} {\bibfnamefont
  {J.}~\bibnamefont {Yoshikawa}}, \bibinfo {author} {\bibfnamefont
  {H.}~\bibnamefont {Yonezawa}}, \bibinfo {author} {\bibfnamefont {N.~C.}\
  \bibnamefont {Menicucci}}, \ and\ \bibinfo {author} {\bibfnamefont
  {A.}~\bibnamefont {Furusawa}},\ }\bibfield  {title} {\enquote {\bibinfo
  {title} {Ultra-large-scale continuous-variable cluster states multiplexed in
  the time domain},}\ }\href@noop {} {\bibfield  {journal} {\bibinfo  {journal}
  {Nat. Photonics}\ }\textbf {\bibinfo {volume} {7}},\ \bibinfo {pages}
  {982--986} (\bibinfo {year} {2013})}\BibitemShut {NoStop}%
\bibitem [{\citenamefont {Yoshikawa}\ \emph {et~al.}(2016)\citenamefont
  {Yoshikawa}, \citenamefont {Yokoyama}, \citenamefont {Kaji}, \citenamefont
  {Sornphiphatphong}, \citenamefont {Shiozawa}, \citenamefont {Makino},\ and\
  \citenamefont {Furusawa}}]{16Yoshikawa}%
  \BibitemOpen
  \bibfield  {author} {\bibinfo {author} {\bibfnamefont {J.}~\bibnamefont
  {Yoshikawa}}, \bibinfo {author} {\bibfnamefont {S.}~\bibnamefont {Yokoyama}},
  \bibinfo {author} {\bibfnamefont {T.}~\bibnamefont {Kaji}}, \bibinfo {author}
  {\bibfnamefont {C.}~\bibnamefont {Sornphiphatphong}}, \bibinfo {author}
  {\bibfnamefont {Y.}~\bibnamefont {Shiozawa}}, \bibinfo {author}
  {\bibfnamefont {K.}~\bibnamefont {Makino}}, \ and\ \bibinfo {author}
  {\bibfnamefont {A.}~\bibnamefont {Furusawa}},\ }\bibfield  {title} {\enquote
  {\bibinfo {title} {Invited article: Generation of one-million-mode
  continuous-variable cluster state by unlimited time-domain multiplexing},}\
  }\href {\doibase 10.1063/1.4962732} {\bibfield  {journal} {\bibinfo
  {journal} {APL Photonics}\ }\textbf {\bibinfo {volume} {1}},\ \bibinfo
  {pages} {060801} (\bibinfo {year} {2016})}\BibitemShut {NoStop}%
\bibitem [{\citenamefont {Takeda}\ and\ \citenamefont
  {Furusawa}(2017)}]{17Takeda}%
  \BibitemOpen
  \bibfield  {author} {\bibinfo {author} {\bibfnamefont {S.}~\bibnamefont
  {Takeda}}\ and\ \bibinfo {author} {\bibfnamefont {A.}~\bibnamefont
  {Furusawa}},\ }\bibfield  {title} {\enquote {\bibinfo {title} {Universal
  quantum computing with measurement-induced continuous-variable gate sequence
  in a loop-based architecture},}\ }\href {\doibase
  10.1103/PhysRevLett.119.120504} {\bibfield  {journal} {\bibinfo  {journal}
  {Phys. Rev. Lett.}\ }\textbf {\bibinfo {volume} {119}},\ \bibinfo {pages}
  {120504} (\bibinfo {year} {2017})}\BibitemShut {NoStop}%
\bibitem [{\citenamefont {Takeda}, \citenamefont {Takase},\ and\ \citenamefont
  {Furusawa}(2018)}]{18Takeda}%
  \BibitemOpen
  \bibfield  {author} {\bibinfo {author} {\bibfnamefont {S.}~\bibnamefont
  {Takeda}}, \bibinfo {author} {\bibfnamefont {K.}~\bibnamefont {Takase}}, \
  and\ \bibinfo {author} {\bibfnamefont {A.}~\bibnamefont {Furusawa}},\
  }\href@noop {} {\enquote {\bibinfo {title} {On-demand photonic entanglement
  synthesizer},}\ }\bibinfo {howpublished} {arXiv:1811.10704} (\bibinfo {year}
  {2018})\BibitemShut {NoStop}%
\bibitem [{\citenamefont {Marek}, \citenamefont {Filip},\ and\ \citenamefont
  {Furusawa}(2011)}]{11Marek}%
  \BibitemOpen
  \bibfield  {author} {\bibinfo {author} {\bibfnamefont {P.}~\bibnamefont
  {Marek}}, \bibinfo {author} {\bibfnamefont {R.}~\bibnamefont {Filip}}, \ and\
  \bibinfo {author} {\bibfnamefont {A.}~\bibnamefont {Furusawa}},\ }\bibfield
  {title} {\enquote {\bibinfo {title} {Deterministic implementation of weak
  quantum cubic nonlinearity},}\ }\href@noop {} {\bibfield  {journal} {\bibinfo
   {journal} {Phys. Rev. A}\ }\textbf {\bibinfo {volume} {84}},\ \bibinfo
  {pages} {053802} (\bibinfo {year} {2011})}\BibitemShut {NoStop}%
\bibitem [{\citenamefont {Miyata}\ \emph {et~al.}(2016)\citenamefont {Miyata},
  \citenamefont {Ogawa}, \citenamefont {Marek}, \citenamefont {Filip},
  \citenamefont {Yonezawa}, \citenamefont {Yoshikawa},\ and\ \citenamefont
  {Furusawa}}]{16Miyata}%
  \BibitemOpen
  \bibfield  {author} {\bibinfo {author} {\bibfnamefont {K.}~\bibnamefont
  {Miyata}}, \bibinfo {author} {\bibfnamefont {H.}~\bibnamefont {Ogawa}},
  \bibinfo {author} {\bibfnamefont {P.}~\bibnamefont {Marek}}, \bibinfo
  {author} {\bibfnamefont {R.}~\bibnamefont {Filip}}, \bibinfo {author}
  {\bibfnamefont {H.}~\bibnamefont {Yonezawa}}, \bibinfo {author}
  {\bibfnamefont {J.}~\bibnamefont {Yoshikawa}}, \ and\ \bibinfo {author}
  {\bibfnamefont {A.}~\bibnamefont {Furusawa}},\ }\bibfield  {title} {\enquote
  {\bibinfo {title} {Implementation of a quantum cubic gate by an adaptive
  non-{Gaussian} measurement},}\ }\href {\doibase 10.1103/PhysRevA.93.022301}
  {\bibfield  {journal} {\bibinfo  {journal} {Phys. Rev. A}\ }\textbf {\bibinfo
  {volume} {93}},\ \bibinfo {pages} {022301} (\bibinfo {year}
  {2016})}\BibitemShut {NoStop}%
\bibitem [{\citenamefont {Marek}\ \emph {et~al.}(2018)\citenamefont {Marek},
  \citenamefont {Filip}, \citenamefont {Ogawa}, \citenamefont {Sakaguchi},
  \citenamefont {Takeda}, \citenamefont {Yoshikawa},\ and\ \citenamefont
  {Furusawa}}]{18Marek}%
  \BibitemOpen
  \bibfield  {author} {\bibinfo {author} {\bibfnamefont {P.}~\bibnamefont
  {Marek}}, \bibinfo {author} {\bibfnamefont {R.}~\bibnamefont {Filip}},
  \bibinfo {author} {\bibfnamefont {H.}~\bibnamefont {Ogawa}}, \bibinfo
  {author} {\bibfnamefont {A.}~\bibnamefont {Sakaguchi}}, \bibinfo {author}
  {\bibfnamefont {S.}~\bibnamefont {Takeda}}, \bibinfo {author} {\bibfnamefont
  {J.}~\bibnamefont {Yoshikawa}}, \ and\ \bibinfo {author} {\bibfnamefont
  {A.}~\bibnamefont {Furusawa}},\ }\bibfield  {title} {\enquote {\bibinfo
  {title} {General implementation of arbitrary nonlinear quadrature phase
  gates},}\ }\href {\doibase 10.1103/PhysRevA.97.022329} {\bibfield  {journal}
  {\bibinfo  {journal} {Phys. Rev. A}\ }\textbf {\bibinfo {volume} {97}},\
  \bibinfo {pages} {022329} (\bibinfo {year} {2018})}\BibitemShut {NoStop}%
\bibitem [{\citenamefont {Gottesman}, \citenamefont {Kitaev},\ and\
  \citenamefont {Preskill}(2001)}]{01Gottesman}%
  \BibitemOpen
  \bibfield  {author} {\bibinfo {author} {\bibfnamefont {D.}~\bibnamefont
  {Gottesman}}, \bibinfo {author} {\bibfnamefont {A.}~\bibnamefont {Kitaev}}, \
  and\ \bibinfo {author} {\bibfnamefont {J.}~\bibnamefont {Preskill}},\
  }\bibfield  {title} {\enquote {\bibinfo {title} {Encoding a qubit in an
  oscillator},}\ }\href {\doibase 10.1103/PhysRevA.64.012310} {\bibfield
  {journal} {\bibinfo  {journal} {Phys. Rev. A}\ }\textbf {\bibinfo {volume}
  {64}},\ \bibinfo {pages} {012310} (\bibinfo {year} {2001})}\BibitemShut
  {NoStop}%
\bibitem [{\citenamefont {Cochrane}, \citenamefont {Milburn},\ and\
  \citenamefont {Munro}(1999)}]{99Cochrane}%
  \BibitemOpen
  \bibfield  {author} {\bibinfo {author} {\bibfnamefont {P.~T.}\ \bibnamefont
  {Cochrane}}, \bibinfo {author} {\bibfnamefont {G.~J.}\ \bibnamefont
  {Milburn}}, \ and\ \bibinfo {author} {\bibfnamefont {W.~J.}\ \bibnamefont
  {Munro}},\ }\bibfield  {title} {\enquote {\bibinfo {title} {Macroscopically
  distinct quantum-superposition states as a bosonic code for amplitude
  damping},}\ }\href {\doibase 10.1103/PhysRevA.59.2631} {\bibfield  {journal}
  {\bibinfo  {journal} {Phys. Rev. A}\ }\textbf {\bibinfo {volume} {59}},\
  \bibinfo {pages} {2631--2634} (\bibinfo {year} {1999})}\BibitemShut {NoStop}%
\bibitem [{\citenamefont {Leghtas}\ \emph {et~al.}(2013)\citenamefont
  {Leghtas}, \citenamefont {Kirchmair}, \citenamefont {Vlastakis},
  \citenamefont {Schoelkopf}, \citenamefont {Devoret},\ and\ \citenamefont
  {Mirrahimi}}]{13Leghtas}%
  \BibitemOpen
  \bibfield  {author} {\bibinfo {author} {\bibfnamefont {Z.}~\bibnamefont
  {Leghtas}}, \bibinfo {author} {\bibfnamefont {G.}~\bibnamefont {Kirchmair}},
  \bibinfo {author} {\bibfnamefont {B.}~\bibnamefont {Vlastakis}}, \bibinfo
  {author} {\bibfnamefont {R.~J.}\ \bibnamefont {Schoelkopf}}, \bibinfo
  {author} {\bibfnamefont {M.~H.}\ \bibnamefont {Devoret}}, \ and\ \bibinfo
  {author} {\bibfnamefont {M.}~\bibnamefont {Mirrahimi}},\ }\bibfield  {title}
  {\enquote {\bibinfo {title} {Hardware-efficient autonomous quantum memory
  protection},}\ }\href {\doibase 10.1103/PhysRevLett.111.120501} {\bibfield
  {journal} {\bibinfo  {journal} {Phys. Rev. Lett.}\ }\textbf {\bibinfo
  {volume} {111}},\ \bibinfo {pages} {120501} (\bibinfo {year}
  {2013})}\BibitemShut {NoStop}%
\bibitem [{\citenamefont {Michael}\ \emph {et~al.}(2016)\citenamefont
  {Michael}, \citenamefont {Silveri}, \citenamefont {Brierley}, \citenamefont
  {Albert}, \citenamefont {Salmilehto}, \citenamefont {Jiang},\ and\
  \citenamefont {Girvin}}]{16Michael}%
  \BibitemOpen
  \bibfield  {author} {\bibinfo {author} {\bibfnamefont {M.~H.}\ \bibnamefont
  {Michael}}, \bibinfo {author} {\bibfnamefont {M.}~\bibnamefont {Silveri}},
  \bibinfo {author} {\bibfnamefont {R.~T.}\ \bibnamefont {Brierley}}, \bibinfo
  {author} {\bibfnamefont {V.~V.}\ \bibnamefont {Albert}}, \bibinfo {author}
  {\bibfnamefont {J.}~\bibnamefont {Salmilehto}}, \bibinfo {author}
  {\bibfnamefont {L.}~\bibnamefont {Jiang}}, \ and\ \bibinfo {author}
  {\bibfnamefont {S.~M.}\ \bibnamefont {Girvin}},\ }\bibfield  {title}
  {\enquote {\bibinfo {title} {New class of quantum error-correcting codes for
  a bosonic mode},}\ }\href {\doibase 10.1103/PhysRevX.6.031006} {\bibfield
  {journal} {\bibinfo  {journal} {Phys. Rev. X}\ }\textbf {\bibinfo {volume}
  {6}},\ \bibinfo {pages} {031006} (\bibinfo {year} {2016})}\BibitemShut
  {NoStop}%
\bibitem [{\citenamefont {Shor}(1995)}]{95Shor}%
  \BibitemOpen
  \bibfield  {author} {\bibinfo {author} {\bibfnamefont {P.~W.}\ \bibnamefont
  {Shor}},\ }\bibfield  {title} {\enquote {\bibinfo {title} {Scheme for
  reducing decoherence in quantum computer memory},}\ }\href {\doibase
  10.1103/PhysRevA.52.R2493} {\bibfield  {journal} {\bibinfo  {journal} {Phys.
  Rev. A}\ }\textbf {\bibinfo {volume} {52}},\ \bibinfo {pages} {R2493--R2496}
  (\bibinfo {year} {1995})}\BibitemShut {NoStop}%
\bibitem [{\citenamefont {Steane}(1996)}]{96Steane}%
  \BibitemOpen
  \bibfield  {author} {\bibinfo {author} {\bibfnamefont {A.~M.}\ \bibnamefont
  {Steane}},\ }\bibfield  {title} {\enquote {\bibinfo {title} {Error correcting
  codes in quantum theory},}\ }\href {\doibase 10.1103/PhysRevLett.77.793}
  {\bibfield  {journal} {\bibinfo  {journal} {Phys. Rev. Lett.}\ }\textbf
  {\bibinfo {volume} {77}},\ \bibinfo {pages} {793--797} (\bibinfo {year}
  {1996})}\BibitemShut {NoStop}%
\bibitem [{\citenamefont {Ralph}(1999)}]{99Ralph}%
  \BibitemOpen
  \bibfield  {author} {\bibinfo {author} {\bibfnamefont {T.~C.}\ \bibnamefont
  {Ralph}},\ }\bibfield  {title} {\enquote {\bibinfo {title} {All-optical
  quantum teleportation},}\ }\href {\doibase 10.1364/OL.24.000348} {\bibfield
  {journal} {\bibinfo  {journal} {Opt. Lett.}\ }\textbf {\bibinfo {volume}
  {24}},\ \bibinfo {pages} {348--350} (\bibinfo {year} {1999})}\BibitemShut
  {NoStop}%
\bibitem [{\citenamefont {Knill}, \citenamefont {Laflamme},\ and\ \citenamefont
  {Milburn}(2001)}]{01Knill}%
  \BibitemOpen
  \bibfield  {author} {\bibinfo {author} {\bibfnamefont {E.}~\bibnamefont
  {Knill}}, \bibinfo {author} {\bibfnamefont {R.}~\bibnamefont {Laflamme}}, \
  and\ \bibinfo {author} {\bibfnamefont {G.~J.}\ \bibnamefont {Milburn}},\
  }\bibfield  {title} {\enquote {\bibinfo {title} {A scheme for efficient
  quantum computation with linear optics},}\ }\href@noop {} {\bibfield
  {journal} {\bibinfo  {journal} {Nature}\ }\textbf {\bibinfo {volume} {409}},\
  \bibinfo {pages} {46--52} (\bibinfo {year} {2001})}\BibitemShut {NoStop}%
\bibitem [{\citenamefont {Bennett}\ \emph {et~al.}(1993)\citenamefont
  {Bennett}, \citenamefont {Brassard}, \citenamefont {Cr\'epeau}, \citenamefont
  {Jozsa}, \citenamefont {Peres},\ and\ \citenamefont {Wootters}}]{93Bennett}%
  \BibitemOpen
  \bibfield  {author} {\bibinfo {author} {\bibfnamefont {C.~H.}\ \bibnamefont
  {Bennett}}, \bibinfo {author} {\bibfnamefont {G.}~\bibnamefont {Brassard}},
  \bibinfo {author} {\bibfnamefont {C.}~\bibnamefont {Cr\'epeau}}, \bibinfo
  {author} {\bibfnamefont {R.}~\bibnamefont {Jozsa}}, \bibinfo {author}
  {\bibfnamefont {A.}~\bibnamefont {Peres}}, \ and\ \bibinfo {author}
  {\bibfnamefont {W.~K.}\ \bibnamefont {Wootters}},\ }\bibfield  {title}
  {\enquote {\bibinfo {title} {Teleporting an unknown quantum state via dual
  classical and einstein-podolsky-rosen channels},}\ }\href {\doibase
  10.1103/PhysRevLett.70.1895} {\bibfield  {journal} {\bibinfo  {journal}
  {Phys. Rev. Lett.}\ }\textbf {\bibinfo {volume} {70}},\ \bibinfo {pages}
  {1895--1899} (\bibinfo {year} {1993})}\BibitemShut {NoStop}%
\bibitem [{\citenamefont {Gottesman}\ and\ \citenamefont
  {Chuang}(1999)}]{99Gottesman}%
  \BibitemOpen
  \bibfield  {author} {\bibinfo {author} {\bibfnamefont {D.}~\bibnamefont
  {Gottesman}}\ and\ \bibinfo {author} {\bibfnamefont {I.~L.}\ \bibnamefont
  {Chuang}},\ }\bibfield  {title} {\enquote {\bibinfo {title} {Demonstrating
  the viability of universal quantum computation using teleportation and
  single-qubit operations},}\ }\href@noop {} {\bibfield  {journal} {\bibinfo
  {journal} {Nature}\ }\textbf {\bibinfo {volume} {402}},\ \bibinfo {pages}
  {390--393} (\bibinfo {year} {1999})}\BibitemShut {NoStop}%
\bibitem [{\citenamefont {Bouwmeester}\ \emph {et~al.}(1997)\citenamefont
  {Bouwmeester}, \citenamefont {Pan}, \citenamefont {Mattle}, \citenamefont
  {Eibl}, \citenamefont {Weinfurter},\ and\ \citenamefont
  {Zeilinger}}]{97Bauwmeester}%
  \BibitemOpen
  \bibfield  {author} {\bibinfo {author} {\bibfnamefont {D.}~\bibnamefont
  {Bouwmeester}}, \bibinfo {author} {\bibfnamefont {J.-W.}\ \bibnamefont
  {Pan}}, \bibinfo {author} {\bibfnamefont {K.}~\bibnamefont {Mattle}},
  \bibinfo {author} {\bibfnamefont {M.}~\bibnamefont {Eibl}}, \bibinfo {author}
  {\bibfnamefont {H.}~\bibnamefont {Weinfurter}}, \ and\ \bibinfo {author}
  {\bibfnamefont {A.}~\bibnamefont {Zeilinger}},\ }\bibfield  {title} {\enquote
  {\bibinfo {title} {Experimental quantum teleportation},}\ }\href@noop {}
  {\bibfield  {journal} {\bibinfo  {journal} {Nature}\ }\textbf {\bibinfo
  {volume} {390}},\ \bibinfo {pages} {575--579} (\bibinfo {year}
  {1997})}\BibitemShut {NoStop}%
\bibitem [{\citenamefont {Calsamiglia}\ and\ \citenamefont
  {L{\"u}tkenhaus}(2001)}]{01Calsamiglia}%
  \BibitemOpen
  \bibfield  {author} {\bibinfo {author} {\bibfnamefont {J.}~\bibnamefont
  {Calsamiglia}}\ and\ \bibinfo {author} {\bibfnamefont {N.}~\bibnamefont
  {L{\"u}tkenhaus}},\ }\bibfield  {title} {\enquote {\bibinfo {title} {Maximum
  efficiency of a linear-optical {Bell}-state analyzer},}\ }\href {\doibase
  10.1007/s003400000484} {\bibfield  {journal} {\bibinfo  {journal} {Appl.
  Phys. B}\ }\textbf {\bibinfo {volume} {72}},\ \bibinfo {pages} {67--71}
  (\bibinfo {year} {2001})}\BibitemShut {NoStop}%
\bibitem [{\citenamefont {O'Brien}\ \emph {et~al.}(2003)\citenamefont
  {O'Brien}, \citenamefont {Pryde}, \citenamefont {White}, \citenamefont
  {Ralph},\ and\ \citenamefont {Branning}}]{03OBrien}%
  \BibitemOpen
  \bibfield  {author} {\bibinfo {author} {\bibfnamefont {J.~L.}\ \bibnamefont
  {O'Brien}}, \bibinfo {author} {\bibfnamefont {G.~J.}\ \bibnamefont {Pryde}},
  \bibinfo {author} {\bibfnamefont {A.~G.}\ \bibnamefont {White}}, \bibinfo
  {author} {\bibfnamefont {T.~C.}\ \bibnamefont {Ralph}}, \ and\ \bibinfo
  {author} {\bibfnamefont {D.}~\bibnamefont {Branning}},\ }\bibfield  {title}
  {\enquote {\bibinfo {title} {Demonstration of an all-optical quantum
  controlled-{NOT} gate},}\ }\href@noop {} {\bibfield  {journal} {\bibinfo
  {journal} {Nature}\ }\textbf {\bibinfo {volume} {426}},\ \bibinfo {pages}
  {264--267} (\bibinfo {year} {2003})}\BibitemShut {NoStop}%
\bibitem [{\citenamefont {Pittman}\ \emph {et~al.}(2003)\citenamefont
  {Pittman}, \citenamefont {Fitch}, \citenamefont {Jacobs},\ and\ \citenamefont
  {Franson}}]{03Pittman}%
  \BibitemOpen
  \bibfield  {author} {\bibinfo {author} {\bibfnamefont {T.~B.}\ \bibnamefont
  {Pittman}}, \bibinfo {author} {\bibfnamefont {M.~J.}\ \bibnamefont {Fitch}},
  \bibinfo {author} {\bibfnamefont {B.~C.}\ \bibnamefont {Jacobs}}, \ and\
  \bibinfo {author} {\bibfnamefont {J.~D.}\ \bibnamefont {Franson}},\
  }\bibfield  {title} {\enquote {\bibinfo {title} {Experimental
  controlled-{NOT} logic gate for single photons in the coincidence basis},}\
  }\href {\doibase 10.1103/PhysRevA.68.032316} {\bibfield  {journal} {\bibinfo
  {journal} {Phys. Rev. A}\ }\textbf {\bibinfo {volume} {68}},\ \bibinfo
  {pages} {032316} (\bibinfo {year} {2003})}\BibitemShut {NoStop}%
\bibitem [{\citenamefont {Gasparoni}\ \emph {et~al.}(2004)\citenamefont
  {Gasparoni}, \citenamefont {Pan}, \citenamefont {Walther}, \citenamefont
  {Rudolph},\ and\ \citenamefont {Zeilinger}}]{04Gasparoni}%
  \BibitemOpen
  \bibfield  {author} {\bibinfo {author} {\bibfnamefont {S.}~\bibnamefont
  {Gasparoni}}, \bibinfo {author} {\bibfnamefont {J.-W.}\ \bibnamefont {Pan}},
  \bibinfo {author} {\bibfnamefont {P.}~\bibnamefont {Walther}}, \bibinfo
  {author} {\bibfnamefont {T.}~\bibnamefont {Rudolph}}, \ and\ \bibinfo
  {author} {\bibfnamefont {A.}~\bibnamefont {Zeilinger}},\ }\bibfield  {title}
  {\enquote {\bibinfo {title} {Realization of a photonic controlled-{NOT} gate
  sufficient for quantum computation},}\ }\href {\doibase
  10.1103/PhysRevLett.93.020504} {\bibfield  {journal} {\bibinfo  {journal}
  {Phys. Rev. Lett.}\ }\textbf {\bibinfo {volume} {93}},\ \bibinfo {pages}
  {020504} (\bibinfo {year} {2004})}\BibitemShut {NoStop}%
\bibitem [{\citenamefont {Okamoto}\ \emph {et~al.}(2011)\citenamefont
  {Okamoto}, \citenamefont {O'Brien}, \citenamefont {Hofmann},\ and\
  \citenamefont {Takeuchi}}]{11Okamoto}%
  \BibitemOpen
  \bibfield  {author} {\bibinfo {author} {\bibfnamefont {R.}~\bibnamefont
  {Okamoto}}, \bibinfo {author} {\bibfnamefont {J.~L.}\ \bibnamefont
  {O'Brien}}, \bibinfo {author} {\bibfnamefont {H.~F.}\ \bibnamefont
  {Hofmann}}, \ and\ \bibinfo {author} {\bibfnamefont {S.}~\bibnamefont
  {Takeuchi}},\ }\bibfield  {title} {\enquote {\bibinfo {title} {Realization of
  a knill-laflamme-milburn controlled-{NOT} photonic quantum circuit combining
  effective optical nonlinearities},}\ }\href {\doibase
  10.1073/pnas.1018839108} {\bibfield  {journal} {\bibinfo  {journal} {Proc.
  Natl. Acad. Sci. USA}\ }\textbf {\bibinfo {volume} {108}},\ \bibinfo {pages}
  {10067--10071} (\bibinfo {year} {2011})}\BibitemShut {NoStop}%
\bibitem [{\citenamefont {Lu}\ \emph {et~al.}(2007)\citenamefont {Lu},
  \citenamefont {Browne}, \citenamefont {Yang},\ and\ \citenamefont
  {Pan}}]{07Lu}%
  \BibitemOpen
  \bibfield  {author} {\bibinfo {author} {\bibfnamefont {C.-Y.}\ \bibnamefont
  {Lu}}, \bibinfo {author} {\bibfnamefont {D.~E.}\ \bibnamefont {Browne}},
  \bibinfo {author} {\bibfnamefont {T.}~\bibnamefont {Yang}}, \ and\ \bibinfo
  {author} {\bibfnamefont {J.-W.}\ \bibnamefont {Pan}},\ }\bibfield  {title}
  {\enquote {\bibinfo {title} {Demonstration of a compiled version of shor's
  quantum factoring algorithm using photonic qubits},}\ }\href {\doibase
  10.1103/PhysRevLett.99.250504} {\bibfield  {journal} {\bibinfo  {journal}
  {Phys. Rev. Lett.}\ }\textbf {\bibinfo {volume} {99}},\ \bibinfo {pages}
  {250504} (\bibinfo {year} {2007})}\BibitemShut {NoStop}%
\bibitem [{\citenamefont {Lanyon}\ \emph {et~al.}(2007)\citenamefont {Lanyon},
  \citenamefont {Weinhold}, \citenamefont {Langford}, \citenamefont {Barbieri},
  \citenamefont {James}, \citenamefont {Gilchrist},\ and\ \citenamefont
  {White}}]{07Lanyon}%
  \BibitemOpen
  \bibfield  {author} {\bibinfo {author} {\bibfnamefont {B.~P.}\ \bibnamefont
  {Lanyon}}, \bibinfo {author} {\bibfnamefont {T.~J.}\ \bibnamefont
  {Weinhold}}, \bibinfo {author} {\bibfnamefont {N.~K.}\ \bibnamefont
  {Langford}}, \bibinfo {author} {\bibfnamefont {M.}~\bibnamefont {Barbieri}},
  \bibinfo {author} {\bibfnamefont {D.~F.~V.}\ \bibnamefont {James}}, \bibinfo
  {author} {\bibfnamefont {A.}~\bibnamefont {Gilchrist}}, \ and\ \bibinfo
  {author} {\bibfnamefont {A.~G.}\ \bibnamefont {White}},\ }\bibfield  {title}
  {\enquote {\bibinfo {title} {Experimental demonstration of a compiled version
  of shor's algorithm with quantum entanglement},}\ }\href {\doibase
  10.1103/PhysRevLett.99.250505} {\bibfield  {journal} {\bibinfo  {journal}
  {Phys. Rev. Lett.}\ }\textbf {\bibinfo {volume} {99}},\ \bibinfo {pages}
  {250505} (\bibinfo {year} {2007})}\BibitemShut {NoStop}%
\bibitem [{\citenamefont {Lanyon}\ \emph {et~al.}(2010)\citenamefont {Lanyon},
  \citenamefont {Whitfield}, \citenamefont {Gillett}, \citenamefont {Goggin},
  \citenamefont {Almeida}, \citenamefont {Kassal}, \citenamefont {Biamonte},
  \citenamefont {Mohseni}, \citenamefont {Powell}, \citenamefont {Barbieri},
  \citenamefont {Aspuru-Guzik},\ and\ \citenamefont {White}}]{10Lanyon}%
  \BibitemOpen
  \bibfield  {author} {\bibinfo {author} {\bibfnamefont {B.~P.}\ \bibnamefont
  {Lanyon}}, \bibinfo {author} {\bibfnamefont {J.~D.}\ \bibnamefont
  {Whitfield}}, \bibinfo {author} {\bibfnamefont {G.~G.}\ \bibnamefont
  {Gillett}}, \bibinfo {author} {\bibfnamefont {M.~E.}\ \bibnamefont {Goggin}},
  \bibinfo {author} {\bibfnamefont {M.~P.}\ \bibnamefont {Almeida}}, \bibinfo
  {author} {\bibfnamefont {I.}~\bibnamefont {Kassal}}, \bibinfo {author}
  {\bibfnamefont {J.~D.}\ \bibnamefont {Biamonte}}, \bibinfo {author}
  {\bibfnamefont {M.}~\bibnamefont {Mohseni}}, \bibinfo {author} {\bibfnamefont
  {B.~J.}\ \bibnamefont {Powell}}, \bibinfo {author} {\bibfnamefont
  {M.}~\bibnamefont {Barbieri}}, \bibinfo {author} {\bibfnamefont
  {A.}~\bibnamefont {Aspuru-Guzik}}, \ and\ \bibinfo {author} {\bibfnamefont
  {A.~G.}\ \bibnamefont {White}},\ }\bibfield  {title} {\enquote {\bibinfo
  {title} {Towards quantum chemistry on a quantum computer},}\ }\href@noop {}
  {\bibfield  {journal} {\bibinfo  {journal} {Nat. Chem.}\ }\textbf {\bibinfo
  {volume} {2}},\ \bibinfo {pages} {106--111} (\bibinfo {year}
  {2010})}\BibitemShut {NoStop}%
\bibitem [{\citenamefont {Peruzzo}\ \emph {et~al.}(2014)\citenamefont
  {Peruzzo}, \citenamefont {McClean}, \citenamefont {Shadbolt}, \citenamefont
  {Yung}, \citenamefont {Zhou}, \citenamefont {Love}, \citenamefont
  {Aspuru-Guzik},\ and\ \citenamefont {O'Brien}}]{14Peruzzo}%
  \BibitemOpen
  \bibfield  {author} {\bibinfo {author} {\bibfnamefont {A.}~\bibnamefont
  {Peruzzo}}, \bibinfo {author} {\bibfnamefont {J.}~\bibnamefont {McClean}},
  \bibinfo {author} {\bibfnamefont {P.}~\bibnamefont {Shadbolt}}, \bibinfo
  {author} {\bibfnamefont {M.-H.}\ \bibnamefont {Yung}}, \bibinfo {author}
  {\bibfnamefont {X.-Q.}\ \bibnamefont {Zhou}}, \bibinfo {author}
  {\bibfnamefont {P.~J.}\ \bibnamefont {Love}}, \bibinfo {author}
  {\bibfnamefont {A.}~\bibnamefont {Aspuru-Guzik}}, \ and\ \bibinfo {author}
  {\bibfnamefont {J.~L.}\ \bibnamefont {O'Brien}},\ }\bibfield  {title}
  {\enquote {\bibinfo {title} {A variational eigenvalue solver on a photonic
  quantum processor},}\ }\href@noop {} {\bibfield  {journal} {\bibinfo
  {journal} {Nat. Commun.}\ }\textbf {\bibinfo {volume} {5}},\ \bibinfo {pages}
  {4213} (\bibinfo {year} {2014})}\BibitemShut {NoStop}%
\bibitem [{\citenamefont {Pittman}, \citenamefont {Jacobs},\ and\ \citenamefont
  {Franson}(2005)}]{05Pittman}%
  \BibitemOpen
  \bibfield  {author} {\bibinfo {author} {\bibfnamefont {T.~B.}\ \bibnamefont
  {Pittman}}, \bibinfo {author} {\bibfnamefont {B.~C.}\ \bibnamefont {Jacobs}},
  \ and\ \bibinfo {author} {\bibfnamefont {J.~D.}\ \bibnamefont {Franson}},\
  }\bibfield  {title} {\enquote {\bibinfo {title} {Demonstration of quantum
  error correction using linear optics},}\ }\href {\doibase
  10.1103/PhysRevA.71.052332} {\bibfield  {journal} {\bibinfo  {journal} {Phys.
  Rev. A}\ }\textbf {\bibinfo {volume} {71}},\ \bibinfo {pages} {052332}
  (\bibinfo {year} {2005})}\BibitemShut {NoStop}%
\bibitem [{\citenamefont {O'Brien}\ \emph {et~al.}(2005)\citenamefont
  {O'Brien}, \citenamefont {Pryde}, \citenamefont {White},\ and\ \citenamefont
  {Ralph}}]{05OBrien}%
  \BibitemOpen
  \bibfield  {author} {\bibinfo {author} {\bibfnamefont {J.~L.}\ \bibnamefont
  {O'Brien}}, \bibinfo {author} {\bibfnamefont {G.~J.}\ \bibnamefont {Pryde}},
  \bibinfo {author} {\bibfnamefont {A.~G.}\ \bibnamefont {White}}, \ and\
  \bibinfo {author} {\bibfnamefont {T.~C.}\ \bibnamefont {Ralph}},\ }\bibfield
  {title} {\enquote {\bibinfo {title} {High-fidelity $z$-measurement error
  encoding of optical qubits},}\ }\href {\doibase 10.1103/PhysRevA.71.060303}
  {\bibfield  {journal} {\bibinfo  {journal} {Phys. Rev. A}\ }\textbf {\bibinfo
  {volume} {71}},\ \bibinfo {pages} {060303} (\bibinfo {year}
  {2005})}\BibitemShut {NoStop}%
\bibitem [{\citenamefont {Lu}\ \emph {et~al.}(2008)\citenamefont {Lu},
  \citenamefont {Gao}, \citenamefont {Zhang}, \citenamefont {Zhou},
  \citenamefont {Yang},\ and\ \citenamefont {Pan}}]{08Lu}%
  \BibitemOpen
  \bibfield  {author} {\bibinfo {author} {\bibfnamefont {C.-Y.}\ \bibnamefont
  {Lu}}, \bibinfo {author} {\bibfnamefont {W.-B.}\ \bibnamefont {Gao}},
  \bibinfo {author} {\bibfnamefont {J.}~\bibnamefont {Zhang}}, \bibinfo
  {author} {\bibfnamefont {X.-Q.}\ \bibnamefont {Zhou}}, \bibinfo {author}
  {\bibfnamefont {T.}~\bibnamefont {Yang}}, \ and\ \bibinfo {author}
  {\bibfnamefont {J.-W.}\ \bibnamefont {Pan}},\ }\bibfield  {title} {\enquote
  {\bibinfo {title} {Experimental quantum coding against qubit loss error},}\
  }\href {\doibase 10.1073/pnas.0800740105} {\bibfield  {journal} {\bibinfo
  {journal} {Proc. Natl. Acad. Sci. USA}\ }\textbf {\bibinfo {volume} {105}},\
  \bibinfo {pages} {11050--11054} (\bibinfo {year} {2008})}\BibitemShut
  {NoStop}%
\bibitem [{\citenamefont {Raussendorf}\ and\ \citenamefont
  {Briegel}(2001)}]{01Raussendorf}%
  \BibitemOpen
  \bibfield  {author} {\bibinfo {author} {\bibfnamefont {R.}~\bibnamefont
  {Raussendorf}}\ and\ \bibinfo {author} {\bibfnamefont {H.~J.}\ \bibnamefont
  {Briegel}},\ }\bibfield  {title} {\enquote {\bibinfo {title} {A one-way
  quantum computer},}\ }\href {\doibase 10.1103/PhysRevLett.86.5188} {\bibfield
   {journal} {\bibinfo  {journal} {Phys. Rev. Lett.}\ }\textbf {\bibinfo
  {volume} {86}},\ \bibinfo {pages} {5188--5191} (\bibinfo {year}
  {2001})}\BibitemShut {NoStop}%
\bibitem [{\citenamefont {Nielsen}(2004)}]{04Nielsen}%
  \BibitemOpen
  \bibfield  {author} {\bibinfo {author} {\bibfnamefont {M.~A.}\ \bibnamefont
  {Nielsen}},\ }\bibfield  {title} {\enquote {\bibinfo {title} {Optical quantum
  computation using cluster states},}\ }\href {\doibase
  10.1103/PhysRevLett.93.040503} {\bibfield  {journal} {\bibinfo  {journal}
  {Phys. Rev. Lett.}\ }\textbf {\bibinfo {volume} {93}},\ \bibinfo {pages}
  {040503} (\bibinfo {year} {2004})}\BibitemShut {NoStop}%
\bibitem [{\citenamefont {Walther}\ \emph {et~al.}(2005)\citenamefont
  {Walther}, \citenamefont {Resch}, \citenamefont {Rudolph}, \citenamefont
  {Schenck}, \citenamefont {Weinfurter}, \citenamefont {Vedral}, \citenamefont
  {Aspelmeyer},\ and\ \citenamefont {Zeilinger}}]{05Walther}%
  \BibitemOpen
  \bibfield  {author} {\bibinfo {author} {\bibfnamefont {P.}~\bibnamefont
  {Walther}}, \bibinfo {author} {\bibfnamefont {K.~J.}\ \bibnamefont {Resch}},
  \bibinfo {author} {\bibfnamefont {T.}~\bibnamefont {Rudolph}}, \bibinfo
  {author} {\bibfnamefont {E.}~\bibnamefont {Schenck}}, \bibinfo {author}
  {\bibfnamefont {H.}~\bibnamefont {Weinfurter}}, \bibinfo {author}
  {\bibfnamefont {V.}~\bibnamefont {Vedral}}, \bibinfo {author} {\bibfnamefont
  {M.}~\bibnamefont {Aspelmeyer}}, \ and\ \bibinfo {author} {\bibfnamefont
  {A.}~\bibnamefont {Zeilinger}},\ }\bibfield  {title} {\enquote {\bibinfo
  {title} {Experimental one-way quantum computing},}\ }\href@noop {} {\bibfield
   {journal} {\bibinfo  {journal} {Nature}\ }\textbf {\bibinfo {volume}
  {434}},\ \bibinfo {pages} {169} (\bibinfo {year} {2005})}\BibitemShut
  {NoStop}%
\bibitem [{\citenamefont {Kiesel}\ \emph {et~al.}(2005)\citenamefont {Kiesel},
  \citenamefont {Schmid}, \citenamefont {Weber}, \citenamefont {T\'oth},
  \citenamefont {G\"uhne}, \citenamefont {Ursin},\ and\ \citenamefont
  {Weinfurter}}]{05Kiesel}%
  \BibitemOpen
  \bibfield  {author} {\bibinfo {author} {\bibfnamefont {N.}~\bibnamefont
  {Kiesel}}, \bibinfo {author} {\bibfnamefont {C.}~\bibnamefont {Schmid}},
  \bibinfo {author} {\bibfnamefont {U.}~\bibnamefont {Weber}}, \bibinfo
  {author} {\bibfnamefont {G.}~\bibnamefont {T\'oth}}, \bibinfo {author}
  {\bibfnamefont {O.}~\bibnamefont {G\"uhne}}, \bibinfo {author} {\bibfnamefont
  {R.}~\bibnamefont {Ursin}}, \ and\ \bibinfo {author} {\bibfnamefont
  {H.}~\bibnamefont {Weinfurter}},\ }\bibfield  {title} {\enquote {\bibinfo
  {title} {Experimental analysis of a four-qubit photon cluster state},}\
  }\href {\doibase 10.1103/PhysRevLett.95.210502} {\bibfield  {journal}
  {\bibinfo  {journal} {Phys. Rev. Lett.}\ }\textbf {\bibinfo {volume} {95}},\
  \bibinfo {pages} {210502} (\bibinfo {year} {2005})}\BibitemShut {NoStop}%
\bibitem [{\citenamefont {Prevedel}\ \emph {et~al.}(2007)\citenamefont
  {Prevedel}, \citenamefont {Walther}, \citenamefont {Tiefenbacher},
  \citenamefont {B\"{o}hi}, \citenamefont {Kaltenbaek}, \citenamefont
  {Jennewein},\ and\ \citenamefont {Zeilinger}}]{07Prevedel}%
  \BibitemOpen
  \bibfield  {author} {\bibinfo {author} {\bibfnamefont {R.}~\bibnamefont
  {Prevedel}}, \bibinfo {author} {\bibfnamefont {P.}~\bibnamefont {Walther}},
  \bibinfo {author} {\bibfnamefont {F.}~\bibnamefont {Tiefenbacher}}, \bibinfo
  {author} {\bibfnamefont {P.}~\bibnamefont {B\"{o}hi}}, \bibinfo {author}
  {\bibfnamefont {R.}~\bibnamefont {Kaltenbaek}}, \bibinfo {author}
  {\bibfnamefont {T.}~\bibnamefont {Jennewein}}, \ and\ \bibinfo {author}
  {\bibfnamefont {A.}~\bibnamefont {Zeilinger}},\ }\bibfield  {title} {\enquote
  {\bibinfo {title} {High-speed linear optics quantum computing using active
  feed-forward},}\ }\href@noop {} {\bibfield  {journal} {\bibinfo  {journal}
  {Nature}\ }\textbf {\bibinfo {volume} {445}},\ \bibinfo {pages} {65}
  (\bibinfo {year} {2007})}\BibitemShut {NoStop}%
\bibitem [{\citenamefont {Turchette}\ \emph {et~al.}(1995)\citenamefont
  {Turchette}, \citenamefont {Hood}, \citenamefont {Lange}, \citenamefont
  {Mabuchi},\ and\ \citenamefont {Kimble}}]{95Turchette}%
  \BibitemOpen
  \bibfield  {author} {\bibinfo {author} {\bibfnamefont {Q.~A.}\ \bibnamefont
  {Turchette}}, \bibinfo {author} {\bibfnamefont {C.~J.}\ \bibnamefont {Hood}},
  \bibinfo {author} {\bibfnamefont {W.}~\bibnamefont {Lange}}, \bibinfo
  {author} {\bibfnamefont {H.}~\bibnamefont {Mabuchi}}, \ and\ \bibinfo
  {author} {\bibfnamefont {H.~J.}\ \bibnamefont {Kimble}},\ }\bibfield  {title}
  {\enquote {\bibinfo {title} {Measurement of conditional phase shifts for
  quantum logic},}\ }\href {\doibase 10.1103/PhysRevLett.75.4710} {\bibfield
  {journal} {\bibinfo  {journal} {Phys. Rev. Lett.}\ }\textbf {\bibinfo
  {volume} {75}},\ \bibinfo {pages} {4710--4713} (\bibinfo {year}
  {1995})}\BibitemShut {NoStop}%
\bibitem [{\citenamefont {Duan}\ and\ \citenamefont {Kimble}(2004)}]{04Duan}%
  \BibitemOpen
  \bibfield  {author} {\bibinfo {author} {\bibfnamefont {L.-M.}\ \bibnamefont
  {Duan}}\ and\ \bibinfo {author} {\bibfnamefont {H.~J.}\ \bibnamefont
  {Kimble}},\ }\bibfield  {title} {\enquote {\bibinfo {title} {Scalable
  photonic quantum computation through cavity-assisted interactions},}\ }\href
  {\doibase 10.1103/PhysRevLett.92.127902} {\bibfield  {journal} {\bibinfo
  {journal} {Phys. Rev. Lett.}\ }\textbf {\bibinfo {volume} {92}},\ \bibinfo
  {pages} {127902} (\bibinfo {year} {2004})}\BibitemShut {NoStop}%
\bibitem [{\citenamefont {Hacker}\ \emph {et~al.}(2016)\citenamefont {Hacker},
  \citenamefont {Welte}, \citenamefont {Rempe},\ and\ \citenamefont
  {Ritter}}]{16Hacker}%
  \BibitemOpen
  \bibfield  {author} {\bibinfo {author} {\bibfnamefont {B.}~\bibnamefont
  {Hacker}}, \bibinfo {author} {\bibfnamefont {S.}~\bibnamefont {Welte}},
  \bibinfo {author} {\bibfnamefont {G.}~\bibnamefont {Rempe}}, \ and\ \bibinfo
  {author} {\bibfnamefont {S.}~\bibnamefont {Ritter}},\ }\bibfield  {title}
  {\enquote {\bibinfo {title} {A photon-photon quantum gate based on a single
  atom in an optical resonator},}\ }\href@noop {} {\bibfield  {journal}
  {\bibinfo  {journal} {Nature}\ }\textbf {\bibinfo {volume} {536}},\ \bibinfo
  {pages} {193--196} (\bibinfo {year} {2016})}\BibitemShut {NoStop}%
\bibitem [{\citenamefont {Vaidman}(1994)}]{94Vaidman}%
  \BibitemOpen
  \bibfield  {author} {\bibinfo {author} {\bibfnamefont {L.}~\bibnamefont
  {Vaidman}},\ }\bibfield  {title} {\enquote {\bibinfo {title} {Teleportation
  of quantum states},}\ }\href {\doibase 10.1103/PhysRevA.49.1473} {\bibfield
  {journal} {\bibinfo  {journal} {Phys. Rev. A}\ }\textbf {\bibinfo {volume}
  {49}},\ \bibinfo {pages} {1473--1476} (\bibinfo {year} {1994})}\BibitemShut
  {NoStop}%
\bibitem [{\citenamefont {Furusawa}\ \emph {et~al.}(1998)\citenamefont
  {Furusawa}, \citenamefont {S{\o}rensen}, \citenamefont {Braunstein},
  \citenamefont {Fuchs}, \citenamefont {Kimble},\ and\ \citenamefont
  {Polzik}}]{98Furusawa}%
  \BibitemOpen
  \bibfield  {author} {\bibinfo {author} {\bibfnamefont {A.}~\bibnamefont
  {Furusawa}}, \bibinfo {author} {\bibfnamefont {J.~L.}\ \bibnamefont
  {S{\o}rensen}}, \bibinfo {author} {\bibfnamefont {S.~L.}\ \bibnamefont
  {Braunstein}}, \bibinfo {author} {\bibfnamefont {C.~A.}\ \bibnamefont
  {Fuchs}}, \bibinfo {author} {\bibfnamefont {H.~J.}\ \bibnamefont {Kimble}}, \
  and\ \bibinfo {author} {\bibfnamefont {E.~S.}\ \bibnamefont {Polzik}},\
  }\bibfield  {title} {\enquote {\bibinfo {title} {Unconditional quantum
  teleportation},}\ }\href {\doibase 10.1126/science.282.5389.706} {\bibfield
  {journal} {\bibinfo  {journal} {Science}\ }\textbf {\bibinfo {volume}
  {282}},\ \bibinfo {pages} {706--709} (\bibinfo {year} {1998})}\BibitemShut
  {NoStop}%
\bibitem [{\citenamefont {Filip}, \citenamefont {Marek},\ and\ \citenamefont
  {Andersen}(2005)}]{05Filip}%
  \BibitemOpen
  \bibfield  {author} {\bibinfo {author} {\bibfnamefont {R.}~\bibnamefont
  {Filip}}, \bibinfo {author} {\bibfnamefont {P.}~\bibnamefont {Marek}}, \ and\
  \bibinfo {author} {\bibfnamefont {U.~L.}\ \bibnamefont {Andersen}},\
  }\bibfield  {title} {\enquote {\bibinfo {title} {Measurement-induced
  continuous-variable quantum interactions},}\ }\href {\doibase
  10.1103/PhysRevA.71.042308} {\bibfield  {journal} {\bibinfo  {journal} {Phys.
  Rev. A}\ }\textbf {\bibinfo {volume} {71}},\ \bibinfo {pages} {042308}
  (\bibinfo {year} {2005})}\BibitemShut {NoStop}%
\bibitem [{\citenamefont {Lloyd}\ and\ \citenamefont
  {Braunstein}(1999)}]{99Lloyd}%
  \BibitemOpen
  \bibfield  {author} {\bibinfo {author} {\bibfnamefont {S.}~\bibnamefont
  {Lloyd}}\ and\ \bibinfo {author} {\bibfnamefont {S.~L.}\ \bibnamefont
  {Braunstein}},\ }\bibfield  {title} {\enquote {\bibinfo {title} {Quantum
  computation over continuous variables},}\ }\href {\doibase
  10.1103/PhysRevLett.82.1784} {\bibfield  {journal} {\bibinfo  {journal}
  {Phys. Rev. Lett.}\ }\textbf {\bibinfo {volume} {82}},\ \bibinfo {pages}
  {1784--1787} (\bibinfo {year} {1999})}\BibitemShut {NoStop}%
\bibitem [{\citenamefont {Ukai}\ \emph
  {et~al.}(2011{\natexlab{a}})\citenamefont {Ukai}, \citenamefont {Iwata},
  \citenamefont {Shimokawa}, \citenamefont {Armstrong}, \citenamefont {Politi},
  \citenamefont {Yoshikawa}, \citenamefont {van Loock},\ and\ \citenamefont
  {Furusawa}}]{11Ukai1}%
  \BibitemOpen
  \bibfield  {author} {\bibinfo {author} {\bibfnamefont {R.}~\bibnamefont
  {Ukai}}, \bibinfo {author} {\bibfnamefont {N.}~\bibnamefont {Iwata}},
  \bibinfo {author} {\bibfnamefont {Y.}~\bibnamefont {Shimokawa}}, \bibinfo
  {author} {\bibfnamefont {S.~C.}\ \bibnamefont {Armstrong}}, \bibinfo {author}
  {\bibfnamefont {A.}~\bibnamefont {Politi}}, \bibinfo {author} {\bibfnamefont
  {J.}~\bibnamefont {Yoshikawa}}, \bibinfo {author} {\bibfnamefont
  {P.}~\bibnamefont {van Loock}}, \ and\ \bibinfo {author} {\bibfnamefont
  {A.}~\bibnamefont {Furusawa}},\ }\bibfield  {title} {\enquote {\bibinfo
  {title} {Demonstration of unconditional one-way quantum computations for
  continuous variables},}\ }\href {\doibase 10.1103/PhysRevLett.106.240504}
  {\bibfield  {journal} {\bibinfo  {journal} {Phys. Rev. Lett.}\ }\textbf
  {\bibinfo {volume} {106}},\ \bibinfo {pages} {240504} (\bibinfo {year}
  {2011}{\natexlab{a}})}\BibitemShut {NoStop}%
\bibitem [{\citenamefont {Bartlett}\ and\ \citenamefont
  {Munro}(2003)}]{03Bartlett}%
  \BibitemOpen
  \bibfield  {author} {\bibinfo {author} {\bibfnamefont {S.~D.}\ \bibnamefont
  {Bartlett}}\ and\ \bibinfo {author} {\bibfnamefont {W.~J.}\ \bibnamefont
  {Munro}},\ }\bibfield  {title} {\enquote {\bibinfo {title} {Quantum
  teleportation of optical quantum gates},}\ }\href {\doibase
  10.1103/PhysRevLett.90.117901} {\bibfield  {journal} {\bibinfo  {journal}
  {Phys. Rev. Lett.}\ }\textbf {\bibinfo {volume} {90}},\ \bibinfo {pages}
  {117901} (\bibinfo {year} {2003})}\BibitemShut {NoStop}%
\bibitem [{\citenamefont {Yoshikawa}\ \emph {et~al.}(2007)\citenamefont
  {Yoshikawa}, \citenamefont {Hayashi}, \citenamefont {Akiyama}, \citenamefont
  {Takei}, \citenamefont {Huck}, \citenamefont {Andersen},\ and\ \citenamefont
  {Furusawa}}]{07Yoshikawa}%
  \BibitemOpen
  \bibfield  {author} {\bibinfo {author} {\bibfnamefont {J.}~\bibnamefont
  {Yoshikawa}}, \bibinfo {author} {\bibfnamefont {T.}~\bibnamefont {Hayashi}},
  \bibinfo {author} {\bibfnamefont {T.}~\bibnamefont {Akiyama}}, \bibinfo
  {author} {\bibfnamefont {N.}~\bibnamefont {Takei}}, \bibinfo {author}
  {\bibfnamefont {A.}~\bibnamefont {Huck}}, \bibinfo {author} {\bibfnamefont
  {U.~L.}\ \bibnamefont {Andersen}}, \ and\ \bibinfo {author} {\bibfnamefont
  {A.}~\bibnamefont {Furusawa}},\ }\bibfield  {title} {\enquote {\bibinfo
  {title} {Demonstration of deterministic and high fidelity squeezing of
  quantum information},}\ }\href {\doibase 10.1103/PhysRevA.76.060301}
  {\bibfield  {journal} {\bibinfo  {journal} {Phys. Rev. A}\ }\textbf {\bibinfo
  {volume} {76}},\ \bibinfo {pages} {060301} (\bibinfo {year}
  {2007})}\BibitemShut {NoStop}%
\bibitem [{\citenamefont {Yoshikawa}\ \emph {et~al.}(2008)\citenamefont
  {Yoshikawa}, \citenamefont {Miwa}, \citenamefont {Huck}, \citenamefont
  {Andersen}, \citenamefont {van Loock},\ and\ \citenamefont
  {Furusawa}}]{08Yoshikawa}%
  \BibitemOpen
  \bibfield  {author} {\bibinfo {author} {\bibfnamefont {J.}~\bibnamefont
  {Yoshikawa}}, \bibinfo {author} {\bibfnamefont {Y.}~\bibnamefont {Miwa}},
  \bibinfo {author} {\bibfnamefont {A.}~\bibnamefont {Huck}}, \bibinfo {author}
  {\bibfnamefont {U.~L.}\ \bibnamefont {Andersen}}, \bibinfo {author}
  {\bibfnamefont {P.}~\bibnamefont {van Loock}}, \ and\ \bibinfo {author}
  {\bibfnamefont {A.}~\bibnamefont {Furusawa}},\ }\bibfield  {title} {\enquote
  {\bibinfo {title} {Demonstration of a quantum nondemolition sum gate},}\
  }\href {\doibase 10.1103/PhysRevLett.101.250501} {\bibfield  {journal}
  {\bibinfo  {journal} {Phys. Rev. Lett.}\ }\textbf {\bibinfo {volume} {101}},\
  \bibinfo {pages} {250501} (\bibinfo {year} {2008})}\BibitemShut {NoStop}%
\bibitem [{\citenamefont {Yukawa}\ \emph {et~al.}(2013)\citenamefont {Yukawa},
  \citenamefont {Miyata}, \citenamefont {Yonezawa}, \citenamefont {Marek},
  \citenamefont {Filip},\ and\ \citenamefont {Furusawa}}]{13Yukawa}%
  \BibitemOpen
  \bibfield  {author} {\bibinfo {author} {\bibfnamefont {M.}~\bibnamefont
  {Yukawa}}, \bibinfo {author} {\bibfnamefont {K.}~\bibnamefont {Miyata}},
  \bibinfo {author} {\bibfnamefont {H.}~\bibnamefont {Yonezawa}}, \bibinfo
  {author} {\bibfnamefont {P.}~\bibnamefont {Marek}}, \bibinfo {author}
  {\bibfnamefont {R.}~\bibnamefont {Filip}}, \ and\ \bibinfo {author}
  {\bibfnamefont {A.}~\bibnamefont {Furusawa}},\ }\bibfield  {title} {\enquote
  {\bibinfo {title} {Emulating quantum cubic nonlinearity},}\ }\href {\doibase
  10.1103/PhysRevA.88.053816} {\bibfield  {journal} {\bibinfo  {journal} {Phys.
  Rev. A}\ }\textbf {\bibinfo {volume} {88}},\ \bibinfo {pages} {053816}
  (\bibinfo {year} {2013})}\BibitemShut {NoStop}%
\bibitem [{\citenamefont {Yoshikawa}\ \emph {et~al.}(2013)\citenamefont
  {Yoshikawa}, \citenamefont {Makino}, \citenamefont {Kurata}, \citenamefont
  {van Loock},\ and\ \citenamefont {Furusawa}}]{13Yoshikawa}%
  \BibitemOpen
  \bibfield  {author} {\bibinfo {author} {\bibfnamefont {J.}~\bibnamefont
  {Yoshikawa}}, \bibinfo {author} {\bibfnamefont {K.}~\bibnamefont {Makino}},
  \bibinfo {author} {\bibfnamefont {S.}~\bibnamefont {Kurata}}, \bibinfo
  {author} {\bibfnamefont {P.}~\bibnamefont {van Loock}}, \ and\ \bibinfo
  {author} {\bibfnamefont {A.}~\bibnamefont {Furusawa}},\ }\bibfield  {title}
  {\enquote {\bibinfo {title} {Creation, storage, and on-demand release of
  optical quantum states with a negative {W}igner function},}\ }\href {\doibase
  10.1103/PhysRevX.3.041028} {\bibfield  {journal} {\bibinfo  {journal} {Phys.
  Rev. X}\ }\textbf {\bibinfo {volume} {3}},\ \bibinfo {pages} {041028}
  (\bibinfo {year} {2013})}\BibitemShut {NoStop}%
\bibitem [{\citenamefont {Hashimoto}\ \emph {et~al.}(2018)\citenamefont
  {Hashimoto}, \citenamefont {Toyama}, \citenamefont {Yoshikawa}, \citenamefont
  {Makino}, \citenamefont {Okamoto}, \citenamefont {Sakakibara}, \citenamefont
  {Takeda}, \citenamefont {van Loock},\ and\ \citenamefont
  {Furusawa}}]{18Hashimoto}%
  \BibitemOpen
  \bibfield  {author} {\bibinfo {author} {\bibfnamefont {Y.}~\bibnamefont
  {Hashimoto}}, \bibinfo {author} {\bibfnamefont {T.}~\bibnamefont {Toyama}},
  \bibinfo {author} {\bibfnamefont {J.}~\bibnamefont {Yoshikawa}}, \bibinfo
  {author} {\bibfnamefont {K.}~\bibnamefont {Makino}}, \bibinfo {author}
  {\bibfnamefont {F.}~\bibnamefont {Okamoto}}, \bibinfo {author} {\bibfnamefont
  {R.}~\bibnamefont {Sakakibara}}, \bibinfo {author} {\bibfnamefont
  {S.}~\bibnamefont {Takeda}}, \bibinfo {author} {\bibfnamefont
  {P.}~\bibnamefont {van Loock}}, \ and\ \bibinfo {author} {\bibfnamefont
  {A.}~\bibnamefont {Furusawa}},\ }\href@noop {} {\enquote {\bibinfo {title}
  {All-optical storage of phase-sensitive quantum states of light},}\ }\bibinfo
  {howpublished} {arXiv:1810.10741} (\bibinfo {year} {2018})\BibitemShut
  {NoStop}%
\bibitem [{\citenamefont {Miyata}\ \emph {et~al.}(2014)\citenamefont {Miyata},
  \citenamefont {Ogawa}, \citenamefont {Marek}, \citenamefont {Filip},
  \citenamefont {Yonezawa}, \citenamefont {Yoshikawa},\ and\ \citenamefont
  {Furusawa}}]{14Miyata}%
  \BibitemOpen
  \bibfield  {author} {\bibinfo {author} {\bibfnamefont {K.}~\bibnamefont
  {Miyata}}, \bibinfo {author} {\bibfnamefont {H.}~\bibnamefont {Ogawa}},
  \bibinfo {author} {\bibfnamefont {P.}~\bibnamefont {Marek}}, \bibinfo
  {author} {\bibfnamefont {R.}~\bibnamefont {Filip}}, \bibinfo {author}
  {\bibfnamefont {H.}~\bibnamefont {Yonezawa}}, \bibinfo {author}
  {\bibfnamefont {J.}~\bibnamefont {Yoshikawa}}, \ and\ \bibinfo {author}
  {\bibfnamefont {A.}~\bibnamefont {Furusawa}},\ }\bibfield  {title} {\enquote
  {\bibinfo {title} {Experimental realization of a dynamic squeezing gate},}\
  }\href {\doibase 10.1103/PhysRevA.90.060302} {\bibfield  {journal} {\bibinfo
  {journal} {Phys. Rev. A}\ }\textbf {\bibinfo {volume} {90}},\ \bibinfo
  {pages} {060302} (\bibinfo {year} {2014})}\BibitemShut {NoStop}%
\bibitem [{\citenamefont {Menicucci}\ \emph {et~al.}(2006)\citenamefont
  {Menicucci}, \citenamefont {van Loock}, \citenamefont {Gu}, \citenamefont
  {Weedbrook}, \citenamefont {Ralph},\ and\ \citenamefont
  {Nielsen}}]{06Menicucci}%
  \BibitemOpen
  \bibfield  {author} {\bibinfo {author} {\bibfnamefont {N.~C.}\ \bibnamefont
  {Menicucci}}, \bibinfo {author} {\bibfnamefont {P.}~\bibnamefont {van
  Loock}}, \bibinfo {author} {\bibfnamefont {M.}~\bibnamefont {Gu}}, \bibinfo
  {author} {\bibfnamefont {C.}~\bibnamefont {Weedbrook}}, \bibinfo {author}
  {\bibfnamefont {T.~C.}\ \bibnamefont {Ralph}}, \ and\ \bibinfo {author}
  {\bibfnamefont {M.~A.}\ \bibnamefont {Nielsen}},\ }\bibfield  {title}
  {\enquote {\bibinfo {title} {Universal quantum computation with
  continuous-variable cluster states},}\ }\href {\doibase
  10.1103/PhysRevLett.97.110501} {\bibfield  {journal} {\bibinfo  {journal}
  {Phys. Rev. Lett.}\ }\textbf {\bibinfo {volume} {97}},\ \bibinfo {pages}
  {110501} (\bibinfo {year} {2006})}\BibitemShut {NoStop}%
\bibitem [{\citenamefont {van Loock}, \citenamefont {Weedbrook},\ and\
  \citenamefont {Gu}(2007)}]{07vanLoock}%
  \BibitemOpen
  \bibfield  {author} {\bibinfo {author} {\bibfnamefont {P.}~\bibnamefont {van
  Loock}}, \bibinfo {author} {\bibfnamefont {C.}~\bibnamefont {Weedbrook}}, \
  and\ \bibinfo {author} {\bibfnamefont {M.}~\bibnamefont {Gu}},\ }\bibfield
  {title} {\enquote {\bibinfo {title} {Building {Gaussian} cluster states by
  linear optics},}\ }\href {\doibase 10.1103/PhysRevA.76.032321} {\bibfield
  {journal} {\bibinfo  {journal} {Phys. Rev. A}\ }\textbf {\bibinfo {volume}
  {76}},\ \bibinfo {pages} {032321} (\bibinfo {year} {2007})}\BibitemShut
  {NoStop}%
\bibitem [{\citenamefont {Gu}\ \emph {et~al.}(2009)\citenamefont {Gu},
  \citenamefont {Weedbrook}, \citenamefont {Menicucci}, \citenamefont {Ralph},\
  and\ \citenamefont {van Loock}}]{09Gu}%
  \BibitemOpen
  \bibfield  {author} {\bibinfo {author} {\bibfnamefont {M.}~\bibnamefont
  {Gu}}, \bibinfo {author} {\bibfnamefont {C.}~\bibnamefont {Weedbrook}},
  \bibinfo {author} {\bibfnamefont {N.~C.}\ \bibnamefont {Menicucci}}, \bibinfo
  {author} {\bibfnamefont {T.~C.}\ \bibnamefont {Ralph}}, \ and\ \bibinfo
  {author} {\bibfnamefont {P.}~\bibnamefont {van Loock}},\ }\bibfield  {title}
  {\enquote {\bibinfo {title} {Quantum computing with continuous-variable
  clusters},}\ }\href {\doibase 10.1103/PhysRevA.79.062318} {\bibfield
  {journal} {\bibinfo  {journal} {Phys. Rev. A}\ }\textbf {\bibinfo {volume}
  {79}},\ \bibinfo {pages} {062318} (\bibinfo {year} {2009})}\BibitemShut
  {NoStop}%
\bibitem [{\citenamefont {Alexander}\ \emph {et~al.}(2018)\citenamefont
  {Alexander}, \citenamefont {Yokoyama}, \citenamefont {Furusawa},\ and\
  \citenamefont {Menicucci}}]{18Alexander}%
  \BibitemOpen
  \bibfield  {author} {\bibinfo {author} {\bibfnamefont {R.~N.}\ \bibnamefont
  {Alexander}}, \bibinfo {author} {\bibfnamefont {S.}~\bibnamefont {Yokoyama}},
  \bibinfo {author} {\bibfnamefont {A.}~\bibnamefont {Furusawa}}, \ and\
  \bibinfo {author} {\bibfnamefont {N.~C.}\ \bibnamefont {Menicucci}},\
  }\bibfield  {title} {\enquote {\bibinfo {title} {Universal quantum
  computation with temporal-mode bilayer square lattices},}\ }\href {\doibase
  10.1103/PhysRevA.97.032302} {\bibfield  {journal} {\bibinfo  {journal} {Phys.
  Rev. A}\ }\textbf {\bibinfo {volume} {97}},\ \bibinfo {pages} {032302}
  (\bibinfo {year} {2018})}\BibitemShut {NoStop}%
\bibitem [{\citenamefont {Yukawa}\ \emph {et~al.}(2008)\citenamefont {Yukawa},
  \citenamefont {Ukai}, \citenamefont {van Loock},\ and\ \citenamefont
  {Furusawa}}]{08Yukawa}%
  \BibitemOpen
  \bibfield  {author} {\bibinfo {author} {\bibfnamefont {M.}~\bibnamefont
  {Yukawa}}, \bibinfo {author} {\bibfnamefont {R.}~\bibnamefont {Ukai}},
  \bibinfo {author} {\bibfnamefont {P.}~\bibnamefont {van Loock}}, \ and\
  \bibinfo {author} {\bibfnamefont {A.}~\bibnamefont {Furusawa}},\ }\bibfield
  {title} {\enquote {\bibinfo {title} {Experimental generation of four-mode
  continuous-variable cluster states},}\ }\href {\doibase
  10.1103/PhysRevA.78.012301} {\bibfield  {journal} {\bibinfo  {journal} {Phys.
  Rev. A}\ }\textbf {\bibinfo {volume} {78}},\ \bibinfo {pages} {012301}
  (\bibinfo {year} {2008})}\BibitemShut {NoStop}%
\bibitem [{\citenamefont {Ukai}\ \emph
  {et~al.}(2011{\natexlab{b}})\citenamefont {Ukai}, \citenamefont {Yokoyama},
  \citenamefont {Yoshikawa}, \citenamefont {van Loock},\ and\ \citenamefont
  {Furusawa}}]{11Ukai2}%
  \BibitemOpen
  \bibfield  {author} {\bibinfo {author} {\bibfnamefont {R.}~\bibnamefont
  {Ukai}}, \bibinfo {author} {\bibfnamefont {S.}~\bibnamefont {Yokoyama}},
  \bibinfo {author} {\bibfnamefont {J.}~\bibnamefont {Yoshikawa}}, \bibinfo
  {author} {\bibfnamefont {P.}~\bibnamefont {van Loock}}, \ and\ \bibinfo
  {author} {\bibfnamefont {A.}~\bibnamefont {Furusawa}},\ }\bibfield  {title}
  {\enquote {\bibinfo {title} {Demonstration of a controlled-phase gate for
  continuous-variable one-way quantum computation},}\ }\href {\doibase
  10.1103/PhysRevLett.107.250501} {\bibfield  {journal} {\bibinfo  {journal}
  {Phys. Rev. Lett.}\ }\textbf {\bibinfo {volume} {107}},\ \bibinfo {pages}
  {250501} (\bibinfo {year} {2011}{\natexlab{b}})}\BibitemShut {NoStop}%
\bibitem [{\citenamefont {Takeda}\ \emph
  {et~al.}(2013{\natexlab{a}})\citenamefont {Takeda}, \citenamefont {Mizuta},
  \citenamefont {Fuwa}, \citenamefont {van Loock},\ and\ \citenamefont
  {Furusawa}}]{13Takeda2}%
  \BibitemOpen
  \bibfield  {author} {\bibinfo {author} {\bibfnamefont {S.}~\bibnamefont
  {Takeda}}, \bibinfo {author} {\bibfnamefont {T.}~\bibnamefont {Mizuta}},
  \bibinfo {author} {\bibfnamefont {M.}~\bibnamefont {Fuwa}}, \bibinfo {author}
  {\bibfnamefont {P.}~\bibnamefont {van Loock}}, \ and\ \bibinfo {author}
  {\bibfnamefont {A.}~\bibnamefont {Furusawa}},\ }\bibfield  {title} {\enquote
  {\bibinfo {title} {Deterministic quantum teleportation of photonic quantum
  bits by a hybrid technique},}\ }\href@noop {} {\bibfield  {journal} {\bibinfo
   {journal} {Nature}\ }\textbf {\bibinfo {volume} {500}},\ \bibinfo {pages}
  {315--318} (\bibinfo {year} {2013}{\natexlab{a}})}\BibitemShut {NoStop}%
\bibitem [{\citenamefont {Andersen}\ and\ \citenamefont
  {Ralph}(2013)}]{13Andersen}%
  \BibitemOpen
  \bibfield  {author} {\bibinfo {author} {\bibfnamefont {U.~L.}\ \bibnamefont
  {Andersen}}\ and\ \bibinfo {author} {\bibfnamefont {T.~C.}\ \bibnamefont
  {Ralph}},\ }\bibfield  {title} {\enquote {\bibinfo {title} {High-fidelity
  teleportation of continuous-variable quantum states using delocalized single
  photons},}\ }\href {\doibase 10.1103/PhysRevLett.111.050504} {\bibfield
  {journal} {\bibinfo  {journal} {Phys. Rev. Lett.}\ }\textbf {\bibinfo
  {volume} {111}},\ \bibinfo {pages} {050504} (\bibinfo {year}
  {2013})}\BibitemShut {NoStop}%
\bibitem [{\citenamefont {Sefi}, \citenamefont {Vaibhav},\ and\ \citenamefont
  {van Loock}(2013)}]{13Sefi}%
  \BibitemOpen
  \bibfield  {author} {\bibinfo {author} {\bibfnamefont {S.}~\bibnamefont
  {Sefi}}, \bibinfo {author} {\bibfnamefont {V.}~\bibnamefont {Vaibhav}}, \
  and\ \bibinfo {author} {\bibfnamefont {P.}~\bibnamefont {van Loock}},\
  }\bibfield  {title} {\enquote {\bibinfo {title} {Measurement-induced optical
  kerr interaction},}\ }\href {\doibase 10.1103/PhysRevA.88.012303} {\bibfield
  {journal} {\bibinfo  {journal} {Phys. Rev. A}\ }\textbf {\bibinfo {volume}
  {88}},\ \bibinfo {pages} {012303} (\bibinfo {year} {2013})}\BibitemShut
  {NoStop}%
\bibitem [{\citenamefont {Lee}\ \emph {et~al.}(2011)\citenamefont {Lee},
  \citenamefont {Benichi}, \citenamefont {Takeno}, \citenamefont {Takeda},
  \citenamefont {Webb}, \citenamefont {Huntington},\ and\ \citenamefont
  {Furusawa}}]{11Lee}%
  \BibitemOpen
  \bibfield  {author} {\bibinfo {author} {\bibfnamefont {N.}~\bibnamefont
  {Lee}}, \bibinfo {author} {\bibfnamefont {H.}~\bibnamefont {Benichi}},
  \bibinfo {author} {\bibfnamefont {Y.}~\bibnamefont {Takeno}}, \bibinfo
  {author} {\bibfnamefont {S.}~\bibnamefont {Takeda}}, \bibinfo {author}
  {\bibfnamefont {J.}~\bibnamefont {Webb}}, \bibinfo {author} {\bibfnamefont
  {E.}~\bibnamefont {Huntington}}, \ and\ \bibinfo {author} {\bibfnamefont
  {A.}~\bibnamefont {Furusawa}},\ }\bibfield  {title} {\enquote {\bibinfo
  {title} {Teleportation of nonclassical wave packets of light},}\ }\href
  {\doibase 10.1126/science.1201034} {\bibfield  {journal} {\bibinfo  {journal}
  {Science}\ }\textbf {\bibinfo {volume} {332}},\ \bibinfo {pages} {330--333}
  (\bibinfo {year} {2011})}\BibitemShut {NoStop}%
\bibitem [{\citenamefont {Takeda}\ \emph {et~al.}(2012)\citenamefont {Takeda},
  \citenamefont {Benichi}, \citenamefont {Mizuta}, \citenamefont {Lee},
  \citenamefont {Yoshikawa},\ and\ \citenamefont {Furusawa}}]{12Takeda}%
  \BibitemOpen
  \bibfield  {author} {\bibinfo {author} {\bibfnamefont {S.}~\bibnamefont
  {Takeda}}, \bibinfo {author} {\bibfnamefont {H.}~\bibnamefont {Benichi}},
  \bibinfo {author} {\bibfnamefont {T.}~\bibnamefont {Mizuta}}, \bibinfo
  {author} {\bibfnamefont {N.}~\bibnamefont {Lee}}, \bibinfo {author}
  {\bibfnamefont {J.}~\bibnamefont {Yoshikawa}}, \ and\ \bibinfo {author}
  {\bibfnamefont {A.}~\bibnamefont {Furusawa}},\ }\bibfield  {title} {\enquote
  {\bibinfo {title} {Quantum mode filtering of non-{Gaussian} states for
  teleportation-based quantum information processing},}\ }\href {\doibase
  10.1103/PhysRevA.85.053824} {\bibfield  {journal} {\bibinfo  {journal} {Phys.
  Rev. A}\ }\textbf {\bibinfo {volume} {85}},\ \bibinfo {pages} {053824}
  (\bibinfo {year} {2012})}\BibitemShut {NoStop}%
\bibitem [{\citenamefont {Takeda}\ \emph
  {et~al.}(2013{\natexlab{b}})\citenamefont {Takeda}, \citenamefont {Mizuta},
  \citenamefont {Fuwa}, \citenamefont {Yoshikawa}, \citenamefont {Yonezawa},\
  and\ \citenamefont {Furusawa}}]{13Takeda1}%
  \BibitemOpen
  \bibfield  {author} {\bibinfo {author} {\bibfnamefont {S.}~\bibnamefont
  {Takeda}}, \bibinfo {author} {\bibfnamefont {T.}~\bibnamefont {Mizuta}},
  \bibinfo {author} {\bibfnamefont {M.}~\bibnamefont {Fuwa}}, \bibinfo {author}
  {\bibfnamefont {J.}~\bibnamefont {Yoshikawa}}, \bibinfo {author}
  {\bibfnamefont {H.}~\bibnamefont {Yonezawa}}, \ and\ \bibinfo {author}
  {\bibfnamefont {A.}~\bibnamefont {Furusawa}},\ }\bibfield  {title} {\enquote
  {\bibinfo {title} {Generation and eight-port homodyne characterization of
  time-bin qubits for continuous-variable quantum information processing},}\
  }\href {\doibase 10.1103/PhysRevA.87.043803} {\bibfield  {journal} {\bibinfo
  {journal} {Phys. Rev. A}\ }\textbf {\bibinfo {volume} {87}},\ \bibinfo
  {pages} {043803} (\bibinfo {year} {2013}{\natexlab{b}})}\BibitemShut
  {NoStop}%
\bibitem [{\citenamefont {Takeda}\ \emph {et~al.}(2015)\citenamefont {Takeda},
  \citenamefont {Fuwa}, \citenamefont {van Loock},\ and\ \citenamefont
  {Furusawa}}]{15Takeda}%
  \BibitemOpen
  \bibfield  {author} {\bibinfo {author} {\bibfnamefont {S.}~\bibnamefont
  {Takeda}}, \bibinfo {author} {\bibfnamefont {M.}~\bibnamefont {Fuwa}},
  \bibinfo {author} {\bibfnamefont {P.}~\bibnamefont {van Loock}}, \ and\
  \bibinfo {author} {\bibfnamefont {A.}~\bibnamefont {Furusawa}},\ }\bibfield
  {title} {\enquote {\bibinfo {title} {Entanglement swapping between discrete
  and continuous variables},}\ }\href {\doibase 10.1103/PhysRevLett.114.100501}
  {\bibfield  {journal} {\bibinfo  {journal} {Phys. Rev. Lett.}\ }\textbf
  {\bibinfo {volume} {114}},\ \bibinfo {pages} {100501} (\bibinfo {year}
  {2015})}\BibitemShut {NoStop}%
\bibitem [{\citenamefont {Miwa}\ \emph {et~al.}(2014)\citenamefont {Miwa},
  \citenamefont {Yoshikawa}, \citenamefont {Iwata}, \citenamefont {Endo},
  \citenamefont {Marek}, \citenamefont {Filip}, \citenamefont {van Loock},\
  and\ \citenamefont {Furusawa}}]{14Miwa}%
  \BibitemOpen
  \bibfield  {author} {\bibinfo {author} {\bibfnamefont {Y.}~\bibnamefont
  {Miwa}}, \bibinfo {author} {\bibfnamefont {J.}~\bibnamefont {Yoshikawa}},
  \bibinfo {author} {\bibfnamefont {N.}~\bibnamefont {Iwata}}, \bibinfo
  {author} {\bibfnamefont {M.}~\bibnamefont {Endo}}, \bibinfo {author}
  {\bibfnamefont {P.}~\bibnamefont {Marek}}, \bibinfo {author} {\bibfnamefont
  {R.}~\bibnamefont {Filip}}, \bibinfo {author} {\bibfnamefont
  {P.}~\bibnamefont {van Loock}}, \ and\ \bibinfo {author} {\bibfnamefont
  {A.}~\bibnamefont {Furusawa}},\ }\bibfield  {title} {\enquote {\bibinfo
  {title} {Exploring a new regime for processing optical qubits: Squeezing and
  unsqueezing single photons},}\ }\href {\doibase
  10.1103/PhysRevLett.113.013601} {\bibfield  {journal} {\bibinfo  {journal}
  {Phys. Rev. Lett.}\ }\textbf {\bibinfo {volume} {113}},\ \bibinfo {pages}
  {013601} (\bibinfo {year} {2014})}\BibitemShut {NoStop}%
\bibitem [{\citenamefont {Okada}\ \emph {et~al.}(2017)\citenamefont {Okada},
  \citenamefont {Takase}, \citenamefont {Fuwa}, \citenamefont {Takeda},
  \citenamefont {Yoshikawa}, \citenamefont {van Loock},\ and\ \citenamefont
  {Furusawa}}]{17Okada}%
  \BibitemOpen
  \bibfield  {author} {\bibinfo {author} {\bibfnamefont {M.}~\bibnamefont
  {Okada}}, \bibinfo {author} {\bibfnamefont {K.}~\bibnamefont {Takase}},
  \bibinfo {author} {\bibfnamefont {M.}~\bibnamefont {Fuwa}}, \bibinfo {author}
  {\bibfnamefont {S.}~\bibnamefont {Takeda}}, \bibinfo {author} {\bibfnamefont
  {J.}~\bibnamefont {Yoshikawa}}, \bibinfo {author} {\bibfnamefont
  {P.}~\bibnamefont {van Loock}}, \ and\ \bibinfo {author} {\bibfnamefont
  {A.}~\bibnamefont {Furusawa}},\ }\href@noop {} {\enquote {\bibinfo {title}
  {Quantum teleportation of an optical qutrit},}\ }\bibinfo {howpublished}
  {Conference on Lasers and Electro-Optics Europe \& European Quantum
  Electronics Conference} (\bibinfo {year} {2017}),\ \bibinfo {note}
  {eB-4.3}\BibitemShut {NoStop}%
\bibitem [{\citenamefont {Albert}\ \emph {et~al.}(2018)\citenamefont {Albert},
  \citenamefont {Noh}, \citenamefont {Duivenvoorden}, \citenamefont {Young},
  \citenamefont {Brierley}, \citenamefont {Reinhold}, \citenamefont {Vuillot},
  \citenamefont {Li}, \citenamefont {Shen}, \citenamefont {Girvin},
  \citenamefont {Terhal},\ and\ \citenamefont {Jiang}}]{18Albert}%
  \BibitemOpen
  \bibfield  {author} {\bibinfo {author} {\bibfnamefont {V.~V.}\ \bibnamefont
  {Albert}}, \bibinfo {author} {\bibfnamefont {K.}~\bibnamefont {Noh}},
  \bibinfo {author} {\bibfnamefont {K.}~\bibnamefont {Duivenvoorden}}, \bibinfo
  {author} {\bibfnamefont {D.~J.}\ \bibnamefont {Young}}, \bibinfo {author}
  {\bibfnamefont {R.~T.}\ \bibnamefont {Brierley}}, \bibinfo {author}
  {\bibfnamefont {P.}~\bibnamefont {Reinhold}}, \bibinfo {author}
  {\bibfnamefont {C.}~\bibnamefont {Vuillot}}, \bibinfo {author} {\bibfnamefont
  {L.}~\bibnamefont {Li}}, \bibinfo {author} {\bibfnamefont {C.}~\bibnamefont
  {Shen}}, \bibinfo {author} {\bibfnamefont {S.~M.}\ \bibnamefont {Girvin}},
  \bibinfo {author} {\bibfnamefont {B.~M.}\ \bibnamefont {Terhal}}, \ and\
  \bibinfo {author} {\bibfnamefont {L.}~\bibnamefont {Jiang}},\ }\bibfield
  {title} {\enquote {\bibinfo {title} {Performance and structure of single-mode
  bosonic codes},}\ }\href {\doibase 10.1103/PhysRevA.97.032346} {\bibfield
  {journal} {\bibinfo  {journal} {Phys. Rev. A}\ }\textbf {\bibinfo {volume}
  {97}},\ \bibinfo {pages} {032346} (\bibinfo {year} {2018})}\BibitemShut
  {NoStop}%
\bibitem [{\citenamefont {{Noh}}, \citenamefont {{Albert}},\ and\ \citenamefont
  {{Jiang}}(2019)}]{19Noh}%
  \BibitemOpen
  \bibfield  {author} {\bibinfo {author} {\bibfnamefont {K.}~\bibnamefont
  {{Noh}}}, \bibinfo {author} {\bibfnamefont {V.~V.}\ \bibnamefont {{Albert}}},
  \ and\ \bibinfo {author} {\bibfnamefont {L.}~\bibnamefont {{Jiang}}},\
  }\bibfield  {title} {\enquote {\bibinfo {title} {Quantum capacity bounds of
  gaussian thermal loss channels and achievable rates with
  gottesman-kitaev-preskill codes},}\ }\href@noop {} {\bibfield  {journal}
  {\bibinfo  {journal} {IEEE Trans. Inf. Theory}\ }\textbf {\bibinfo {volume}
  {65}},\ \bibinfo {pages} {2563--2582} (\bibinfo {year} {2019})}\BibitemShut
  {NoStop}%
\bibitem [{\citenamefont {Menicucci}(2014)}]{14Menicucci}%
  \BibitemOpen
  \bibfield  {author} {\bibinfo {author} {\bibfnamefont {N.~C.}\ \bibnamefont
  {Menicucci}},\ }\bibfield  {title} {\enquote {\bibinfo {title}
  {Fault-tolerant measurement-based quantum computing with continuous-variable
  cluster states},}\ }\href {\doibase 10.1103/PhysRevLett.112.120504}
  {\bibfield  {journal} {\bibinfo  {journal} {Phys. Rev. Lett.}\ }\textbf
  {\bibinfo {volume} {112}},\ \bibinfo {pages} {120504} (\bibinfo {year}
  {2014})}\BibitemShut {NoStop}%
\bibitem [{\citenamefont {Walshe}\ \emph {et~al.}(2019)\citenamefont {Walshe},
  \citenamefont {Mensen}, \citenamefont {Baragiola},\ and\ \citenamefont
  {Menicucci}}]{19Walshe}%
  \BibitemOpen
  \bibfield  {author} {\bibinfo {author} {\bibfnamefont {B.~W.}\ \bibnamefont
  {Walshe}}, \bibinfo {author} {\bibfnamefont {L.~J.}\ \bibnamefont {Mensen}},
  \bibinfo {author} {\bibfnamefont {B.~Q.}\ \bibnamefont {Baragiola}}, \ and\
  \bibinfo {author} {\bibfnamefont {N.~C.}\ \bibnamefont {Menicucci}},\
  }\href@noop {} {\enquote {\bibinfo {title} {Robust fault tolerance for
  continuous-variable cluster states with excess anti-squeezing},}\ }\bibinfo
  {howpublished} {arXiv:1903.02162} (\bibinfo {year} {2019})\BibitemShut
  {NoStop}%
\bibitem [{\citenamefont {Vahlbruch}\ \emph {et~al.}(2016)\citenamefont
  {Vahlbruch}, \citenamefont {Mehmet}, \citenamefont {Danzmann},\ and\
  \citenamefont {Schnabel}}]{16Vahlbruch}%
  \BibitemOpen
  \bibfield  {author} {\bibinfo {author} {\bibfnamefont {H.}~\bibnamefont
  {Vahlbruch}}, \bibinfo {author} {\bibfnamefont {M.}~\bibnamefont {Mehmet}},
  \bibinfo {author} {\bibfnamefont {K.}~\bibnamefont {Danzmann}}, \ and\
  \bibinfo {author} {\bibfnamefont {R.}~\bibnamefont {Schnabel}},\ }\bibfield
  {title} {\enquote {\bibinfo {title} {Detection of 15 d{B} squeezed states of
  light and their application for the absolute calibration of photoelectric
  quantum efficiency},}\ }\href@noop {} {\bibfield  {journal} {\bibinfo
  {journal} {Phys. Rev. Lett.}\ }\textbf {\bibinfo {volume} {117}},\ \bibinfo
  {pages} {110801} (\bibinfo {year} {2016})}\BibitemShut {NoStop}%
\bibitem [{\citenamefont {Fukui}, \citenamefont {Tomita},\ and\ \citenamefont
  {Okamoto}(2017)}]{17Fukui}%
  \BibitemOpen
  \bibfield  {author} {\bibinfo {author} {\bibfnamefont {K.}~\bibnamefont
  {Fukui}}, \bibinfo {author} {\bibfnamefont {A.}~\bibnamefont {Tomita}}, \
  and\ \bibinfo {author} {\bibfnamefont {A.}~\bibnamefont {Okamoto}},\
  }\bibfield  {title} {\enquote {\bibinfo {title} {Analog quantum error
  correction with encoding a qubit into an oscillator},}\ }\href {\doibase
  10.1103/PhysRevLett.119.180507} {\bibfield  {journal} {\bibinfo  {journal}
  {Phys. Rev. Lett.}\ }\textbf {\bibinfo {volume} {119}},\ \bibinfo {pages}
  {180507} (\bibinfo {year} {2017})}\BibitemShut {NoStop}%
\bibitem [{\citenamefont {Fukui}\ \emph {et~al.}(2018)\citenamefont {Fukui},
  \citenamefont {Tomita}, \citenamefont {Okamoto},\ and\ \citenamefont
  {Fujii}}]{18Fukui}%
  \BibitemOpen
  \bibfield  {author} {\bibinfo {author} {\bibfnamefont {K.}~\bibnamefont
  {Fukui}}, \bibinfo {author} {\bibfnamefont {A.}~\bibnamefont {Tomita}},
  \bibinfo {author} {\bibfnamefont {A.}~\bibnamefont {Okamoto}}, \ and\
  \bibinfo {author} {\bibfnamefont {K.}~\bibnamefont {Fujii}},\ }\bibfield
  {title} {\enquote {\bibinfo {title} {High-threshold fault-tolerant quantum
  computation with analog quantum error correction},}\ }\href {\doibase
  10.1103/PhysRevX.8.021054} {\bibfield  {journal} {\bibinfo  {journal} {Phys.
  Rev. X}\ }\textbf {\bibinfo {volume} {8}},\ \bibinfo {pages} {021054}
  (\bibinfo {year} {2018})}\BibitemShut {NoStop}%
\bibitem [{\citenamefont {Vasconcelos}, \citenamefont {Sanz},\ and\
  \citenamefont {Glancy}(2010)}]{10Vasconcelos}%
  \BibitemOpen
  \bibfield  {author} {\bibinfo {author} {\bibfnamefont {H.~M.}\ \bibnamefont
  {Vasconcelos}}, \bibinfo {author} {\bibfnamefont {L.}~\bibnamefont {Sanz}}, \
  and\ \bibinfo {author} {\bibfnamefont {S.}~\bibnamefont {Glancy}},\
  }\bibfield  {title} {\enquote {\bibinfo {title} {All-optical generation of
  states for ``{E}ncoding a qubit in an oscillator''},}\ }\href {\doibase
  10.1364/OL.35.003261} {\bibfield  {journal} {\bibinfo  {journal} {Opt.
  Lett.}\ }\textbf {\bibinfo {volume} {35}},\ \bibinfo {pages} {3261--3263}
  (\bibinfo {year} {2010})}\BibitemShut {NoStop}%
\bibitem [{\citenamefont {Weigand}\ and\ \citenamefont
  {Terhal}(2018)}]{18Weigand}%
  \BibitemOpen
  \bibfield  {author} {\bibinfo {author} {\bibfnamefont {D.~J.}\ \bibnamefont
  {Weigand}}\ and\ \bibinfo {author} {\bibfnamefont {B.~M.}\ \bibnamefont
  {Terhal}},\ }\bibfield  {title} {\enquote {\bibinfo {title} {Generating grid
  states from schr\"odinger-cat states without postselection},}\ }\href
  {\doibase 10.1103/PhysRevA.97.022341} {\bibfield  {journal} {\bibinfo
  {journal} {Phys. Rev. A}\ }\textbf {\bibinfo {volume} {97}},\ \bibinfo
  {pages} {022341} (\bibinfo {year} {2018})}\BibitemShut {NoStop}%
\bibitem [{\citenamefont {Motes}\ \emph {et~al.}(2017)\citenamefont {Motes},
  \citenamefont {Baragiola}, \citenamefont {Gilchrist},\ and\ \citenamefont
  {Menicucci}}]{17Motes}%
  \BibitemOpen
  \bibfield  {author} {\bibinfo {author} {\bibfnamefont {K.~R.}\ \bibnamefont
  {Motes}}, \bibinfo {author} {\bibfnamefont {B.~Q.}\ \bibnamefont
  {Baragiola}}, \bibinfo {author} {\bibfnamefont {A.}~\bibnamefont
  {Gilchrist}}, \ and\ \bibinfo {author} {\bibfnamefont {N.~C.}\ \bibnamefont
  {Menicucci}},\ }\bibfield  {title} {\enquote {\bibinfo {title} {Encoding
  qubits into oscillators with atomic ensembles and squeezed light},}\ }\href
  {\doibase 10.1103/PhysRevA.95.053819} {\bibfield  {journal} {\bibinfo
  {journal} {Phys. Rev. A}\ }\textbf {\bibinfo {volume} {95}},\ \bibinfo
  {pages} {053819} (\bibinfo {year} {2017})}\BibitemShut {NoStop}%
\bibitem [{\citenamefont {Eaton}, \citenamefont {Nehra},\ and\ \citenamefont
  {Pfister}(2019)}]{19Eaton}%
  \BibitemOpen
  \bibfield  {author} {\bibinfo {author} {\bibfnamefont {M.}~\bibnamefont
  {Eaton}}, \bibinfo {author} {\bibfnamefont {R.}~\bibnamefont {Nehra}}, \ and\
  \bibinfo {author} {\bibfnamefont {O.}~\bibnamefont {Pfister}},\ }\href@noop
  {} {\enquote {\bibinfo {title} {Gottesman-kitaev-preskill state preparation
  by photon catalysis},}\ }\bibinfo {howpublished} {arXiv:1903.01925} (\bibinfo
  {year} {2019})\BibitemShut {NoStop}%
\bibitem [{\citenamefont {Arrazola}\ \emph {et~al.}(2018)\citenamefont
  {Arrazola}, \citenamefont {Bromley}, \citenamefont {Izaac}, \citenamefont
  {Myers}, \citenamefont {Br\'{a}dler},\ and\ \citenamefont
  {Killoran}}]{18Arrazola}%
  \BibitemOpen
  \bibfield  {author} {\bibinfo {author} {\bibfnamefont {J.~M.}\ \bibnamefont
  {Arrazola}}, \bibinfo {author} {\bibfnamefont {T.~R.}\ \bibnamefont
  {Bromley}}, \bibinfo {author} {\bibfnamefont {J.}~\bibnamefont {Izaac}},
  \bibinfo {author} {\bibfnamefont {C.~R.}\ \bibnamefont {Myers}}, \bibinfo
  {author} {\bibfnamefont {K.}~\bibnamefont {Br\'{a}dler}}, \ and\ \bibinfo
  {author} {\bibfnamefont {N.}~\bibnamefont {Killoran}},\ }\href@noop {}
  {\enquote {\bibinfo {title} {Machine learning method for state preparation
  and gate synthesis on photonic quantum computers},}\ }\bibinfo {howpublished}
  {arXiv:1807.10781} (\bibinfo {year} {2018})\BibitemShut {NoStop}%
\bibitem [{\citenamefont {Masada}\ \emph {et~al.}(2015)\citenamefont {Masada},
  \citenamefont {Miyata}, \citenamefont {Politi}, \citenamefont {Hashimoto},
  \citenamefont {O'Brien},\ and\ \citenamefont {Furusawa}}]{15Masada}%
  \BibitemOpen
  \bibfield  {author} {\bibinfo {author} {\bibfnamefont {G.}~\bibnamefont
  {Masada}}, \bibinfo {author} {\bibfnamefont {K.}~\bibnamefont {Miyata}},
  \bibinfo {author} {\bibfnamefont {A.}~\bibnamefont {Politi}}, \bibinfo
  {author} {\bibfnamefont {T.}~\bibnamefont {Hashimoto}}, \bibinfo {author}
  {\bibfnamefont {J.~L.}\ \bibnamefont {O'Brien}}, \ and\ \bibinfo {author}
  {\bibfnamefont {A.}~\bibnamefont {Furusawa}},\ }\bibfield  {title} {\enquote
  {\bibinfo {title} {Continuous-variable entanglement on a chip},}\ }\href@noop
  {} {\bibfield  {journal} {\bibinfo  {journal} {Nat. Photonics}\ }\textbf
  {\bibinfo {volume} {9}},\ \bibinfo {pages} {316--319} (\bibinfo {year}
  {2015})}\BibitemShut {NoStop}%
\bibitem [{\citenamefont {Politi}\ \emph {et~al.}(2008)\citenamefont {Politi},
  \citenamefont {Cryan}, \citenamefont {Rarity}, \citenamefont {Yu},\ and\
  \citenamefont {O{\textquoteright}Brien}}]{08Politi}%
  \BibitemOpen
  \bibfield  {author} {\bibinfo {author} {\bibfnamefont {A.}~\bibnamefont
  {Politi}}, \bibinfo {author} {\bibfnamefont {M.~J.}\ \bibnamefont {Cryan}},
  \bibinfo {author} {\bibfnamefont {J.~G.}\ \bibnamefont {Rarity}}, \bibinfo
  {author} {\bibfnamefont {S.}~\bibnamefont {Yu}}, \ and\ \bibinfo {author}
  {\bibfnamefont {J.~L.}\ \bibnamefont {O{\textquoteright}Brien}},\ }\bibfield
  {title} {\enquote {\bibinfo {title} {Silica-on-silicon waveguide quantum
  circuits},}\ }\href {\doibase 10.1126/science.1155441} {\bibfield  {journal}
  {\bibinfo  {journal} {Science}\ }\textbf {\bibinfo {volume} {320}},\ \bibinfo
  {pages} {646--649} (\bibinfo {year} {2008})}\BibitemShut {NoStop}%
\bibitem [{\citenamefont {Carolan}\ \emph {et~al.}(2015)\citenamefont
  {Carolan}, \citenamefont {Harrold}, \citenamefont {Sparrow}, \citenamefont
  {Mart{\'\i}n-L{\'o}pez}, \citenamefont {Russell}, \citenamefont
  {Silverstone}, \citenamefont {Shadbolt}, \citenamefont {Matsuda},
  \citenamefont {Oguma}, \citenamefont {Itoh}, \citenamefont {Marshall},
  \citenamefont {Thompson}, \citenamefont {Matthews}, \citenamefont
  {Hashimoto}, \citenamefont {O{\textquoteright}Brien},\ and\ \citenamefont
  {Laing}}]{15Carolan}%
  \BibitemOpen
  \bibfield  {author} {\bibinfo {author} {\bibfnamefont {J.}~\bibnamefont
  {Carolan}}, \bibinfo {author} {\bibfnamefont {C.}~\bibnamefont {Harrold}},
  \bibinfo {author} {\bibfnamefont {C.}~\bibnamefont {Sparrow}}, \bibinfo
  {author} {\bibfnamefont {E.}~\bibnamefont {Mart{\'\i}n-L{\'o}pez}}, \bibinfo
  {author} {\bibfnamefont {N.~J.}\ \bibnamefont {Russell}}, \bibinfo {author}
  {\bibfnamefont {J.~W.}\ \bibnamefont {Silverstone}}, \bibinfo {author}
  {\bibfnamefont {P.~J.}\ \bibnamefont {Shadbolt}}, \bibinfo {author}
  {\bibfnamefont {N.}~\bibnamefont {Matsuda}}, \bibinfo {author} {\bibfnamefont
  {M.}~\bibnamefont {Oguma}}, \bibinfo {author} {\bibfnamefont
  {M.}~\bibnamefont {Itoh}}, \bibinfo {author} {\bibfnamefont {G.~D.}\
  \bibnamefont {Marshall}}, \bibinfo {author} {\bibfnamefont {M.~G.}\
  \bibnamefont {Thompson}}, \bibinfo {author} {\bibfnamefont {J.~C.~F.}\
  \bibnamefont {Matthews}}, \bibinfo {author} {\bibfnamefont {T.}~\bibnamefont
  {Hashimoto}}, \bibinfo {author} {\bibfnamefont {J.~L.}\ \bibnamefont
  {O{\textquoteright}Brien}}, \ and\ \bibinfo {author} {\bibfnamefont
  {A.}~\bibnamefont {Laing}},\ }\bibfield  {title} {\enquote {\bibinfo {title}
  {Universal linear optics},}\ }\href {\doibase 10.1126/science.aab3642}
  {\bibfield  {journal} {\bibinfo  {journal} {Science}\ }\textbf {\bibinfo
  {volume} {349}},\ \bibinfo {pages} {711--716} (\bibinfo {year}
  {2015})}\BibitemShut {NoStop}%
\bibitem [{\citenamefont {Lenzini}\ \emph {et~al.}(2018)\citenamefont
  {Lenzini}, \citenamefont {Janousek}, \citenamefont {Thearle}, \citenamefont
  {Villa}, \citenamefont {Haylock}, \citenamefont {Kasture}, \citenamefont
  {Cui}, \citenamefont {Phan}, \citenamefont {Dao}, \citenamefont {Yonezawa},
  \citenamefont {Lam}, \citenamefont {Huntington},\ and\ \citenamefont
  {Lobino}}]{18Lenzini}%
  \BibitemOpen
  \bibfield  {author} {\bibinfo {author} {\bibfnamefont {F.}~\bibnamefont
  {Lenzini}}, \bibinfo {author} {\bibfnamefont {J.}~\bibnamefont {Janousek}},
  \bibinfo {author} {\bibfnamefont {O.}~\bibnamefont {Thearle}}, \bibinfo
  {author} {\bibfnamefont {M.}~\bibnamefont {Villa}}, \bibinfo {author}
  {\bibfnamefont {B.}~\bibnamefont {Haylock}}, \bibinfo {author} {\bibfnamefont
  {S.}~\bibnamefont {Kasture}}, \bibinfo {author} {\bibfnamefont
  {L.}~\bibnamefont {Cui}}, \bibinfo {author} {\bibfnamefont {H.-P.}\
  \bibnamefont {Phan}}, \bibinfo {author} {\bibfnamefont {D.~V.}\ \bibnamefont
  {Dao}}, \bibinfo {author} {\bibfnamefont {H.}~\bibnamefont {Yonezawa}},
  \bibinfo {author} {\bibfnamefont {P.~K.}\ \bibnamefont {Lam}}, \bibinfo
  {author} {\bibfnamefont {E.~H.}\ \bibnamefont {Huntington}}, \ and\ \bibinfo
  {author} {\bibfnamefont {M.}~\bibnamefont {Lobino}},\ }\bibfield  {title}
  {\enquote {\bibinfo {title} {Integrated photonic platform for quantum
  information with continuous variables},}\ }\href@noop {} {\bibfield
  {journal} {\bibinfo  {journal} {Sci. Adv.}\ }\textbf {\bibinfo {volume} {4}}
  (\bibinfo {year} {2018})}\BibitemShut {NoStop}%
\bibitem [{\citenamefont {Asavanant}\ \emph {et~al.}(2019)\citenamefont
  {Asavanant}, \citenamefont {Shiozawa}, \citenamefont {Yokoyama},
  \citenamefont {Charoensombutamon}, \citenamefont {Emura}, \citenamefont
  {Alexander}, \citenamefont {Takeda}, \citenamefont {Yoshikawa}, \citenamefont
  {Menicucci}, \citenamefont {Yonezawa},\ and\ \citenamefont
  {Furusawa}}]{19Asavanant}%
  \BibitemOpen
  \bibfield  {author} {\bibinfo {author} {\bibfnamefont {W.}~\bibnamefont
  {Asavanant}}, \bibinfo {author} {\bibfnamefont {Y.}~\bibnamefont {Shiozawa}},
  \bibinfo {author} {\bibfnamefont {S.}~\bibnamefont {Yokoyama}}, \bibinfo
  {author} {\bibfnamefont {B.}~\bibnamefont {Charoensombutamon}}, \bibinfo
  {author} {\bibfnamefont {H.}~\bibnamefont {Emura}}, \bibinfo {author}
  {\bibfnamefont {R.~N.}\ \bibnamefont {Alexander}}, \bibinfo {author}
  {\bibfnamefont {S.}~\bibnamefont {Takeda}}, \bibinfo {author} {\bibfnamefont
  {J.}~\bibnamefont {Yoshikawa}}, \bibinfo {author} {\bibfnamefont {N.~C.}\
  \bibnamefont {Menicucci}}, \bibinfo {author} {\bibfnamefont {H.}~\bibnamefont
  {Yonezawa}}, \ and\ \bibinfo {author} {\bibfnamefont {A.}~\bibnamefont
  {Furusawa}},\ }\href@noop {} {\enquote {\bibinfo {title} {Time-domain
  multiplexed 2-dimensional cluster state: Universal quantum computing
  platform},}\ }\bibinfo {howpublished} {arXiv:1903.03918} (\bibinfo {year}
  {2019})\BibitemShut {NoStop}%
\bibitem [{\citenamefont {Menicucci}, \citenamefont {Ma},\ and\ \citenamefont
  {Ralph}(2010)}]{10Menicucci}%
  \BibitemOpen
  \bibfield  {author} {\bibinfo {author} {\bibfnamefont {N.~C.}\ \bibnamefont
  {Menicucci}}, \bibinfo {author} {\bibfnamefont {X.}~\bibnamefont {Ma}}, \
  and\ \bibinfo {author} {\bibfnamefont {T.~C.}\ \bibnamefont {Ralph}},\
  }\bibfield  {title} {\enquote {\bibinfo {title} {Arbitrarily large
  continuous-variable cluster states from a single quantum nondemolition
  gate},}\ }\href {\doibase 10.1103/PhysRevLett.104.250503} {\bibfield
  {journal} {\bibinfo  {journal} {Phys. Rev. Lett.}\ }\textbf {\bibinfo
  {volume} {104}},\ \bibinfo {pages} {250503} (\bibinfo {year}
  {2010})}\BibitemShut {NoStop}%
\bibitem [{\citenamefont {Menicucci}(2011)}]{11Menicucci}%
  \BibitemOpen
  \bibfield  {author} {\bibinfo {author} {\bibfnamefont {N.~C.}\ \bibnamefont
  {Menicucci}},\ }\bibfield  {title} {\enquote {\bibinfo {title} {Temporal-mode
  continuous-variable cluster states using linear optics},}\ }\href {\doibase
  10.1103/PhysRevA.83.062314} {\bibfield  {journal} {\bibinfo  {journal} {Phys.
  Rev. A}\ }\textbf {\bibinfo {volume} {83}},\ \bibinfo {pages} {062314}
  (\bibinfo {year} {2011})}\BibitemShut {NoStop}%
\bibitem [{\citenamefont {Menicucci}, \citenamefont {Flammia},\ and\
  \citenamefont {Pfister}(2008)}]{08Menicucci}%
  \BibitemOpen
  \bibfield  {author} {\bibinfo {author} {\bibfnamefont {N.~C.}\ \bibnamefont
  {Menicucci}}, \bibinfo {author} {\bibfnamefont {S.~T.}\ \bibnamefont
  {Flammia}}, \ and\ \bibinfo {author} {\bibfnamefont {O.}~\bibnamefont
  {Pfister}},\ }\bibfield  {title} {\enquote {\bibinfo {title} {One-way quantum
  computing in the optical frequency comb},}\ }\href {\doibase
  10.1103/PhysRevLett.101.130501} {\bibfield  {journal} {\bibinfo  {journal}
  {Phys. Rev. Lett.}\ }\textbf {\bibinfo {volume} {101}},\ \bibinfo {pages}
  {130501} (\bibinfo {year} {2008})}\BibitemShut {NoStop}%
\bibitem [{\citenamefont {Pysher}\ \emph {et~al.}(2011)\citenamefont {Pysher},
  \citenamefont {Miwa}, \citenamefont {Shahrokhshahi}, \citenamefont
  {Bloomer},\ and\ \citenamefont {Pfister}}]{11Pysher}%
  \BibitemOpen
  \bibfield  {author} {\bibinfo {author} {\bibfnamefont {M.}~\bibnamefont
  {Pysher}}, \bibinfo {author} {\bibfnamefont {Y.}~\bibnamefont {Miwa}},
  \bibinfo {author} {\bibfnamefont {R.}~\bibnamefont {Shahrokhshahi}}, \bibinfo
  {author} {\bibfnamefont {R.}~\bibnamefont {Bloomer}}, \ and\ \bibinfo
  {author} {\bibfnamefont {O.}~\bibnamefont {Pfister}},\ }\bibfield  {title}
  {\enquote {\bibinfo {title} {Parallel generation of quadripartite cluster
  entanglement in the optical frequency comb},}\ }\href {\doibase
  10.1103/PhysRevLett.107.030505} {\bibfield  {journal} {\bibinfo  {journal}
  {Phys. Rev. Lett.}\ }\textbf {\bibinfo {volume} {107}},\ \bibinfo {pages}
  {030505} (\bibinfo {year} {2011})}\BibitemShut {NoStop}%
\bibitem [{\citenamefont {Chen}, \citenamefont {Menicucci},\ and\ \citenamefont
  {Pfister}(2014)}]{14Chen}%
  \BibitemOpen
  \bibfield  {author} {\bibinfo {author} {\bibfnamefont {M.}~\bibnamefont
  {Chen}}, \bibinfo {author} {\bibfnamefont {N.~C.}\ \bibnamefont {Menicucci}},
  \ and\ \bibinfo {author} {\bibfnamefont {O.}~\bibnamefont {Pfister}},\
  }\bibfield  {title} {\enquote {\bibinfo {title} {Experimental realization of
  multipartite entanglement of 60 modes of a quantum optical frequency comb},}\
  }\href {\doibase 10.1103/PhysRevLett.112.120505} {\bibfield  {journal}
  {\bibinfo  {journal} {Phys. Rev. Lett.}\ }\textbf {\bibinfo {volume} {112}},\
  \bibinfo {pages} {120505} (\bibinfo {year} {2014})}\BibitemShut {NoStop}%
\bibitem [{\citenamefont {Roslund}\ \emph {et~al.}(2014)\citenamefont
  {Roslund}, \citenamefont {de~Araújo}, \citenamefont {Jiang}, \citenamefont
  {Fabre},\ and\ \citenamefont {Treps}}]{13Roslund}%
  \BibitemOpen
  \bibfield  {author} {\bibinfo {author} {\bibfnamefont {J.}~\bibnamefont
  {Roslund}}, \bibinfo {author} {\bibfnamefont {R.~M.}\ \bibnamefont
  {de~Araújo}}, \bibinfo {author} {\bibfnamefont {S.}~\bibnamefont {Jiang}},
  \bibinfo {author} {\bibfnamefont {C.}~\bibnamefont {Fabre}}, \ and\ \bibinfo
  {author} {\bibfnamefont {N.}~\bibnamefont {Treps}},\ }\bibfield  {title}
  {\enquote {\bibinfo {title} {Wavelength-multiplexed quantum networks with
  ultrafast frequency combs},}\ }\href@noop {} {\bibfield  {journal} {\bibinfo
  {journal} {Nat. Photonics}\ }\textbf {\bibinfo {volume} {8}},\ \bibinfo
  {pages} {109--112} (\bibinfo {year} {2014})}\BibitemShut {NoStop}%
\bibitem [{\citenamefont {Lassen}, \citenamefont {Leuchs},\ and\ \citenamefont
  {Andersen}(2009)}]{09Lassen}%
  \BibitemOpen
  \bibfield  {author} {\bibinfo {author} {\bibfnamefont {M.}~\bibnamefont
  {Lassen}}, \bibinfo {author} {\bibfnamefont {G.}~\bibnamefont {Leuchs}}, \
  and\ \bibinfo {author} {\bibfnamefont {U.~L.}\ \bibnamefont {Andersen}},\
  }\bibfield  {title} {\enquote {\bibinfo {title} {Continuous variable
  entanglement and squeezing of orbital angular momentum states},}\ }\href
  {\doibase 10.1103/PhysRevLett.102.163602} {\bibfield  {journal} {\bibinfo
  {journal} {Phys. Rev. Lett.}\ }\textbf {\bibinfo {volume} {102}},\ \bibinfo
  {pages} {163602} (\bibinfo {year} {2009})}\BibitemShut {NoStop}%
\bibitem [{\citenamefont {Armstrong}\ \emph {et~al.}(2012)\citenamefont
  {Armstrong}, \citenamefont {Morizur}, \citenamefont {Janousek}, \citenamefont
  {Hage}, \citenamefont {Treps}, \citenamefont {Lam},\ and\ \citenamefont
  {Bachor}}]{12Armstrong}%
  \BibitemOpen
  \bibfield  {author} {\bibinfo {author} {\bibfnamefont {S.}~\bibnamefont
  {Armstrong}}, \bibinfo {author} {\bibfnamefont {J.-F.}\ \bibnamefont
  {Morizur}}, \bibinfo {author} {\bibfnamefont {J.}~\bibnamefont {Janousek}},
  \bibinfo {author} {\bibfnamefont {B.}~\bibnamefont {Hage}}, \bibinfo {author}
  {\bibfnamefont {N.}~\bibnamefont {Treps}}, \bibinfo {author} {\bibfnamefont
  {P.~K.}\ \bibnamefont {Lam}}, \ and\ \bibinfo {author} {\bibfnamefont
  {H.-A.}\ \bibnamefont {Bachor}},\ }\bibfield  {title} {\enquote {\bibinfo
  {title} {Programmable multimode quantum networks},}\ }\href@noop {}
  {\bibfield  {journal} {\bibinfo  {journal} {Nat. Commun.}\ }\textbf {\bibinfo
  {volume} {3}},\ \bibinfo {pages} {1026} (\bibinfo {year} {2012})}\BibitemShut
  {NoStop}%
\bibitem [{\citenamefont {Miwa}\ \emph {et~al.}(2010)\citenamefont {Miwa},
  \citenamefont {Ukai}, \citenamefont {Yoshikawa}, \citenamefont {Filip},
  \citenamefont {van Loock},\ and\ \citenamefont {Furusawa}}]{10Miwa}%
  \BibitemOpen
  \bibfield  {author} {\bibinfo {author} {\bibfnamefont {Y.}~\bibnamefont
  {Miwa}}, \bibinfo {author} {\bibfnamefont {R.}~\bibnamefont {Ukai}}, \bibinfo
  {author} {\bibfnamefont {J.}~\bibnamefont {Yoshikawa}}, \bibinfo {author}
  {\bibfnamefont {R.}~\bibnamefont {Filip}}, \bibinfo {author} {\bibfnamefont
  {P.}~\bibnamefont {van Loock}}, \ and\ \bibinfo {author} {\bibfnamefont
  {A.}~\bibnamefont {Furusawa}},\ }\bibfield  {title} {\enquote {\bibinfo
  {title} {Demonstration of cluster-state shaping and quantum erasure for
  continuous variables},}\ }\href {\doibase 10.1103/PhysRevA.82.032305}
  {\bibfield  {journal} {\bibinfo  {journal} {Phys. Rev. A}\ }\textbf {\bibinfo
  {volume} {82}},\ \bibinfo {pages} {032305} (\bibinfo {year}
  {2010})}\BibitemShut {NoStop}%
\bibitem [{\citenamefont {Motes}\ \emph {et~al.}(2014)\citenamefont {Motes},
  \citenamefont {Gilchrist}, \citenamefont {Dowling},\ and\ \citenamefont
  {Rohde}}]{14Motes}%
  \BibitemOpen
  \bibfield  {author} {\bibinfo {author} {\bibfnamefont {K.~R.}\ \bibnamefont
  {Motes}}, \bibinfo {author} {\bibfnamefont {A.}~\bibnamefont {Gilchrist}},
  \bibinfo {author} {\bibfnamefont {J.~P.}\ \bibnamefont {Dowling}}, \ and\
  \bibinfo {author} {\bibfnamefont {P.~P.}\ \bibnamefont {Rohde}},\ }\bibfield
  {title} {\enquote {\bibinfo {title} {Scalable boson sampling with time-bin
  encoding using a loop-based architecture},}\ }\href {\doibase
  10.1103/PhysRevLett.113.120501} {\bibfield  {journal} {\bibinfo  {journal}
  {Phys. Rev. Lett.}\ }\textbf {\bibinfo {volume} {113}},\ \bibinfo {pages}
  {120501} (\bibinfo {year} {2014})}\BibitemShut {NoStop}%
\bibitem [{\citenamefont {Rohde}(2015)}]{15Rohde}%
  \BibitemOpen
  \bibfield  {author} {\bibinfo {author} {\bibfnamefont {P.~P.}\ \bibnamefont
  {Rohde}},\ }\bibfield  {title} {\enquote {\bibinfo {title} {Simple scheme for
  universal linear-optics quantum computing with constant experimental
  complexity using fiber loops},}\ }\href {\doibase 10.1103/PhysRevA.91.012306}
  {\bibfield  {journal} {\bibinfo  {journal} {Phys. Rev. A}\ }\textbf {\bibinfo
  {volume} {91}},\ \bibinfo {pages} {012306} (\bibinfo {year}
  {2015})}\BibitemShut {NoStop}%
\bibitem [{\citenamefont {Schreiber}\ \emph {et~al.}(2012)\citenamefont
  {Schreiber}, \citenamefont {G{\'a}bris}, \citenamefont {Rohde}, \citenamefont
  {Laiho}, \citenamefont {{\v S}tefa{\v n}{\'a}k}, \citenamefont {Poto{\v
  c}ek}, \citenamefont {Hamilton}, \citenamefont {Jex},\ and\ \citenamefont
  {Silberhorn}}]{12Schreiber}%
  \BibitemOpen
  \bibfield  {author} {\bibinfo {author} {\bibfnamefont {A.}~\bibnamefont
  {Schreiber}}, \bibinfo {author} {\bibfnamefont {A.}~\bibnamefont
  {G{\'a}bris}}, \bibinfo {author} {\bibfnamefont {P.~P.}\ \bibnamefont
  {Rohde}}, \bibinfo {author} {\bibfnamefont {K.}~\bibnamefont {Laiho}},
  \bibinfo {author} {\bibfnamefont {M.}~\bibnamefont {{\v S}tefa{\v n}{\'a}k}},
  \bibinfo {author} {\bibfnamefont {V.}~\bibnamefont {Poto{\v c}ek}}, \bibinfo
  {author} {\bibfnamefont {C.}~\bibnamefont {Hamilton}}, \bibinfo {author}
  {\bibfnamefont {I.}~\bibnamefont {Jex}}, \ and\ \bibinfo {author}
  {\bibfnamefont {C.}~\bibnamefont {Silberhorn}},\ }\bibfield  {title}
  {\enquote {\bibinfo {title} {A 2{D} quantum walk simulation of two-particle
  dynamics},}\ }\href@noop {} {\bibfield  {journal} {\bibinfo  {journal}
  {Science}\ }\textbf {\bibinfo {volume} {336}},\ \bibinfo {pages} {55--58}
  (\bibinfo {year} {2012})}\BibitemShut {NoStop}%
\bibitem [{\citenamefont {He}\ \emph {et~al.}(2017)\citenamefont {He},
  \citenamefont {Ding}, \citenamefont {Su}, \citenamefont {Huang},
  \citenamefont {Qin}, \citenamefont {Wang}, \citenamefont {Unsleber},
  \citenamefont {Chen}, \citenamefont {Wang}, \citenamefont {He}, \citenamefont
  {Wang}, \citenamefont {Zhang}, \citenamefont {Chen}, \citenamefont
  {Schneider}, \citenamefont {Kamp}, \citenamefont {You}, \citenamefont {Wang},
  \citenamefont {H\"ofling}, \citenamefont {Lu},\ and\ \citenamefont
  {Pan}}]{17He}%
  \BibitemOpen
  \bibfield  {author} {\bibinfo {author} {\bibfnamefont {Y.}~\bibnamefont
  {He}}, \bibinfo {author} {\bibfnamefont {X.}~\bibnamefont {Ding}}, \bibinfo
  {author} {\bibfnamefont {Z.-E.}\ \bibnamefont {Su}}, \bibinfo {author}
  {\bibfnamefont {H.-L.}\ \bibnamefont {Huang}}, \bibinfo {author}
  {\bibfnamefont {J.}~\bibnamefont {Qin}}, \bibinfo {author} {\bibfnamefont
  {C.}~\bibnamefont {Wang}}, \bibinfo {author} {\bibfnamefont {S.}~\bibnamefont
  {Unsleber}}, \bibinfo {author} {\bibfnamefont {C.}~\bibnamefont {Chen}},
  \bibinfo {author} {\bibfnamefont {H.}~\bibnamefont {Wang}}, \bibinfo {author}
  {\bibfnamefont {Y.-M.}\ \bibnamefont {He}}, \bibinfo {author} {\bibfnamefont
  {X.-L.}\ \bibnamefont {Wang}}, \bibinfo {author} {\bibfnamefont {W.-J.}\
  \bibnamefont {Zhang}}, \bibinfo {author} {\bibfnamefont {S.-J.}\ \bibnamefont
  {Chen}}, \bibinfo {author} {\bibfnamefont {C.}~\bibnamefont {Schneider}},
  \bibinfo {author} {\bibfnamefont {M.}~\bibnamefont {Kamp}}, \bibinfo {author}
  {\bibfnamefont {L.-X.}\ \bibnamefont {You}}, \bibinfo {author} {\bibfnamefont
  {Z.}~\bibnamefont {Wang}}, \bibinfo {author} {\bibfnamefont {S.}~\bibnamefont
  {H\"ofling}}, \bibinfo {author} {\bibfnamefont {C.-Y.}\ \bibnamefont {Lu}}, \
  and\ \bibinfo {author} {\bibfnamefont {J.-W.}\ \bibnamefont {Pan}},\
  }\bibfield  {title} {\enquote {\bibinfo {title} {Time-bin-encoded boson
  sampling with a single-photon device},}\ }\href {\doibase
  10.1103/PhysRevLett.118.190501} {\bibfield  {journal} {\bibinfo  {journal}
  {Phys. Rev. Lett.}\ }\textbf {\bibinfo {volume} {118}},\ \bibinfo {pages}
  {190501} (\bibinfo {year} {2017})}\BibitemShut {NoStop}%
\bibitem [{\citenamefont {Qi}\ \emph {et~al.}(2018)\citenamefont {Qi},
  \citenamefont {Helt}, \citenamefont {Su}, \citenamefont {Vernon},\ and\
  \citenamefont {Br\'{a}dler}}]{18Qi}%
  \BibitemOpen
  \bibfield  {author} {\bibinfo {author} {\bibfnamefont {H.}~\bibnamefont
  {Qi}}, \bibinfo {author} {\bibfnamefont {L.}~\bibnamefont {Helt}}, \bibinfo
  {author} {\bibfnamefont {G.~D.}\ \bibnamefont {Su}}, \bibinfo {author}
  {\bibfnamefont {Z.}~\bibnamefont {Vernon}}, \ and\ \bibinfo {author}
  {\bibfnamefont {K.}~\bibnamefont {Br\'{a}dler}},\ }\href@noop {} {\enquote
  {\bibinfo {title} {Linear multiport photonic interferometers: loss analysis
  of temporally-encoded architectures},}\ }\bibinfo {howpublished}
  {arXiv:1812.07015} (\bibinfo {year} {2018})\BibitemShut {NoStop}%
\bibitem [{\citenamefont {Su}\ \emph {et~al.}(2018)\citenamefont {Su},
  \citenamefont {Dhand}, \citenamefont {Helt}, \citenamefont {Vernon},\ and\
  \citenamefont {Br\'{a}dler}}]{18Su}%
  \BibitemOpen
  \bibfield  {author} {\bibinfo {author} {\bibfnamefont {D.}~\bibnamefont
  {Su}}, \bibinfo {author} {\bibfnamefont {I.}~\bibnamefont {Dhand}}, \bibinfo
  {author} {\bibfnamefont {L.~G.}\ \bibnamefont {Helt}}, \bibinfo {author}
  {\bibfnamefont {Z.}~\bibnamefont {Vernon}}, \ and\ \bibinfo {author}
  {\bibfnamefont {K.}~\bibnamefont {Br\'{a}dler}},\ }\href@noop {} {\enquote
  {\bibinfo {title} {Hybrid spatio-temporal architectures for universal linear
  optics},}\ }\bibinfo {howpublished} {arXiv:1812.07939} (\bibinfo {year}
  {2018})\BibitemShut {NoStop}%
\bibitem [{\citenamefont {Kaiser}\ \emph {et~al.}(2016)\citenamefont {Kaiser},
  \citenamefont {Fedrici}, \citenamefont {Zavatta}, \citenamefont {D'Auria},\
  and\ \citenamefont {Tanzilli}}]{16Kaiser}%
  \BibitemOpen
  \bibfield  {author} {\bibinfo {author} {\bibfnamefont {F.}~\bibnamefont
  {Kaiser}}, \bibinfo {author} {\bibfnamefont {B.}~\bibnamefont {Fedrici}},
  \bibinfo {author} {\bibfnamefont {A.}~\bibnamefont {Zavatta}}, \bibinfo
  {author} {\bibfnamefont {V.}~\bibnamefont {D'Auria}}, \ and\ \bibinfo
  {author} {\bibfnamefont {S.}~\bibnamefont {Tanzilli}},\ }\bibfield  {title}
  {\enquote {\bibinfo {title} {A fully guided-wave squeezing experiment for
  fiber quantum networks},}\ }\href {\doibase 10.1364/OPTICA.3.000362}
  {\bibfield  {journal} {\bibinfo  {journal} {Optica}\ }\textbf {\bibinfo
  {volume} {3}},\ \bibinfo {pages} {362--365} (\bibinfo {year}
  {2016})}\BibitemShut {NoStop}%
\bibitem [{\citenamefont {Larsen}\ \emph {et~al.}(2018)\citenamefont {Larsen},
  \citenamefont {Guo}, \citenamefont {Breum}, \citenamefont
  {Neergaard-Nielsen},\ and\ \citenamefont {Andersen}}]{18Larsen}%
  \BibitemOpen
  \bibfield  {author} {\bibinfo {author} {\bibfnamefont {M.~V.}\ \bibnamefont
  {Larsen}}, \bibinfo {author} {\bibfnamefont {X.}~\bibnamefont {Guo}},
  \bibinfo {author} {\bibfnamefont {C.~R.}\ \bibnamefont {Breum}}, \bibinfo
  {author} {\bibfnamefont {J.~S.}\ \bibnamefont {Neergaard-Nielsen}}, \ and\
  \bibinfo {author} {\bibfnamefont {U.~L.}\ \bibnamefont {Andersen}},\
  }\href@noop {} {\enquote {\bibinfo {title} {Fiber coupled epr-state
  generation using a single temporally multiplexed squeezed light source},}\
  }\bibinfo {howpublished} {arXiv:1812.05358} (\bibinfo {year}
  {2018})\BibitemShut {NoStop}%
\bibitem [{\citenamefont {Shiozawa}\ \emph {et~al.}(2018)\citenamefont
  {Shiozawa}, \citenamefont {Yoshikawa}, \citenamefont {Yokoyama},
  \citenamefont {Kaji}, \citenamefont {Makino}, \citenamefont {Serikawa},
  \citenamefont {Nakamura}, \citenamefont {Suzuki}, \citenamefont {Yamazaki},
  \citenamefont {Asavanant}, \citenamefont {Takeda}, \citenamefont {van
  Loock},\ and\ \citenamefont {Furusawa}}]{18Shiozawa}%
  \BibitemOpen
  \bibfield  {author} {\bibinfo {author} {\bibfnamefont {Y.}~\bibnamefont
  {Shiozawa}}, \bibinfo {author} {\bibfnamefont {J.}~\bibnamefont {Yoshikawa}},
  \bibinfo {author} {\bibfnamefont {S.}~\bibnamefont {Yokoyama}}, \bibinfo
  {author} {\bibfnamefont {T.}~\bibnamefont {Kaji}}, \bibinfo {author}
  {\bibfnamefont {K.}~\bibnamefont {Makino}}, \bibinfo {author} {\bibfnamefont
  {T.}~\bibnamefont {Serikawa}}, \bibinfo {author} {\bibfnamefont
  {R.}~\bibnamefont {Nakamura}}, \bibinfo {author} {\bibfnamefont
  {S.}~\bibnamefont {Suzuki}}, \bibinfo {author} {\bibfnamefont
  {S.}~\bibnamefont {Yamazaki}}, \bibinfo {author} {\bibfnamefont
  {W.}~\bibnamefont {Asavanant}}, \bibinfo {author} {\bibfnamefont
  {S.}~\bibnamefont {Takeda}}, \bibinfo {author} {\bibfnamefont
  {P.}~\bibnamefont {van Loock}}, \ and\ \bibinfo {author} {\bibfnamefont
  {A.}~\bibnamefont {Furusawa}},\ }\bibfield  {title} {\enquote {\bibinfo
  {title} {Quantum nondemolition gate operations and measurements in real time
  on fluctuating signals},}\ }\href {\doibase 10.1103/PhysRevA.98.052311}
  {\bibfield  {journal} {\bibinfo  {journal} {Phys. Rev. A}\ }\textbf {\bibinfo
  {volume} {98}},\ \bibinfo {pages} {052311} (\bibinfo {year}
  {2018})}\BibitemShut {NoStop}%
\bibitem [{\citenamefont {Kumar}\ \emph {et~al.}(2012)\citenamefont {Kumar},
  \citenamefont {Barrios}, \citenamefont {MacRae}, \citenamefont {Cairns},
  \citenamefont {Huntington},\ and\ \citenamefont {Lvovsky}}]{12Kumar}%
  \BibitemOpen
  \bibfield  {author} {\bibinfo {author} {\bibfnamefont {R.}~\bibnamefont
  {Kumar}}, \bibinfo {author} {\bibfnamefont {E.}~\bibnamefont {Barrios}},
  \bibinfo {author} {\bibfnamefont {A.}~\bibnamefont {MacRae}}, \bibinfo
  {author} {\bibfnamefont {E.}~\bibnamefont {Cairns}}, \bibinfo {author}
  {\bibfnamefont {E.}~\bibnamefont {Huntington}}, \ and\ \bibinfo {author}
  {\bibfnamefont {A.}~\bibnamefont {Lvovsky}},\ }\bibfield  {title} {\enquote
  {\bibinfo {title} {Versatile wideband balanced detector for quantum optical
  homodyne tomography},}\ }\href {\doibase
  https://doi.org/10.1016/j.optcom.2012.07.103} {\bibfield  {journal} {\bibinfo
   {journal} {Opt. Commun.}\ }\textbf {\bibinfo {volume} {285}},\ \bibinfo
  {pages} {5259 -- 5267} (\bibinfo {year} {2012})}\BibitemShut {NoStop}%
\bibitem [{\citenamefont {Serikawa}\ \emph {et~al.}(2016)\citenamefont
  {Serikawa}, \citenamefont {Yoshikawa}, \citenamefont {Makino},\ and\
  \citenamefont {Frusawa}}]{16Serikawa}%
  \BibitemOpen
  \bibfield  {author} {\bibinfo {author} {\bibfnamefont {T.}~\bibnamefont
  {Serikawa}}, \bibinfo {author} {\bibfnamefont {J.}~\bibnamefont {Yoshikawa}},
  \bibinfo {author} {\bibfnamefont {K.}~\bibnamefont {Makino}}, \ and\ \bibinfo
  {author} {\bibfnamefont {A.}~\bibnamefont {Frusawa}},\ }\bibfield  {title}
  {\enquote {\bibinfo {title} {Creation and measurement of broadband squeezed
  vacuum from a ring optical parametric oscillator},}\ }\href {\doibase
  10.1364/OE.24.028383} {\bibfield  {journal} {\bibinfo  {journal} {Opt.
  Express}\ }\textbf {\bibinfo {volume} {24}},\ \bibinfo {pages} {28383--28391}
  (\bibinfo {year} {2016})}\BibitemShut {NoStop}%
\bibitem [{\citenamefont {Yoshino}, \citenamefont {Aoki},\ and\ \citenamefont
  {Furusawa}(2007)}]{07Yoshino}%
  \BibitemOpen
  \bibfield  {author} {\bibinfo {author} {\bibfnamefont {K.}~\bibnamefont
  {Yoshino}}, \bibinfo {author} {\bibfnamefont {T.}~\bibnamefont {Aoki}}, \
  and\ \bibinfo {author} {\bibfnamefont {A.}~\bibnamefont {Furusawa}},\
  }\bibfield  {title} {\enquote {\bibinfo {title} {Generation of
  continuous-wave broadband entangled beams using periodically poled lithium
  niobate waveguides},}\ }\href {\doibase 10.1063/1.2437057} {\bibfield
  {journal} {\bibinfo  {journal} {Appl. Phys. Lett.}\ }\textbf {\bibinfo
  {volume} {90}},\ \bibinfo {pages} {041111} (\bibinfo {year}
  {2007})}\BibitemShut {NoStop}%
\bibitem [{\citenamefont {Shaked}\ \emph {et~al.}(2018)\citenamefont {Shaked},
  \citenamefont {Michael}, \citenamefont {Vered}, \citenamefont {Bello},
  \citenamefont {Rosenbluh},\ and\ \citenamefont {Pe'er}}]{18Shaked}%
  \BibitemOpen
  \bibfield  {author} {\bibinfo {author} {\bibfnamefont {Y.}~\bibnamefont
  {Shaked}}, \bibinfo {author} {\bibfnamefont {Y.}~\bibnamefont {Michael}},
  \bibinfo {author} {\bibfnamefont {R.~Z.}\ \bibnamefont {Vered}}, \bibinfo
  {author} {\bibfnamefont {L.}~\bibnamefont {Bello}}, \bibinfo {author}
  {\bibfnamefont {M.}~\bibnamefont {Rosenbluh}}, \ and\ \bibinfo {author}
  {\bibfnamefont {A.}~\bibnamefont {Pe'er}},\ }\bibfield  {title} {\enquote
  {\bibinfo {title} {Lifting the bandwidth limit of optical homodyne
  measurement with broadband parametric amplification},}\ }\href@noop {}
  {\bibfield  {journal} {\bibinfo  {journal} {Nat. Commun.}\ }\textbf {\bibinfo
  {volume} {9}},\ \bibinfo {pages} {609} (\bibinfo {year} {2018})}\BibitemShut
  {NoStop}%
\bibitem [{\citenamefont {Wu}\ \emph {et~al.}(2016)\citenamefont {Wu},
  \citenamefont {Jiang}, \citenamefont {Ma}, \citenamefont {Qi}, \citenamefont
  {Yu}, \citenamefont {Bi},\ and\ \citenamefont {Ma}}]{16Lifei}%
  \BibitemOpen
  \bibfield  {author} {\bibinfo {author} {\bibfnamefont {L.}~\bibnamefont
  {Wu}}, \bibinfo {author} {\bibfnamefont {Y.}~\bibnamefont {Jiang}}, \bibinfo
  {author} {\bibfnamefont {C.}~\bibnamefont {Ma}}, \bibinfo {author}
  {\bibfnamefont {W.}~\bibnamefont {Qi}}, \bibinfo {author} {\bibfnamefont
  {H.}~\bibnamefont {Yu}}, \bibinfo {author} {\bibfnamefont {Z.}~\bibnamefont
  {Bi}}, \ and\ \bibinfo {author} {\bibfnamefont {L.}~\bibnamefont {Ma}},\
  }\bibfield  {title} {\enquote {\bibinfo {title} {0.26-{H}z-linewidth
  ultrastable lasers at 1557 nm},}\ }\href@noop {} {\bibfield  {journal}
  {\bibinfo  {journal} {Scientific Reports}\ }\textbf {\bibinfo {volume} {6}},\
  \bibinfo {pages} {24969} (\bibinfo {year} {2016})}\BibitemShut {NoStop}%
\bibitem [{\citenamefont {{The LIGO Scientific Collaboration}}(2013)}]{13LIGO}%
  \BibitemOpen
  \bibfield  {author} {\bibinfo {author} {\bibnamefont {{The LIGO Scientific
  Collaboration}}},\ }\bibfield  {title} {\enquote {\bibinfo {title} {Enhanced
  sensitivity of the ligo gravitational wave detector by using squeezed states
  of light},}\ }\href@noop {} {\bibfield  {journal} {\bibinfo  {journal} {Nat.
  Photonics}\ }\textbf {\bibinfo {volume} {7}},\ \bibinfo {pages} {613}
  (\bibinfo {year} {2013})}\BibitemShut {NoStop}%
\bibitem [{\citenamefont {Huo}\ \emph {et~al.}(2018)\citenamefont {Huo},
  \citenamefont {Qin}, \citenamefont {Cheng}, \citenamefont {Yan},
  \citenamefont {Qin}, \citenamefont {Su}, \citenamefont {Jia}, \citenamefont
  {Xie},\ and\ \citenamefont {Peng}}]{18Huo}%
  \BibitemOpen
  \bibfield  {author} {\bibinfo {author} {\bibfnamefont {M.}~\bibnamefont
  {Huo}}, \bibinfo {author} {\bibfnamefont {J.}~\bibnamefont {Qin}}, \bibinfo
  {author} {\bibfnamefont {J.}~\bibnamefont {Cheng}}, \bibinfo {author}
  {\bibfnamefont {Z.}~\bibnamefont {Yan}}, \bibinfo {author} {\bibfnamefont
  {Z.}~\bibnamefont {Qin}}, \bibinfo {author} {\bibfnamefont {X.}~\bibnamefont
  {Su}}, \bibinfo {author} {\bibfnamefont {X.}~\bibnamefont {Jia}}, \bibinfo
  {author} {\bibfnamefont {C.}~\bibnamefont {Xie}}, \ and\ \bibinfo {author}
  {\bibfnamefont {K.}~\bibnamefont {Peng}},\ }\bibfield  {title} {\enquote
  {\bibinfo {title} {Deterministic quantum teleportation through fiber
  channels},}\ }\href@noop {} {\bibfield  {journal} {\bibinfo  {journal} {Sci.
  Adv.}\ }\textbf {\bibinfo {volume} {4}} (\bibinfo {year} {2018})}\BibitemShut
  {NoStop}%
\bibitem [{\citenamefont {Herriott}, \citenamefont {Kogelnik},\ and\
  \citenamefont {Kompfner}(1964)}]{64Herriott}%
  \BibitemOpen
  \bibfield  {author} {\bibinfo {author} {\bibfnamefont {D.}~\bibnamefont
  {Herriott}}, \bibinfo {author} {\bibfnamefont {H.}~\bibnamefont {Kogelnik}},
  \ and\ \bibinfo {author} {\bibfnamefont {R.}~\bibnamefont {Kompfner}},\
  }\bibfield  {title} {\enquote {\bibinfo {title} {Off-axis paths in spherical
  mirror interferometers},}\ }\href@noop {} {\bibfield  {journal} {\bibinfo
  {journal} {Appl. Opt.}\ }\textbf {\bibinfo {volume} {3}},\ \bibinfo {pages}
  {523--526} (\bibinfo {year} {1964})}\BibitemShut {NoStop}%
\end{thebibliography}%

\end{document}